\documentclass[12pt]{article}

\usepackage{natbib}
\usepackage[english]{babel}
\usepackage{csquotes}
\usepackage{setspace}
\usepackage{booktabs}
\usepackage{multirow}
\usepackage{multicol}
\usepackage{tabu}
\usepackage[T1]{fontenc}
\usepackage{amsmath}
\usepackage{amsthm}
\usepackage{XCharter}
\usepackage[charter]{newtxmath}
\usepackage{bbm}
\usepackage{microtype}
\usepackage{enumerate}
\usepackage{tikz}
\usetikzlibrary{arrows}
\usepackage[labelfont=bf,format=plain,justification=raggedright,singlelinecheck=false]{caption}
\usepackage{subcaption}
\usepackage{makecell} % in preamble
\usepackage{tabularx}
\usepackage{siunitx}
\usepackage{adjustbox} % For adjusting the box size

\usepackage{comment}

\usepackage[a4paper,top=2.54cm,bottom=2.54cm,left=2.58cm,right=2.58cm,marginparwidth=1.75cm]{geometry}

\usepackage{thmtools, thm-restate}
\usepackage{graphicx}
\usepackage[colorlinks=true, allcolors=blue]{hyperref}
\usepackage{floatrow}
\usepackage{float}
\usepackage{longtable}

\DeclareNewFloatType{mygame}%
	{name=Game}

\makeatletter
\renewenvironment{abstract}{%
    \if@twocolumn
      \section*{\abstractname}%
    \else %% <- here I've removed \small
      \begin{center}%
        {\bfseries \Large\abstractname\vspace{\z@}}%  %% <- here I've added \Large
      \end{center}%
      \quotation
    \fi}
    {\if@twocolumn\else\endquotation\fi}
\makeatother

\theoremstyle{plain}
\newtheorem{theorem}{Theorem}

\newtheorem{lemma}{Lemma}
\newtheorem{prop}{Proposition}
\theoremstyle{definition}
\newtheorem{definition}{Definition}[section]
\theoremstyle{remark}
\newtheorem*{remark}{Remark}

\theoremstyle{definition}
\newtheorem{assumption}{Assumption}
\newtheorem*{assumption*}{Assumption}
\newtheorem*{linhu}{Theorem 1 of \citet{Lin2024}}

\newenvironment{customproof}[1]{%
  \par\noindent\textit{Proof of #1:} \rmfamily\ignorespaces
}{\hfill$\square$\par}

\begin{document}

\title{When the Norm is Wrong: 
\\ Social Norm Misspecification and Targeting with an Application to Microfinance Adoption}

\author{Shantanu Chadha\thanks{Department of Economics, University of Warwick. E-mail: \url{Shantanu.Chadha@warwick.ac.uk}.}}

% Generate title
\maketitle

\begin{abstract}
Structural models of peer effects typically capture peer influence through the average, aggregate, or in some cases through extreme peer actions. Such specifications, however, are imposed ex ante and are typically ad hoc. We study the consequences of social norm misspecification by extending the constant elasticity of substitution (CES) norm from continuous to discrete choices. Considering a binary-choice game with incomplete information, we derive results on existence and uniqueness of a rational expectations equilibrium and establish conditions for identification of the structural parameters. We estimate the model by the Nested Pseudo-Likelihood estimator, implemented using the spectral algorithm. Monte Carlo simulations show that the CES estimates of the peer effects approach their true values as the sample grows, whereas imposing a linear norm biases them even in large samples. We apply the framework to microfinance adoption in India and find a strongly max-leaning norm, with significant heterogeneity in curvature and peer effects across villages grouped by caste composition and wealth. In a counterfactual exercise that allocates a fixed subsidy budget to maximise adoption, the estimated and linear norms target very different households, with the former favouring village ``leaders'', who are more likely to adopt and were the first to be informed about microfinance.
\end{abstract}

\textbf{Keywords}: Peer Effects, Social Norms, Misspecification, Discrete Choice

\newpage
\doublespacing

\section{Introduction}\label{S1}

Economic and social interactions are rarely confined to choices of isolated agents. Whether it be firms, households, or individuals, an agent’s actions often generate indirect costs or benefits that shape the incentives and behaviours of others. Peer effects are one important class of such externalities, generated when agents interact through a social or economic network that mediates the flow of information, influence or norms that shape their decisions. Substantial empirical evidence supports their importance.\footnote{For example, \citet{mas2009peers} document peer effects in worker productivity and \citet{calvo2009peer} in educational performance. See \citet{bramoulle2020peer} for a recent survey.} 

A particularly salient class of problems with peer effects involves discrete choices, such as whether to adopt microfinance \citep{banerjee2013}, smoke \citep{nakajima2007}, buy a new car \citep{joshua2022}, take up disaster insurance \citep{Cai2015} or install solar panels \citep{bollinger2012peer}. At the same time, discrete choice models raise complications that differ from those with continuous outcomes, from existence and uniqueness of an equilibrium to identification and estimation. The policy relevance of these choices nevertheless motivates incorporating peer effects into structural models, enabling counterfactual and welfare analysis. Any model-based prediction, however, depends critically on correctly specifying model primitives. Omitting or misspecifying peer effects where they exist can therefore bias policy recommendations.\footnote{For instance, \citet{yakovlev2018} studies peer influence and habit formation in alcohol consumption, finding that accounting for peer effects makes demand more elastic and more responsive to tax increases.}

Typically, structural models capture peer effects through a benefit from \emph{complementarity} between an agent's own action and the social norm, and a \emph{conformity} cost from deviating from said norm.\footnote{The most  general form, as studied in \citet{liu2014}, \citet{lambotte2025peer} and \citet{boucher2024peer}, is
\[
U_i\bigl(y_i, y_{-i}, \mathbf{G}\bigr) 
= \alpha_i\,y_i 
+ \underbrace{\theta_1\,y_i\,\tilde{y}_{-i}}_{\text{Complementarity}} 
- \tfrac{1}{2}\underbrace{\theta_2\bigl(y_i - \tilde{y}_{-i}\bigr)^2}_{\text{Conformity}}.
\]
$\alpha_i$ captures the intrinsic utility of choosing action $y_i$, while $\tilde{y}_{-i}$ represents the local social norm determined by the actions of others ($y_{-i}$) and the network $\mathbf{G}$. $\theta_1$ and $\theta_2$ capture complementarity and conformity.} Two kinds of misspecification can arise in such models. The first concerns the functional form of
the utility, usually lacking an axiomatic foundation,\footnote{\citet{chambers2022}, for instance, provide axioms embedding the linear peer effects model within the Luce framework.} while the second concerns the specification of the social norm itself. This paper focuses on the latter, studying how peer influence is aggregated in discrete choice settings.

Since \citet{manski1993identification}, most peer effect studies assume that peer influence is captured by the mean or aggregate behaviour of the group \citep{boucher2016}. Alternative specifications use extreme peer actions—for example, the maximum \citep{Carrell2010}, the minimum \citep{Hahn2020Friendship}, or both \citep{Tatsi2015},\footnote{See also \citet{BrockDurlauf2001b} and \citet{blume2010} for related peer effect models.} but there is little \emph{a priori} reason to prefer any one. For continuous outcomes, recent work accommodates more flexible norms, with \citet{boucher2024peer} adopting a CES aggregator and \citet{Tchuente2026} treating the aggregator itself as the object of identification. The assumption of a linear norm, however, persists and remains ad hoc even in discrete choice models, where the analysis for continuous outcomes does not carry over directly.

The main contribution of this paper is to extend the CES aggregator of \citet{boucher2024peer},\footnote{Their social norm is represented by the constant elasticity of substitution (CES) norm given by:
\begin{align}\label{eq:intro_social_norm}
    \tilde{y}_{-i}(\beta) = \left( \sum_{j=1}^{N} \tilde{g}_{ij} y_j^\beta \right)^{\frac{1}{\beta}}.
\end{align}
Here, $\tilde{g}_{ij}$ denotes the $ij^{\text{th}}$ element of the row-normalised adjacency matrix. When $\beta = 1$, the norm reduces to the average, while extreme values ($\beta \to \pm \infty$) imply that the highest or lowest action in an agent’s neighbourhood determines the norm.} nesting the mean, maximum, and minimum norms, to discrete choice models with both complementarity and conformity. Consistent with other discrete choice models of peer effects, we define a simultaneous game of incomplete information in which agents form rational expectations over their neighbours' actions and best-respond to those beliefs \citep{BrockDurlauf2001,lee2014binarychoice}. We present results that address the specific theoretical and econometric challenges that discrete choice raises for equilibrium, identification and estimation.

Existence and uniqueness of equilibrium are essential for yielding consistent and empirically meaningful counterfactual predictions without an equilibrium selection rule. Furthermore, multiple equilibria may also hamper identification and estimation. Unlike similar models, our system is neither linear nor, in general, a contraction, and lacks global curvature. Therefore, standard arguments such as those in \citet{boucher2024peer,lambotte2025peer} fail to apply directly. We instead impose supermodularity\footnote{A supermodular game is one in which actions are strategic complements; see \citet{Topkis1998}.} for existence, and use the order structure of the set of fixed points together with a bound on the elasticity of best responses to the norm for uniqueness. Under this bound, best responses move less than proportionally with the norm, making relative differences in beliefs contract. In this regard, our proof techniques are novel in the peer effects literature, providing an alternative methodology to \citet{boucher2024peer}. 

While models with continuous outcomes face the reflection problem of \citet{manski1993identification}, since outcomes and the norm are determined simultaneously, the non-linearity of the incomplete-information structure breaks this link. On the other hand, the CES aggregator introduces a countervailing difficulty for global identification. Using the observational equivalence strategy (see \citealp{BrockDurlauf2007} and \citealp{lambotte2025peer}), we show that under rank conditions and variation in neighbours’ covariates, discrete or continuous, the CES elasticity, and hence the full parameter vector, is point identified.\footnote{In the absence of peer effects, the curvature is not identified so the model is only partially identified.} 

Lastly, we jointly estimate the unobserved equilibrium choice probabilities and the parameters as a fixed point of the system using the nested pseudo-likelihood (NPL) estimator of \citet{aguirregabiria2007sequential}.\footnote{The algorithm has been used in peer effect applications like \citet{Lin2024} and \citet{lambotte2025peer}.} Despite guarantees of a unique equilibrium for a fixed vector of parameters, the joint fixed point in conditional choice probabilities and parameters need not be unique, nor is convergence to a fixed point guaranteed. Together with the absence of a contraction mapping in our setting, this complicates estimation. To address these issues, we adopt the spectral algorithm of \citet{Aguirregabiria2021}, which overcomes challenges with convergence without arbitrary tuning parameters (as in \citealp{Kasahara2012}) and without substantially increasing the computational burden.

We run Monte Carlo simulations to assess the finite-sample properties of the estimator under different CES elasticities and adoption rates. We find that imposing linearity distorts the composition of peer effects and this bias persists as the sample grows. In the largest sample, the restricted model under-states conformity in every design, by up to about $50\%$, while complementarity is underestimated by about $50\%$ when the norm is min-like and overestimated between $27-93\%$ when it is max-like. Under the CES norm, the peer effect estimates approach their true values as the sample grows although the elasticity itself is very noisy with nearly flat likelihood in small samples. For extreme elasticities, the estimator cannot pin down the elasticity but still recovers the correct regime and the peer effects for large samples, particularly for the max-like norm.

We apply our framework to the microfinance-adoption data of \citet{banerjee2013}. Leading evaluations of microcredit find only modest average effects \citep{banerjee2015miracle,banerjeekarlanzinman2015}, shifting attention towards increasing take-up, and adoption spreads through village networks, so the shape of the norm matters for which households a programme should target. While \citet{banerjee2013} model the diffusion of information, we, like \citet{lambotte2025peer}, structurally model adoption as an equilibrium choice, with peer effects entering households' utility directly. The data reject the linear norm ($p=0.03$) in favour of a strongly max-leaning one and imposing linearity distorts the peer mechanism. It overstates complementarity more than threefold, with non-overlapping confidence intervals, and understates conformity by about $25\%$, with overlapping ones. For a household of average degree, the peer effects are about $9\%$ larger under the linear norm.

The villages in the dataset are structurally distinct,\footnote{No cross-village ties exist and the villages differ systematically in caste composition and wealth.} and imposing a single norm and peer parameters ignores important dimensions of heterogeneity in caste composition and wealth.\footnote{\citet{munshirosenzweig2006} show that caste networks are relevant for economic interactions in India. The social norm therefore plausibly varies with the caste composition of a neighbourhood, while wealth governs responsiveness to the product.} Partitioning villages by caste (proportion of Scheduled Castes/Tribes) and wealth (bed(s)/latrine ownership), we estimate segment-specific norms and peer effects. Under one partition (bed ownership), the norm lines up with wealth, max-leaning in richer villages and min-leaning in poorer ones. The peer parameters echo this pattern, with poorer villages showing stronger complementarity and weaker conformity. A plausible explanation is that adoption is less risky for richer households, therefore, one informed neighbour suffices, whereas poorer households may require many adopting peers to participate. These patterns, including how conformity varies with caste, are sensitive to the wealth proxy.

Using the estimated model, we then ask \emph{whom} a planner should target, characterising the allocation of a fixed subsidy budget that maximises aggregate expected adoption. The planner funds the households whose subsidy raises adoption most, directly and through network feedback, and equates these marginal returns across recipients. The marginal return is a generalised Katz--Bonacich centrality in the Jacobian of the norm, similar to \citet{ballester2006} when the norm is linear but weighted toward each household's most- or least-adopting neighbours as the norm becomes max- or min-leaning. The shape of the norm therefore governs \emph{whom} to subsidise, and by how much.

\begin{figure}[!ht]
  \centering
  \includegraphics[width=\linewidth]{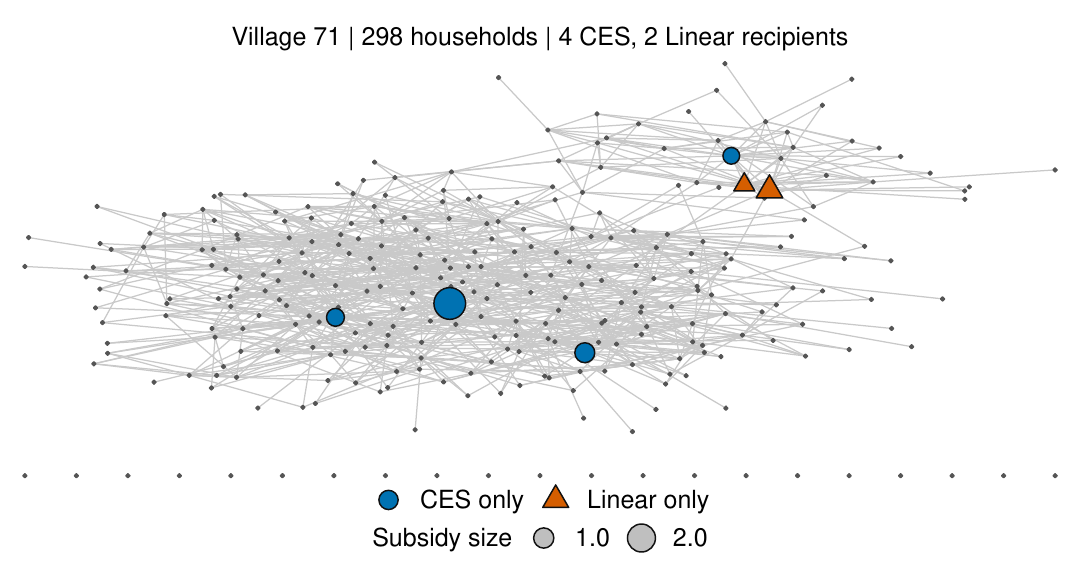}
  \caption{Optimal subsidy recipients under the estimated CES and linear norms in an
  illustrative village.}
  \label{fig:pooled-village-subsidy-comparison}
  \begin{minipage}{\linewidth}
    \footnotesize
    \textit{Notes:} Village 71 ($298$ households), chosen to make the contrast visible. The estimated norm is strongly max-leaning ($\hat\beta\approx491$) and the linear norm imposes $\beta=1$. Each model uses its own estimates and equilibrium to allocate a common budget of $100$ utility-index units. The two norms fund disjoint sets for this village, four households under CES and two under the linear norm, with no overlap, and yield different incremental adoption rates.
  \end{minipage}
\end{figure}

Comparing the optimal allocation under the estimated versus linear norm, we find that the two fund almost disjoint sets of households, with their rankings negatively correlated (Spearman $-0.52$). Figure \ref{fig:pooled-village-subsidy-comparison} presents this difference clearly in an illustrative village, where the two norms share no common recipients. Moreover, the linear norm spreads the budget thinly, whereas the estimated max-leaning norm concentrates it on village ``Leaders'' ($62\%$ of its unique recipients, against $12\%$ under the linear norm). These are exactly the households that \citet{banerjee2013} find central to information diffusion, and that imposing linearity would miss.  Misspecifying the norm or ignoring underlying heterogeneity can therefore misdirect the intervention, targeting the wrong households and incorrectly forecasting the changes in adoption it achieves.

The rest of the paper is organised as follows: Section \ref{S_lit} provides a brief review of literature, while Section \ref{S_model} describes the model and the theoretical results of the underlying game. Section~\ref{S_id} covers the data structure and identification, Section~\ref{S_est} presents the estimation procedure and Monte Carlo evidence for its finite sample performance. Finally, Section \ref{S_emp} presents the empirical results while optimal targeting policy is presented in Section \ref{S_seed} before we conclude in Section \ref{S_con}.

\section{Related Literature}\label{S_lit}

The availability of network data and advances in network econometrics have led to a steady increase in both theoretical and empirical models of peer effects  \citep[see][for a survey]{bramoulle2020peer}. This work belongs to the strand that models peer effects with discrete choices. Early theoretical models of peer effects include \citet{BrockDurlauf2001b} and \citet{BrockDurlauf2001}, who study binary-choice under homogeneous beliefs, extended by \citet{lee2014binarychoice} to heterogeneous beliefs that depend on peers' characteristics. More recently, \citet{boucherbramoulle2026binary} establish coherency and micro-foundations for linear interactions in binary outcomes under incomplete information. \citet{liu2014}, \citet{boucher2024peer}, and \citet{lambotte2025peer} have explored identification of more general utility functions incorporating conformity and complementarity effects. \citet{volpe2025discrete} generalises the discrete choice model of \citet{BrockDurlauf2001} to let interactions differ across types. Recent work by \citet{houndetoungan2026count} models count outcomes as a game of incomplete information, identified through friends of friends, while \citet{lambotte2025heterogeneity} lets conformity in binary outcomes depend on the agent's own action. Our generalisation is orthogonal, wherein we keep a common interaction structure but generalise the social norm itself for discrete choice models.

The vast majority of the empirical literature relies on a linear social norm \citep[e.g.,][]{calvo2009peer, yakovlev2018, guerra2022, lambotte2025peer}, under which an agent's norm is the mean or aggregate behaviour of their peers. Some studies instead use the maximum peer action \citep{Carrell2010, tao2014, diaz2021, islam2021}, the minimum \citep{bietenbeck2020, Hahn2020Friendship}, or both \citep{hoxby2005, Tatsi2015}, but these are case-specific choices rather than a general theory \citep[see][for overviews]{BrockDurlauf2001b, blume2010}. \citet{houndetoungan2025quantile} lets payoffs depend on quantiles of peer outcomes, and \citet{herstad2026heterogeneous} make peer influence rank-dependent; both allow non-monotonic peer influence but concern continuous outcomes. \citet{boucher2024peer} propose a non-linear CES social norm, governed by an elasticity parameter, that nests averaging and extreme behaviour for continuous outcomes. We extend the CES social norm to discrete choice peer effect models, combining the two strands of literature. Similar to \citet{boucher2024peer} and \citet{lambotte2025peer}, we model spillover and conformity effects and give conditions for their separate identification.

\citet{Tchuente2026} treats the peer aggregator, including the CES power mean, as a central primitive. Its identification, estimation and application concern continuous outcomes, although the binary-choice extension (Section~2.6.3 and Appendix~A.3) proves uniqueness with Lipschitz bounds that shrink differences in probability levels, which for CES depend on curvature and require probabilities bounded away from zero. We instead control proportional differences through a bound on the best-response elasticity, exploiting the monotonicity and degree-one homogeneity of the CES norm, and develop identification and estimation, which \citet{Tchuente2026} does not. Our contribution therefore is the complete structural and econometric development of the CES norm for binary decisions under incomplete information and application to a relevant setting.

In line with \citet{BrockDurlauf2001}, our norm is belief-based: agents best-respond to an aggregation of their beliefs about peers' choices, not to the expectation of an aggregate of realised actions. Since utility is belief-dependent, the model belongs to the literature on psychological games and belief-dependent motivations \citep[surveyed in][]{BattigalliDufwenberg2022}, building on \citet{GeanakoplosPearceStacchetti1989} and \citet{BattigalliDufwenberg2009}. The equilibrium is a psychological Nash equilibrium, in which the belief-dependent payoffs are evaluated at correct beliefs, equivalently a self-consistent rational-expectations equilibrium. Imposing this self-consistency while letting beliefs enter utility directly parallels the personal equilibrium of \citet{koszegi2006reference} in individual decision theory, and gives the conformity motive a behavioural foundation for which \citet{charness2019opportunistic} provide experimental evidence. Consistent with a causal role for perceived norms, \citet{BursztynGonzalezYanagizawaDrott2020} show experimentally that correcting beliefs about others' attitudes shifts economically relevant choices.

Estimation of non-linear models with conditional choice probabilities defined as fixed points of best responses has been studied in \citep{aguirregabiria2007sequential, Pesendorfer2008, Pesendorfer2010, Kasahara2012, Aguirregabiria2021} among others with previous application to the peer effect models such as \citep{Lin2024, lambotte2025peer}. We leverage results from \citep{Menzel2016, Lin2024} on the asymptotic properties of this estimator in the peer effect setting. However, unlike existing work, we employ the spectral algorithm of \citet{Aguirregabiria2021}, which overcomes convergence issues of \citet{aguirregabiria2007sequential} without requiring the selection of an arbitrary contraction parameter, as in \citet{Kasahara2012}. To the best of our knowledge, this represents the first application of this algorithm to a peer effects model.

Finally, we contribute to the literature on targeting interventions in networks \citep{ballester2006,galeotti2020targeting}. Whereas existing work studies targeting under local-average or asymmetric conformity \citep{meng2025,houndetoungan2026asymmetries} or the seeding of information \citep{banerjee2019gossip,beaman2021network}, we characterise adoption-maximising subsidies under a non-linear CES norm, showing how its curvature changes whom to target. We find that the CES-based allocations disproportionately target village leaders, the households informed first in \citet{banerjee2013}, while imposing a linear norm changes the recommended recipients.

\section{Binary-Choice Peer Effects Model with CES Social Norm}\label{S_model}

\subsection{Setting}

Let there be $N$ individuals who interact over a known and fixed network $\mathbf{G}(V, E)$, where $V$ is the set of nodes and $E$ the set of edges. The associated $N \times N$ adjacency matrix is $\mathbf{G} = [g_{ij}]$. We assume the network is undirected ($g_{ij} = g_{ji}$) and unweighted, with $g_{ij} \in \{0,1\}$ indicating whether agents $i$ and $j$ are connected.\footnote{Our framework can be extended to directed and weighted networks.}  For each agent $i$, let $N_i = \{j \mid g_{ij} = 1\}$ denote the set of neighbours, excluding self-loops ($i \notin N_i$). The degree of node $i$ (number of connections) is given by $d_i = |N_i|$, the cardinality of $N_i$. We also define the row-normalised adjacency matrix $\tilde{\mathbf{G}} = [\tilde{g}_{ij}]$, where
\[
\tilde{g}_{ij} = 
\begin{cases} 
\dfrac{g_{ij}}{d_i}, & d_i > 0, \\
0, & d_i = 0,
\end{cases}
\]
and the diagonal degree matrix $\mathbf{D}$ with entries $D_{ii} = d_i$.

We consider a simultaneous-move game with incomplete information. Each agent knows their own type but does not observe the types of other players, including their neighbours. This setup follows the standard discrete choice literature on network games \citep[e.g.,][]{BrockDurlauf2001,lee2014binarychoice}.

The action space of each agent is binary, i.e. $Y_i := \{0,1\}$. Here, $y_i = 1$ corresponds to the ``high'' action (example, adopting a new technology), while $y_i = 0$ corresponds to the ``low'' action (example, non-adoption). An outcome of the game is a strategy profile $y = (y_1, \dots, y_N)' \in Y = \times_{i=1}^N Y_i$, with $y_{-i}$ denoting the actions of all players except agent $i$. The utility function, in line with \citet{lambotte2025peer} and \citet{boucher2024peer}, incorporates two distinct social interaction effects: (i) \emph{complementarity} capturing strategic spillovers, and (ii) \emph{conformity} capturing the pressure to match peers' actions.

Formally,\footnote{Unlike the continuous case of \citet{boucher2024peer}, separate identification of the two effects in discrete choice models relies on scaling the complementarity term by $d_i$. This normalisation follows \citet{lambotte2025peer}. In the continuous case, isolated individuals can help identify them (see \citet{boucher2024peer}), but this strategy fails in discrete models due to scale and location normalisations for the distribution of shocks.}
\begin{align}\label{eq:pay-offs}
    U_i\bigl(y_i, y_{-i}, \mathbf{G}\bigr) \;=\; \alpha_i\,y_i \;+\;
\underbrace{ \theta_1 d_i\,y_i\,\tilde{y}_{-i}}_{\text{Complementarity}} 
\;-\;\frac{1}{2}\underbrace{\Bigl[\theta_2 \,\bigl(y_i - \tilde{y}_{-i}\bigr)^2\Bigr]}_{\text{Conformity}}
\;+\; \epsilon_i(y_i),
\end{align}
where $\epsilon_i(y_i)$ is an idiosyncratic preference shock.  

For isolated nodes (no neighbours, i.e.\ $d_i = 0$), the utility reduces to
\begin{align}\label{eq:pay-offs_iso}
    U_i\bigl(y_i, y_{-i}, \mathbf{G}\bigr) \;=\; \alpha_i\,y_i + \epsilon_i(y_i).
\end{align}
Here $\alpha_i$ captures the intrinsic utility of agent $i$ from selecting the high action ($y_i=1$).\footnote{This intrinsic term depends on observable individual characteristics, group-level factors, and correlated effects; we will specify the exact functional form when turning to the empirical model.} Peer influence is captured by the \emph{social norm function}, which we define as the CES norm of \citet{boucher2024peer}, allowing for flexible aggregation of peer behaviours:
\begin{align}\label{eq:social_norm}
    \tilde{y}_{-i}(\beta) \;=\; \left( \sum_{j=1}^{N} \tilde{g}_{ij}\, y_j^\beta \right)^{\frac{1}{\beta}}.
\end{align}
 When $\beta = 1$, it reduces to the usual linear norm and captures the proportion of peers that adopt the high action. For $\beta < 0$, it is defined as:
\begin{align}
    \tilde{y}_{-i}(\beta) = 
\begin{cases} 
\left(\sum\limits_{j=1}^N \widetilde{g}_{ij} y_j^{\beta}\right)^{\frac{1}{\beta}}, & \text{if } y_j > 0 \;\;\forall j \in N_i, \\[6pt]
0, & \text{if } y_j = 0 \;\;\text{for some } j \in N_i.
\end{cases}
\end{align}

Here $\epsilon_i(y_i)$ represents the private information of agent $i$, which is observed only by them and not by other players (or by the econometrician). We assume that the distribution of types is common knowledge, i.i.d.\ across players, and that the network $\mathbf{G}$ is exogenous.\footnote{Unlike \citet{lambotte2025peer}, we treat the network as exogenous, since our focus is on the consequences of misspecifying the norm on peer effect  estimates and counterfactuals.} 
\begin{assumption}[A1]\label{asum:A1}
We assume that $\epsilon(y_i=0)$ and $\epsilon(y_i=1)$ are i.i.d.\ Gumbel random variables with cumulative distribution function $G(x) = \exp\!\bigl(-\exp(-x)\bigr).$
Moreover, we assume that $\mathbf{G}$ and $\alpha$ are independent of the idiosyncratic shocks $\epsilon(y_i) \ \forall i$.
\end{assumption}

The public information of the game therefore, is the vector of intrinsic utilities $\boldsymbol{\alpha} = (\alpha_1,\dots,\alpha_N)$ together with the network $\mathbf{G}$, i.e.\ $\overline{\mathcal{I}} = \{\boldsymbol{\alpha}, \mathbf{G}\}$. The private information set of agent $i$ additionally includes their own preference shocks, $\mathcal{I}_i = \{\boldsymbol{\alpha}, \mathbf{G}, \epsilon_i(y_i)\}$. In line with the literature on discrete choice games with incomplete information, we assume that players form rational expectations about the actions of others and maximise utility given these beliefs. This implies that agents base their decisions on their beliefs about others’ actions instead of the realised actions themselves. Furthermore, we impose the consistency requirements on these beliefs as in \citet{BrockDurlauf2001, lee2014binarychoice, lambotte2025peer} that ensure that these rational expectations of agents are in fact correct in equilibrium.
\begin{assumption}[A2]\label{asum:A2}
Each agent forms rational expectations about others’ actions conditional on their private information. That is, agent $i$’s belief about the probability that agent $j$ chooses the high action is denoted by $p_j(\mathcal{I}_i) \in [0,1]$.
\end{assumption}
Given Assumption~\ref{asum:A1}, the shocks $\epsilon_i(y_i)$ are independent of both $\alpha_i$ and the network $\mathbf{G}$. Hence, agent $i$’s belief about player $j$’s action depends only on the public information set, i.e.\ $p_j(\mathcal{I}_i) = p_j(\overline{\mathcal{I}})$. It follows that all players hold the same belief about player $j$’s action conditional on public information, and we denote this common expectation by $p_j$. We further assume that the norm itself is evaluated at these beliefs.
\begin{assumption}[A3]\label{asum:A3}
Let $p_j$ be the common belief about $j$'s choice, then the \emph{expected social norm} for agent $i$ is
\[
\tilde{y}^e_{-i}(\beta) := \left( \sum_{j=1}^N \tilde{g}_{ij} \, p_j^\beta \right)^{1/\beta},
\]
and agent $i$ maximises the belief-dependent utility,
\begin{align}\label{eq:exp_ut}
U^e_i\bigl(y_i, \mathbf{p}_{-i}, \mathbf{G}\bigr) \;=\; \alpha_i\,y_i \;+\;
\theta_1 \, d_i \, y_i \, \tilde{y}^e_{-i}(\beta)
\;-\;\tfrac{1}{2}\theta_2 \bigl(y_i - \tilde{y}^e_{-i}(\beta)\bigr)^2
\;+\;\epsilon_i(y_i),
\end{align}
where $\mathbf{p}_{-i} = (p_j)_{j \neq i}$ is the vector of common beliefs about the other players.
\end{assumption}
\begin{remark}
Evaluating the norm at beliefs is distinct from taking the expectation of the norm of realised actions. For $\beta\neq1$, $\tilde{y}^e_{-i}(\beta) \neq \mathbb{E}[\,\tilde{y}_{-i}(\beta)]$ (Jensen's inequality), and the two coincide only in the linear case $\beta = 1$. Utility therefore depends directly on beliefs about others' actions.\footnote{\citet{charness2019opportunistic} provide experimental evidence of agents responding to their beliefs about peer behaviour with conformity effects. This places the game among models with belief-dependent payoffs, namely psychological games \citep{GeanakoplosPearceStacchetti1989, BattigalliDufwenberg2009} \citep[see][for a survey]{BattigalliDufwenberg2022} and reference-dependent preferences \citep{koszegi2006reference}, whose equilibrium requires those beliefs to be correct.} Appendix~\ref{SC} presents a more formal discussion on this assumption.
\end{remark}

The parameter $\beta$ therefore governs how much weight the norm places on the peers most and least likely to adopt. The linear and the extreme norms then become special cases depending on the value of $\beta$:
\[
\underbrace{\min_{j\in N_i}p_j}_{\beta\to-\infty}
\;\le\;
\underbrace{\prod_{j}p_j^{\tilde g_{ij}}}_{\beta\to0}
\;\le\;
\underbrace{\sum_{j}\tilde g_{ij}\,p_j}_{\beta=1}
\;\le\;
\underbrace{\max_{j\in N_i}p_j}_{\beta\to+\infty}.
\]
At $\beta=1$ the norm is the believed share of adopting peers. As $\beta\to+\infty$ the peer most likely to adopt sets the norm, and a single likely adopter can sway the neighbourhood. As $\beta\to-\infty$ the norm is high only when all peers are likely to adopt, as when collective agreement is required. Estimating $\beta$ therefore avoids choosing an aggregator ex ante.\footnote{Evaluating the norm at beliefs is what keeps $\beta$ meaningful. For realised binary actions, $y_j^\beta=y_j$, so the CES of realised actions is just a power of the adoption share, and for $\beta\le0$, it only records whether every peer adopts or not. Because beliefs lie strictly in $(0,1)$, $p_j^\beta$ varies with $\beta$, and different curvatures generate distinguishable norms.}

Furthermore, to separate the two peer effect mechanisms, we scale the complementarity term by each agent's degree. At $\beta=1$, $d_i\tilde y^e_{-i}=\sum_j g_{ij}p_j$ is the expected number of $i$'s neighbours who adopt, and in general $d_i\tilde y^e_{-i}(\beta)$ is the effective number of peers expected to adopt. Complementarity $\theta_1$ is therefore the payoff from each expected adopting peer, whatever $i$'s degree \citep{lambotte2025peer}.

Finally, we assume that the maximum degree remains bounded, which controls the degree distribution as the network size grows.
\begin{assumption}[A4]\label{asum:A4}
The maximum degree of the network is bounded, i.e. there exists a $\bar d<\infty$, independent of $N$, such that $d_{\max}\le\bar d$, where $d_{\max} = \max_{1 \leq i \leq N} d_i$.
\end{assumption}
Because the utility function includes the complementarity term which is scaled by the degree, bounding the degree prevents the spillover term from scaling with $N$. Similar restrictions are common in the literature with \citet{Lin2024,lambotte2025peer} imposing bounded degree for existence and uniqueness, while \citet{dePaula2018} use it for identification of preferences.

\subsection{Best Responses}

For the remainder of this section we assume that there are no isolated individuals in the network. We relax this in the empirical analysis. Under Assumptions \ref{asum:A2} and \ref{asum:A3}, agents maximise their belief-dependent utility in \eqref{eq:exp_ut} by choosing the action that yields higher utility. In particular, agent $i$ chooses $y_i = 1$ if and only if
\[
    \Delta U^e_i \equiv U_i^e(1,\mathbf{p}_{-i}, \mathbf{G}) - U_i^e(0,\mathbf{p}_{-i}, \mathbf{G}) \geq 0.
\]
Solving further gives us,
\begin{align}\label{eq:br}
    \Delta U^e_i &= \alpha_i + \theta_1 d_i \tilde{y}^e_{-i}(\beta) - \frac{1}{2} [\theta_2 (1 + \tilde{y}^{e}_{-i}(\beta)^2  - 2 \tilde{y}^e_{-i}(\beta) )] + \epsilon_i(1) \nonumber\\
    &- \left[ - \frac{1}{2} \left[\theta_2 \tilde{y}^{e}_{-i}(\beta)^2   \right] + \epsilon_i(0) \right] \geq 0, \nonumber \\
    &= \alpha_i  +  \theta_1 d_i \tilde{y}^e_{-i}(\beta)   + \theta_2 (\tilde{y}^e_{-i}(\beta)  - \frac{1}{2})  -(\epsilon_i(0) - \epsilon_i(1)) \geq 0, \nonumber \\
    &\iff z_i(\mathbf{p}, \boldsymbol{\theta}) \geq \eta_i.
\end{align}
We define $\eta_i \equiv \epsilon_i(0) - \epsilon_i(1)$, which by Assumption~\ref{asum:A1} follows a logistic distribution with mean $0$, scale $1$, and cumulative distribution function $F_\eta(x)$. Since $\eta_i$ is independent of the public information by Assumption~\ref{asum:A1}, $i$'s conditional choice probability is given by:
\begin{align}\label{eq:choice_prob}
 p_i \equiv \Pr(y_i = 1 \mid \boldsymbol{\theta}, \overline{\mathcal{I}})
 = \Pr(\Delta U_i^e \geq 0)
 = \Pr\big(z_i(\mathbf{p}, \boldsymbol{\theta}) \geq \eta_i\big)
 = F_\eta\!\big(z_i(\mathbf{p}, \boldsymbol{\theta})\big).
\end{align}

\subsection{Existence and Uniqueness}

Let $\Gamma_i(\mathbf{p}; \boldsymbol{\theta})=F_\eta\big(z_i(\mathbf{p},\boldsymbol{\theta})\big)$, with $\Gamma_i:[0,1]^N \to [0,1]$, denote the best-response mapping. A rational expectations equilibrium of the game is defined as
\begin{definition}[Rational Expectations Equilibrium]\label{def:ree}
A Rational Expectations Equilibrium is a vector of beliefs $\mathbf{p}^*\in[0,1]^N$ such that, given these beliefs, each agent $i$ chooses $y_i=1$ if and only if $\Delta U^e_i\ge0$, and the beliefs are correct: $p_j^*=\Pr(y_j=1\mid\boldsymbol\theta,\overline{\mathcal{I}})$ for all $j$.
\end{definition}
Stacking across agents, an equilibrium is equivalently a fixed point of the system
\begin{equation}\label{eq:system}
\mathbf p^*=\Gamma(\mathbf p^*;\boldsymbol\theta)=\mathbf F_\eta\big(Z(\mathbf p^*)\big),
\end{equation}
where $Z(\mathbf p^*)=\boldsymbol\alpha+\theta_1\mathbf D\,\tilde{\mathbf y}^e(\beta)+\theta_2\big(\tilde{\mathbf y}^e(\beta)-\tfrac12\iota\big)$ and $\tilde{\mathbf y}^e(\beta)=\big(\tilde y^e_{-1}(\beta),\dots,\tilde y^e_{-N}(\beta)\big)'$ is the vector of stacked norms.

There is no guarantee that a rational expectations equilibrium exists or if it is unique. Both properties are essential for coherent empirical analysis. Moreover, multiple equilibria complicate identification and estimation, and make counterfactual predictions unreliable without an equilibrium selection mechanism.

Compared to standard problems, the mapping of conditional choice probabilities given by equation \eqref{eq:system} is neither linear nor a contraction. Depending on $\beta$, the derivative of the social-norm aggregator $\tilde{y}^e_{-i}(\beta)$ with respect to $p_j$ can be arbitrarily large, ruling out contraction mapping arguments commonly employed in such models. Instead, under Assumption~\ref{asum:A5} below, the best-response mapping, $\Gamma(\mathbf{p}; \boldsymbol{\theta})$, is monotone in $\mathbf{p}$. We then invoke Tarski’s fixed-point theorem, standard in supermodular games, to establish the existence of a rational expectations equilibrium.\footnote{We can get existence using Brouwer's fixed-point theorem which only requires continuity. However, Tarski's fixed-point theorem provides additional lattice structure that we exploit for uniqueness.}
\begin{assumption}[A5]\label{asum:A5}  
Let $d_{\min} = \min_{i: d_i > 0} d_i$ and $d_{\max} = \max_{i: d_i > 0} d_i$ denote the minimum and maximum degrees of non-isolated individuals in the network. We assume that  
\[
\min_{d \in [d_{\min}, d_{\max}]} \{\theta_1 d + \theta_2\} > 0.
\]  
\end{assumption}  
\begin{remark}  
Assumption \ref{asum:A5} requires peer effects to be \textit{strategic complements}, but it allows either mechanism to be negative. This is needed for the best-response mapping to be monotone, which places the model within the class of supermodular games, a structure commonly assumed in the literature (see, for example, \citet{boucher2024peer}). The assumption is natural in many socio-economic environments where individuals benefit from coordination, imitation, or conformity. Examples include adopting a new technology, consuming the same product brand, or choosing high effort in a task. At the same time, Assumption \ref{asum:A5} rules out settings with \textit{strategic substitutability}, such as competition for scarce resources or congestion effects.\footnote{\citet{volpe2025discrete} is a recent paper that studies heterogeneous peer effects where they allow for strategic substitutability and negative interactions showing that it can yield a unique but unstable equilibrium, or multiple equilibria.}
\end{remark}

Theorem~\ref{thm:existence} below presents results on the existence and on the structure of fixed points of the system of conditional choice probabilities.

\begin{theorem}[Existence of Rational Expectations Equilibrium]\label{thm:existence}
Let \( \mathbf{p} = (p_1, \dots, p_N) \in [0,1]^N \) be the vector of individual choice probabilities and
\[
\Gamma_i(\mathbf{p}; \boldsymbol \theta) = F_{\eta_i}\left( \alpha_i + (\theta_1 d_i + \theta_2) \tilde{y}^e_{-i}(\beta) - \frac{\theta_2}{2} \right).
\]
Under Assumptions~\ref{asum:A1}-\ref{asum:A5}, there exists a Rational Expectations equilibrium $\mathbf{p}^*$ of the generalised peer effects model with the CES social norm such that:
    \[
    \mathbf{p}^* = \Gamma(\mathbf{p}^*; \boldsymbol \theta) = \mathbf{F}_{\eta}(Z(\mathbf{p}^*)).
    \]
Furthermore, the set of fixed points is a complete lattice, with a least element $\underline{\mathbf p}^*$ and a greatest element $\overline{\mathbf p}^*$, such that $\underline{\mathbf p}^*\le\mathbf p^*\le\overline{\mathbf p}^*$ for every fixed point $\mathbf p^*$.
\end{theorem}
\begin{proof}
    Proof in Appendix Section \ref{SA.1}.
\end{proof}

\citet{bajari2010identification,bajari2015dynamic}, among others, handle multiple equilibria in discrete choice games by computing all of them and estimating an equilibrium selection mechanism. However, this becomes infeasible as the number of players and actions grows, since the parameter count rises faster than the available independent conditional choice probabilities. Identification and estimation then demand stronger restrictions.

The fixed point mapping in our model is neither a contraction,\footnote{It fails to be a contraction because its Jacobian,
\[
\frac{\partial \Gamma_i}{\partial p_j} = f_\eta(z_i)\,(\theta_1 d_i + \theta_2)\,\frac{\partial \tilde{y}^e_{-i}(\beta)}{\partial p_j},
\qquad
\frac{\partial \tilde{y}^e_{-i}(\beta)}{\partial p_j} = \tilde{g}_{ij}\left(\frac{\tilde{y}^e_{-i}(\beta)}{p_j}\right)^{1-\beta},
\]
need not have norm below one: its magnitude scales with the strength of peer effects $(\theta_1 d_i + \theta_2)$, which may be large, and the gradient of the CES aggregator diverges as a peer's belief $p_j \to 0$ whenever $0 \le \beta < 1$. Hence no uniform Lipschitz constant below one exists.} nor globally convex or concave.\footnote{Our mapping is contingent on the cumulative distribution function $F_{\eta}$ which is neither globally concave nor globally convex. This is true for most empirically relevant CDFs.} Instead, we use the lattice structure of the fixed points to establish uniqueness and provide an alternative proof to \citet{boucher2024peer}. A common sufficient condition for uniqueness in peer-effect models, both continuous \citep{liu2014,boucher2024peer} and discrete choice \citep{BrockDurlauf2001,lee2014binarychoice,lambotte2025peer}, is that peer effects be sufficiently small. We adopt a related condition, where instead of bounding the peer-effect parameters directly, we bound the elasticity of the best response with respect to the social norm over the range of norm values, which guarantees that the set of fixed points is a singleton.

We define the social norm mapping $\tilde{\mathbf{y}}^e(\cdot;\beta): [0,1]^N \rightarrow [0,1]^N$, that takes $\mathbf{p}$ and returns the social norm for each agent at fixed $\beta$. Writing the best response mapping directly as a function of the norm,
\[
\Gamma_i\big(\mathbf{p};\boldsymbol \theta\big) = \Gamma_i\big(\tilde y^e_{-i};\boldsymbol \theta\big) = F_\eta\!\left(\alpha_i + (\theta_1 d_i + \theta_2)\,\tilde y^e_{-i} - \tfrac{\theta_2}{2}\right),
\]
where $\tilde y^e_{-i} \equiv \tilde y^e_{-i}(\mathbf{p};\beta)$, we define its elasticity with respect to $i$'s norm as,
\begin{equation}\label{eq:log_elasticity}
    \varepsilon_i^{BR} := \frac{\partial\log\Gamma_i(\tilde y^e_{-i};\boldsymbol \theta)}{\partial\log \tilde y^e_{-i}}.
\end{equation}
For the logistic CDF, this elasticity is
\begin{equation}\label{eq:best_response_elasticity}
\varepsilon_i^{BR}=(\theta_1d_i+\theta_2)\,\tilde y^e_{-i}\,\Bigl[1-F_\eta\bigl(\alpha_i-\tfrac{\theta_2}{2}+(\theta_1d_i+\theta_2)\,\tilde y^e_{-i}\bigr)\Bigr].
\end{equation}

Since Theorem~\ref{thm:existence} delivers a least and a greatest fixed point, every equilibrium lies between them, and equilibrium norms lie in a compact sub-interval of $(0,1)$.\footnote{The CES norm is a power mean of the beliefs, so it lies between their smallest and largest values.} Because $\Gamma$ is increasing under Assumption~\ref{asum:A5} and continuous, iterating the best responses from $\mathbf 0$ and from $\mathbf 1$ bounds every equilibrium, $\Gamma^t(\mathbf 0)\le\mathbf p^*\le\Gamma^t(\mathbf 1)$ for every $t\ge1$, and the two sequences converge to the least and greatest fixed points as $t\to\infty$ \citep{Topkis1998}. A single iteration already gives an interval $[a,b]$ that contains every equilibrium norm and does not depend on $\beta$, with $a=\min_{j:\,d_j>0}F_\eta(\alpha_j-\tfrac{\theta_2}{2})$ and $b=\max_{j:\,d_j>0}F_\eta(\alpha_j+\theta_1d_j+\tfrac{\theta_2}{2})$. Assumption~\ref{asum:A6} bounds the best-response elasticity \eqref{eq:best_response_elasticity} uniformly over such an interval.
\begin{assumption}[A6]\label{asum:A6}
There exists a closed interval $[a,b]\subset(0,1)$ such that
\[
\big\{\, \tilde y^e_{-i}(\mathbf p^{*};\beta) \;:\; 1\le i\le N,\ \ \mathbf p^{*}=\Gamma(\mathbf p^{*};\boldsymbol \theta) \,\big\}\ \subseteq\ [a,b],
\]
and the best-response elasticity is uniformly below one on it,
\[
\overline\varepsilon^{BR} :=\max_{1\le i\le N}\ \sup_{\tilde y^e_{-i}\in[a,b]}\ \varepsilon_i^{BR} < 1.
\]
\end{assumption}
\begin{remark}
Assumption~\ref{asum:A6} requires each agent's best response to move less than proportionally to a proportional change in the expected social norm. It makes the map a contraction in log coordinates, equivalently the Thompson metric, which together with homogeneity of degree one of the CES norm gives uniqueness. Peer effects may still be large in levels but cannot generate strong enough proportional feedback to sustain two distinct fixed points. See Appendix \ref{SA.2} for a discussion.
\end{remark}

By \eqref{eq:best_response_elasticity}, $\varepsilon_i^{BR}$ is single-peaked in the norm. It rises until $(\theta_1d_i+\theta_2)\,\tilde y^e_{-i}F_\eta(z_i)=1$ and falls thereafter with its peak value at $e^{-z_i}$, with $z_i$ evaluated at the peak.\footnote{Since $F_\eta'=F_\eta(1-F_\eta)$, the derivative of \eqref{eq:best_response_elasticity} in the norm is $(\theta_1d_i+\theta_2)[1-F_\eta(z_i)][1-(\theta_1d_i+\theta_2)\tilde y^e_{-i}F_\eta(z_i)]$, whose last factor falls from one and so changes sign at most once. At the peak, $(\theta_1d_i+\theta_2)\tilde y^e_{-i}=1/F_\eta(z_i)=1+e^{-z_i}$, so $\varepsilon_i^{BR}=e^{-z_i}$ and $z_i-e^{-z_i}=\alpha_i-\tfrac{\theta_2}{2}+1$.} This gives sufficient conditions for uniqueness, two of which recover restrictions used in the literature.
\begin{itemize}
\item \emph{Floor-probability bound, $\min_i(\alpha_i-\tfrac{\theta_2}{2})>-2$.} The peak value is below one if and only if $\alpha_i-\tfrac{\theta_2}{2}>-2$, so under this bound the elasticity is below one at every norm and Assumption~\ref{asum:A6} holds for every $\beta$ and every strength of peer effects. It requires each agent to adopt with probability above $1/(1+e^2)\approx0.12$ when no peer is expected to adopt.
\item \emph{Below-one bound, $\max_i(\theta_1d_i+\theta_2)\,b<1$.} Since $\varepsilon_i^{BR}<(\theta_1d_i+\theta_2)\,\tilde y^e_{-i}\le(\theta_1d_i+\theta_2)\,b$ on $[a,b]$, it holds whenever $\max_i(\theta_1d_i+\theta_2)<1$, the discrete analogue of the restrictions imposed by \citet{boucher2024peer} for continuous outcomes.
\item \emph{Linear-norm bound, $\max_i(\theta_1 d_i+\theta_2)<4$.} For $\beta\ge1$, including the linear norm, the CES norm is non-expansive in levels. Therefore, since $\sup_z f_\eta(z)=\tfrac14$, this bound makes $\Gamma$ a contraction in levels, giving uniqueness. For any $\beta$, if also $\Gamma_i(\tilde y^e_{-i};\boldsymbol \theta)\ge\tilde y^e_{-i}$ on $[a,b]$, the bound gives $\varepsilon_i^{BR}\le(\theta_1 d_i+\theta_2)/4<1$, so Assumption~\ref{asum:A6} holds. It is exactly the coefficient restriction \citet{lambotte2025peer} use for the linear social norm.
\end{itemize}
Assumption~\ref{asum:A6} is weaker than the first two bounds, and than the third when $\Gamma_i(\tilde y^e_{-i};\boldsymbol \theta)\ge\tilde y^e_{-i}$ on $[a,b]$. We now state our uniqueness result.

\begin{theorem}[Uniqueness of Fixed Point]\label{thm:uniqueness}
Under Assumptions \ref{asum:A1}--\ref{asum:A6}, the Rational Expectations Equilibrium of the generalised peer effects model with the CES social norm given by the system in equation \eqref{eq:system}, \( \mathbf{p}^* \in [0,1]^N \) of \( \Gamma \), is unique.
\end{theorem}  
\begin{proof}
See Appendix Section \ref{SA.2}.
\end{proof}  
\begin{remark}
Assumptions \ref{asum:A5} and \ref{asum:A6} are sufficient but not necessary for existence and uniqueness. Uniqueness does not strictly require changes in best responses to be small. Under Assumption~\ref{asum:A5}, uniqueness can also be checked directly by iterating the mapping from $\mathbf 0$ and from $\mathbf 1$.\footnote{For our application for the pooled sample, the elasticity at the equilibrium norm exceeds one for some very high degree households (also fails for linear norm and condition of \citet{lambotte2025peer}). However, the two iterations nevertheless agree to within $10^{-10}$, ensuring equilibrium is unique. It does however fail under the estimated parameters for some segments, once we partition the villages.}
\end{remark}

\section{Identification}\label{S_id}

We have provided conditions under which there exists a unique equilibrium of the discrete choice model with the non-linear CES social norm. In this section, we provide results that help establish identification of the structural parameters under reasonable restrictions.

\subsection{Data Structure}

We assume that the intrinsic utility that agents derive from choosing the high action takes an additively separable form:
\[
\alpha_i = M_i \xi  + X_{i} \gamma,
\]
where $X_i$ collects agent $i$'s observable characteristics. In our application these are household covariates,\footnote{We omit contextual effects, the average characteristics $\tilde{G}_i\mathbf{X}$ of $i$'s peers. We assume peers' characteristics affect $i$'s adoption only indirectly, through their influence on peers' own adoption, which already enters $i$'s payoff via the social norm $\tilde{y}^e_{-i}$; they have no separate direct effect. This assumption may not hold in other situations and then identification variation needs to come from higher order neighbourhoods.} and they proxy for income and social standing of households. We assume that we observe an exogenous network with row-normalised adjacency matrix $\mathbf{\tilde{G}}$. In general, $\mathbf{\tilde{G}}$ may instead be weighted, letting the strength of interactions vary across pairs, or directed, capturing asymmetric relationships, although we require it to be exogenous. 

$M_i$ is the $i^{th}$ row of the selection matrix $\mathbf{M}$ ($N \times m$), which indicates the subnetwork (out of $m$ disjoint subnetworks) to which an individual belongs. These could be villages, schools or firms, so $M_i\xi$ is a subnetwork fixed effect, with $\xi_s$ capturing correlated effects common to all agents in subnetwork $s$, such as local infrastructure or institutions.

Thus, the structural utility for agent $i$ can be written as:
\begin{align}\label{eq:ident_main}
    z_i(\mathbf{p}) = M_i \xi + X_i \gamma  + \theta_1 d_i\tilde{y}^e_{-i}(\beta) + \theta_2 \left( \tilde{y}^e_{-i}(\beta) - \tfrac{1}{2} \right).
\end{align}
We now allow for the possibility of isolated individuals in the network:
\begin{align}\label{eq:ident_main_iso}
    z_i(\mathbf{p}) =  M_i \xi + 1\{d_i = 0\} \delta +  X_i \gamma + \theta_1 d_i\tilde{y}^e_{-i}(\beta)+  1\{d_i > 0\} \left(  \theta_2 \left( \tilde{y}^e_{-i}(\beta) - \frac{1}{2} \right) \right),
\end{align}
with $1\{d_i >0\}$ the indicator equal to one if agent $i$ has positive degree and zero otherwise.

There is one additional fixed effect that we add for isolated agents, $\delta$. For a connected agent, $\theta_2\bigl(\tilde y^e_{-i}-\tfrac12\bigr)=\theta_2\tilde y^e_{-i}-\tfrac{\theta_2}{2}$. Therefore, the constant $-\tfrac{\theta_2}{2}$ is present for agents with peers but absent for isolated ones. Without the additional dummy for isolated individuals, the model would assign baseline differences between connected and isolated individuals to $\theta_2$ which is behaviourally incoherent and conflates conformity with being connected. The dummy absorbs the level shift and identifies $\theta_2$ from variation in the social norm among connected agents instead of the connected–isolated gap. We present both specifications, with and without isolated-dummy, in the empirical section.  

Let $\mathbf{1}_{d=0}$ be the indicator vector for isolated nodes. In matrix form, the terms in \eqref{eq:ident_main_iso} that do not involve the social norm are $\mathbf M\xi+\mathbf 1_{d=0}\delta+\mathbf X\gamma$, with regressor matrix $[\,\mathbf M\ \ \mathbf 1_{d=0}\ \ \mathbf X\,]$. Here, what is observable is the network, the individual characteristics $X_i$, the selection matrix $\mathbf{M}$, and the realised choices $\mathbf{y}$. While actual realised actions are observed, we do not observe the conditional choice probabilities $\mathbf{p}$ directly. They are equilibrium objects determined by the parameters, the network and the covariates through the fixed point of \eqref{eq:system}. Moreover, this implies that the expected social norms $\tilde{\mathbf{y}}^e(\beta)$ are unobserved as well. Similarly, the preference shocks $\eta_i$ are unobservable to the econometrician for all $i$ and we only know the distribution of the shocks. Our aim is therefore to guarantee that we can uniquely identify $\boldsymbol \theta = (\xi, \gamma, \delta, \theta_1,\theta_2, \beta)' \in \boldsymbol \Theta$.

\subsection{Identification Strategy}

In contrast to the continuous outcomes case, which suffers from the reflection problem of \citet{manski1993identification} because of simultaneous determination of social and individual outcomes, discrete choice models avoid this issue due to the non-linearity introduced by the cumulative distribution function. Nevertheless, once a CES social norm is incorporated, the resulting non-linearities make identification non-trivial. We utilise variation in degree to separate the two peer effects and variation in neighbours' covariates to identify $\beta$.

While local identification in non-linear models can often be guaranteed by an appropriate rank condition on the Jacobian, global identification is usually more complex (see \citet{Rothenberg1971} and \citet{Bowden1973}). We instead establish global identification using the observational equivalence strategy \citep{BrockDurlauf2007,lambotte2025peer,lee2014binarychoice}. Under Assumptions~\ref{asum:A1} and~\ref{asum:A2}, the distribution of $\mathbf y$ given $(\mathbf G,\mathbf M,\mathbf X)$ is summarised by the conditional choice probabilities $p^*_i=\Pr(y_i=1\mid\mathbf G,\mathbf M,\mathbf X)$, which are therefore what the data identify. Two parameter vectors are then observationally equivalent if they generate the same conditional choice probabilities at every draw of $(\mathbf G,\mathbf M,\mathbf X)$ in the support of the joint distribution. We assume that the admissible parameter space contains $(\theta_1,\theta_2)=(0,0)$.

We prove two main results. When at least one of the peer effect parameters is non-zero, the model is point identified under either of two alternative sets of conditions on the observed covariates. One relies on a continuous covariate, while the other requires only a covariate taking at least three values. The second result is when both peer effect parameters are zero, $\beta$ remains unidentified, and we achieve only partial identification.

We first make the following weak assumption on the observables and parameters.
\begin{assumption}[A7]\label{asum:ident}
We assume that:
\begin{enumerate}
\item \textbf{Finite utility.} The intrinsic utility
$\alpha_i=M_i\xi+ 1\{d_i=0\}\delta +X_i\gamma$ is finite for every individual.
\item \textbf{Payoff relevance.} Some component $x$ of $X_i$ has a common
coefficient $\gamma_x\neq0$.
\item \textbf{Full rank.}
$[\,\mathbf M\ \ \mathbf 1_{d=0}\ \ \mathbf X\,]$ has full column rank.
\end{enumerate}
\end{assumption}
Finite utility ensures that equilibrium probabilities lie strictly in $(0,1)$, and that the norm is well defined for every $\beta$. Unlike typical discrete choice peer effect models, which assume full rank of a matrix that includes regressors built from the equilibrium choice probabilities \citep[e.g.,][]{lambotte2025peer}, we impose no such condition. Instead, payoff relevance ensures that variation in a neighbour's $x$ moves that neighbour's choice probability and, through it, the norm. The full-rank condition involves only the observed regressors and is standard. For simplicity, we assume that the population contains isolated individuals to keep $\delta$ meaningful. This is without loss as otherwise $\delta$ is dropped from $\boldsymbol \theta$, and $\mathbf 1_{d=0}$ from the full-rank condition.

We also assume that Assumptions~\ref{asum:A1}-\ref{asum:A4} hold for all supported configurations and Assumptions~\ref{asum:A5} and \ref{asum:A6} hold at every supported configuration over a common interval, when at least one of $\theta_1$ and $\theta_2$ is non-zero. We now state our main identification result.

\begin{theorem}[Identification]\label{thm:ident}
Let $x$ be the payoff-relevant covariate, $\mathbf X^{-x}$ the other covariates and, for a set $S$ of individuals, $\mathbf x_{-S}=(x_k)_{k\notin S}$. Under Assumptions \ref{asum:A1}-\ref{asum:ident}:
\begin{enumerate}[(i)]
\item \textbf{Point identification with a continuous variable.}
If $(\theta_1,\theta_2)\neq(0,0)$ and:
\begin{enumerate}[(a)]
\item some $i$ with $d_i\geq2$ is such that, for fixed $(\mathbf G,\mathbf M,\mathbf X^{-x},\mathbf x_{-N_i})$, the support of $(x_j)_{j\in N_i}$ contains a non-empty open set;
\item some $k$ with $d_k\neq d_i$ has a neighbour $j$ such that, for fixed $(\mathbf G,\mathbf M,\mathbf X^{-x},\mathbf x_{-j})$, the support of $x_j$ contains at least two points.
\end{enumerate}
Then $\boldsymbol \theta = (\xi, \gamma, \delta, \theta_1,\theta_2, \beta)'$ is point identified.
\item \textbf{Point identification with a discrete variable.}
If $(\theta_1,\theta_2)\neq(0,0)$ and:
\begin{enumerate}[(a)]
\item some $i$ with $d_i=1$ has a neighbour $j$ such that, for fixed $(\mathbf G,\mathbf M,\mathbf X^{-x},\mathbf x_{-j})$, the support of $x_j$ contains at least two points;
\item $[\,\mathbf M\ \ \mathbf X\,]$, restricted to individuals with $d_i=1$, has full column rank;
\item some $i$ with $d_i=2$ has a neighbour $j$ such that, for fixed $(\mathbf G,\mathbf M,\mathbf X^{-x},\mathbf x_{-j})$, the support of $x_j$ contains at least three points.
\end{enumerate}
Then $\boldsymbol \theta = (\xi, \gamma, \delta, \theta_1,\theta_2, \beta)'$ is point identified.
\item \textbf{Partial identification with no peer effects.}
If $\theta_1=\theta_2=0$ and some $i$ has a neighbour $j$ such that, for fixed $(\mathbf G,\mathbf M,\mathbf X^{-x},\mathbf x_{-j})$, the support of $x_j$ contains at least two points, then $(\xi,\gamma,\delta,\theta_1,\theta_2)'$ is point identified, but $\beta$ is not.
\end{enumerate}
\end{theorem}
\begin{proof}
    Proof in Appendix Section \ref{SA.3}.
\end{proof}
Variation in a neighbour's covariate moves their choice probability, and hence $i$'s norm, while holding $i$'s own characteristics fixed, so the resulting change in $i$'s choice isolates the peer effect. In part (i), a payoff-relevant covariate that varies continuously across the neighbours of an individual with at least two neighbours moves their equilibrium choice probabilities over a non-empty open set, so $i$'s choices trace out the shape of the norm, which identifies $\beta$. Neighbour variation at a second degree $d$ then pins down $\theta_1d+\theta_2$. 

Without a continuous covariate, part (ii) relies on individuals with one and two neighbours. At degree one, the norm is independent of $\beta$, so variation in the neighbour's covariate identifies $\theta_1+\theta_2$, and full rank among these individuals identifies $\gamma$, and $\xi$ up to the constant $\theta_2/2$. At degree two, $\beta$ affects the norm only through the spread between the two neighbours' choice probabilities. A neighbour's $x$ taking at least three values changes the spread, which identifies $\beta$ and both $\theta_1$ and $\theta_2$. In parts (i) and (ii), the full-rank condition in Assumption~\ref{asum:ident} identifies the remaining own coefficients once the peer parameters are known. As in \citet{lambotte2025peer} and \citet{liu2014}, two different degrees separate $\theta_1$ from $\theta_2$. 

Finally, in part (iii), when peer effects are absent, variation in a neighbour's covariate moves their choice probability while holding $i$'s own characteristics fixed, and $i$'s choice does not respond, so no admissible alternative can have peer effects either. The full-rank condition in Assumption~\ref{asum:ident} then identifies the remaining parameters, while $\beta$ never affects the choices.

Our application satisfies the sample counterparts of the conditions in part (ii), with the number of rooms as the payoff-relevant discrete covariate. Every village has isolated households (628 in total), households with one neighbour (727) and households with two neighbours (687), and both full-rank conditions hold. The number of rooms takes 11 values at the neighbours of degree-one households, and 10 at the neighbours of degree-two households.

\section{Estimation and Simulations}\label{S_est}

In this section we present an appropriate estimator and present Monte Carlo simulations to illustrate its finite sample properties. We also compare it to the misspecified linear model with $\beta=1$.

\subsection{Nested Pseudo-Likelihood Estimator}\label{Sub_est}

Similar to other discrete choice models, we estimate our model parameters using a maximum likelihood–based approach. However, the conditional choice probabilities are not directly observed and are instead generated endogenously within the model, with equilibrium probabilities determined as the unique fixed point of the system in equation \eqref{eq:system}. This feature makes the \textit{Nested Pseudo-Likelihood Estimation} (NPL) strategy of \citet{aguirregabiria2007sequential} particularly well-suited, and in-fact been previously applied to estimate peer effects under rational expectations.\footnote{See, for example, \citet{lambotte2025peer}, \citet{Lin2024}, and others.} The likelihood is pseudo exactly because the conditional choice probabilities are not directly observed but are instead derived.

By Assumptions \ref{asum:A1}--\ref{asum:A2}, agents' rational expectations depend only on the public information set $\overline{\mathcal{I}}$, so peers' realised choices do not affect $i$'s beliefs. Since the private shocks are also independent across agents, peers' realised choices do not affect $y_i$, and the individual contributions to the likelihood are independent conditional on $\overline{\mathcal{I}}$. Thus, the conditional log-likelihood is expressed as:
\begin{align}\label{eq:cond_like}
    \mathcal{L}_N(\boldsymbol \theta; \mathbf{p}^*) := \mathcal{L}_N(\boldsymbol \theta; \mathbf y | \overline{\mathcal{I}}) = \frac{1}{N} \sum_{i=1}^N \{ y_i  \ln{p^*_i} + (1-y_i) \ln{(1-p^*_i)} \}
\end{align}
where $\boldsymbol \theta \in \boldsymbol \Theta$ is the parameter vector and $p_i^* = \Gamma_i(\mathbf{p}^*; \boldsymbol \theta)$.

If instead the conditional choice probabilities were observable, we could use the standard Maximum Likelihood Estimator (MLE):
\begin{align}
\hat{\boldsymbol \theta}_{MLE} 
= \arg\max_{\boldsymbol \theta} \mathcal{L}_N(\boldsymbol \theta; \mathbf{p}^*) 
\quad \text{s.t. } \mathbf{p}^* = \Gamma(\mathbf{p}^*; \boldsymbol \theta).
\end{align}
where $\Gamma(\mathbf{p}; \boldsymbol \theta) = F_{\eta}(Z(\mathbf{p}))$.
Since the conditional choice probabilities are not directly observed, we require an estimator that consistently recovers both the probabilities and the structural parameters. Given the high dimensionality of the public information set $\overline{\mathcal{I}}$, non-parametric estimation of these probabilities is infeasible. The NPL estimator of \citet{aguirregabiria2007sequential} can instead help iteratively recover the two objects of interest. The pseudo log-likelihood is thus defined as:
\begin{align}\label{eq:pseudo_like}
    \mathcal{Q}_N(\boldsymbol \theta; \mathbf{p}) 
    = \frac{1}{N} \sum_{i=1}^N 
    \Big\{ y_i \ln(\Gamma_i(\mathbf{p}; \boldsymbol{\theta})) + (1-y_i) \ln(1-\Gamma_i(\mathbf{p}; \boldsymbol{\theta})) \Big\},
\end{align}
where $\mathbf{p}$ now need not be the equilibrium probabilities. Let $q_i(\boldsymbol \theta, \mathbf{p}) = y_i \ln(\Gamma_i(\mathbf{p}; \boldsymbol{\theta})) + (1-y_i) \ln(1-\Gamma_i(\mathbf{p}; \boldsymbol{\theta}))$ be agent $i$'s contribution to the pseudo log-likelihood function and let $\mathcal{Q}_0(\boldsymbol \theta, \mathbf{p}) = \mathbb{E}[q_i(\boldsymbol \theta, \mathbf{p})]$. Then define:
\[\tilde{\boldsymbol \theta}_0(\mathbf{p}) = \arg\max_{\boldsymbol \theta} \mathcal{Q}_0(\boldsymbol \theta,\mathbf{p}) \text{ and } \Gamma_0(\mathbf{p}):= \Gamma(\mathbf{p}; \tilde{\boldsymbol \theta}_0(\mathbf{p})) \]
\[ \tilde{\boldsymbol \theta}_N(\mathbf{p}) = \arg\max_{\boldsymbol \theta} \mathcal{Q}_N(\boldsymbol \theta,\mathbf{p}) \text{ and } \Gamma_N(\mathbf{p}):= \Gamma(\mathbf{p};\tilde{\boldsymbol \theta}_N(\mathbf{p}))\]
with $\theta_0$ and $\mathbf{p}^*$ the true values of the parameters and the conditional choice probabilities. Furthermore, let
\[ \Lambda_{0N} \equiv \{(\boldsymbol \theta,\mathbf{p}): \boldsymbol \theta = \tilde{\boldsymbol \theta}_0(\mathbf{p}), \mathbf{p} =  \Gamma_0(\mathbf{p})\}\]
\[ \Lambda_{N} \equiv \{(\boldsymbol \theta,\mathbf{p}): \boldsymbol \theta = \tilde{\boldsymbol \theta}_N(\mathbf{p}), \mathbf{p} =  \Gamma_N(\mathbf{p}) \} \]
For any $\mathbf{p}$, an NPL fixed point is a pair $(\boldsymbol \theta, \mathbf{p})$. $ \Lambda_{0N}$ is the set of fixed points in the population and $\Lambda_{N}$ is the set of fixed points in the sample. Although the fixed point of the system is unique under our assumptions for a given parameter vector, this does not imply that $\Lambda_N$ is a singleton. Multiple parameter–probability pairs may satisfy the fixed point conditions, especially since the mapping is not a contraction. Hence, the NPL estimator is defined as the maximiser of the pseudo log-likelihood over all such fixed points:
\begin{align}\label{eq:npl_est}
    (\hat{\boldsymbol \theta}_{NPL}, \hat{\mathbf{p}}_{NPL})
    = \arg\max_{(\boldsymbol \theta,\mathbf{p}) \in \Lambda_N} 
    \mathcal{Q}_N(\boldsymbol \theta, \mathbf{p}).
\end{align}

We define two matrices $V_1(\theta_0)$ and $V_2(\theta_0)$:
\begin{align}\label{eq:variances}
       \mathbb{E} \Bigg[
\frac{\partial^2 \mathcal{Q}_N(\theta_0, \mathbf{p}^*)}{\partial \boldsymbol \theta \partial \boldsymbol \theta'}
+ 
\frac{\partial^2 \mathcal{Q}_N(\theta_0, \mathbf{p}^*)}{\partial \boldsymbol \theta \partial \mathbf{p}'}
\cdot
\left[ I - \left( \frac{\partial \Gamma(\mathbf{p}^*; \theta_0)}{\partial \mathbf{p}} \right) \right]^{-1}
\cdot
\frac{\partial \Gamma(\mathbf{p}^*; \theta_0)}{\partial \boldsymbol \theta'}
\Bigg| \mathcal{\overline{I}} \Bigg] \rightarrow_p&  V_1(\theta_0)
 \\
\mathbb{E} \left[
\frac{\partial q_i(\theta_0, \mathbf{p}^*)}{\partial \boldsymbol \theta}
\frac{\partial q_i(\theta_0, \mathbf{p}^*)}{\partial \boldsymbol \theta'}
\Bigg| \mathcal{\overline{I}} \right]  \rightarrow_p&  V_2(\theta_0)
    \end{align} 

Under regularity conditions (discussed in Appendix Section \ref{SB.1}), \citet{Lin2024}, relying on \citet{Menzel2016} show that the NPL estimator as described in equation \eqref{eq:npl_est}, is consistent and asymptotically normal.
\begin{linhu}
    Under Assumptions \ref{asum:A1}-\ref{assump:estimation},
    \[\hat{\boldsymbol \theta}_{NPL} \rightarrow_p \theta_0, \qquad \sqrt{N} \left( \hat{\boldsymbol \theta}_{NPL} - \theta_0 \right)  \rightarrow_d \mathcal{N}(0, V_{NPL}), \] and where $V_{NPL} = V_1^{-1}(\theta_0) V_2(\theta_0) V_1^{-1}(\theta_0)^{\top}$ is the variance of the NPL estimator where $V_1(\theta_0)$ and $V_2(\theta_0)$ are defined in equation \eqref{eq:variances}.
\end{linhu}
\begin{proof}
    See \citet{Lin2024} (Appendix A. Proofs, Proof of Theorem 1, p. 14-15).
\end{proof}

For the actual estimation procedure, we use the spectral algorithm of \citet{Aguirregabiria2021}. This algorithm overcomes the non-convergence of the usual iterative estimation of the NPL algorithm of \citet{aguirregabiria2007sequential}. In addition, it does not require selection of an arbitrary parameter such as in the algorithm of \citet{Kasahara2012}. For details on the spectral algorithm, see Appendix Section \ref{SB.1}.

In practice, two issues complicate estimating the CES elasticity jointly with the other parameters. First, the likelihood can be nearly flat in $\beta$, since the norm converges to its max or min very quickly as $|\beta|$ grows. Second, the gradient with respect to $\beta$ can be on a different scale from those of the other parameters, such as $\theta_1$ and $\theta_2$, which makes joint non-linear optimisation ill-conditioned. We address both by exploiting that $\beta$ is the only parameter entering the utility index non-linearly. We fix $\beta$ on a grid and, at each value, estimate the remaining parameters using the spectral NPL, recording the profiled pseudo log-likelihood. The grid is dense near zero, with a geometric tail on each side and the exact $\pm\infty$ (max- and min-aggregator) endpoints. 

We pick the best grid point for $\beta<0$ and for $\beta\ge0$, refine each by a one-dimensional search over $\beta$ between its neighbouring grid points, re-estimating the other parameters by spectral NPL at every candidate, and keep the point with the higher refined likelihood. Although Theorem~\ref{thm:ident} identifies $\beta$ exactly, its numerical recovery becomes fragile as $\beta\to\pm\infty$. When the profiled likelihood is flat, up to a tolerance, out to a $\pm\infty$ limit, the candidate values of $\beta$ are numerically indistinguishable, without affecting the other estimates. We classify that side as a flat tail. In those cases, we report both the point estimate and whether it exhibits tail behaviour, giving the corresponding max- or min-aggregator fit in the latter case.

\subsection{Monte Carlo Simulations}\label{Sub_sim}

We assess the finite-sample behaviour of the NPL estimator and profile-based algorithm by Monte Carlo simulations, comparing the CES norm to the restricted linear norm ($\beta=1$). We calibrate the network and structural parameters to match the setting of the empirical application. Networks are generated from the Chung--Lu model \citep{chunglu2002} to match the mean degree ($5.75$) and share of isolated households ($9.3\%$) (Table~\ref{tab:summary_stats}). We set $\theta_1=0.15$ and $\theta_2=1$, vary $\beta\in\{-8.3,\,3.9,\,22.3\}$, consider a low- and a high-adoption design, and run $500$ replications for each $N\in\{500,1000,2000,4000\}$. Appendix~\ref{SB.2} gives the set-up details, and all simulation tables are in Appendix~\ref{SD}.

The estimator performs well when $\beta$ is not extreme, particularly with high adoption. Under the CES norm, $\hat\theta_1$ is within $0.02$ of its true value at $N=4000$ in every design, and $\hat\theta_2$ is within $0.07$ in four of the six (Tables~\ref{tab:sim_main_CL_a} and~\ref{tab:sim_main_CL_b}). The exceptions are in the low-adoption designs. The restricted linear model, by contrast, stays biased as $N$ grows. At $N=4000$, it under-estimates complementarity by about $50\%$ when $\beta=-8.3$, and when $\beta=22.3$ it over-estimates complementarity by $60$ to $95\%$ and under-estimates conformity by $45$ to $50\%$. In most designs its bias and MSE persist, whereas under the CES norm both shrink towards zero (Tables~\ref{tab:sim_biasmse_main_CL_a}--\ref{tab:sim_biasmse_main_CL_b}).

For draws with an interior $\hat\beta$, the coverage of $\hat\theta_2$ is at or above the nominal level in every design, at $90$ to $100\%$ (Tables~\ref{tab:sim_coverage_main_CL_a} and~\ref{tab:sim_coverage_main_CL_b}). That of $\hat\theta_1$ is lower ($70$ to $95\%$). For tail draws, where the profile likelihood is nearly flat, the local Wald standard error understates the uncertainty and coverage collapses. As $N$ grows, the median $\hat\beta$ moves towards the truth and the share of tail solutions falls, though slowly for $\beta<0$, while the tail draws distort the mean (Tables~\ref{tab:sim_main_CL_a} and~\ref{tab:sim_main_CL_b}). The LR rejection rate rises with $N$; at $N=4000$ it is essentially $100\%$ at $\beta=22.3$ but only $15$ to $32\%$ at $\beta=-8.3$. For the extreme values, the estimator recovers the sign of $\beta$ but not its magnitude, both models under-estimate complementarity when the norm is strongly min-like, and the LR test has power only against the max-like norm (Tables~\ref{tab:sim_extremes_CL}--\ref{tab:sim_coverage_ext_CL_b}). See Appendix~\ref{SB.2} for a more detailed discussion of the results.

\section{Empirical Results}\label{S_emp}

To analyse the effect of norm misspecification on the distortion of optimal targeting policy, we take the framework developed above to the microfinance-adoption dataset of \citet{banerjee2013}. Their original focus was the diffusion of information about the programme through a small set of key agents in each village and its effect on take-up. In line with our framework, and similar to \citet{lambotte2025peer}, we instead take awareness of the programme as given and model the adoption choice itself as a simultaneous game of incomplete information.

The programme was operated by Bharatha Swamukti Samsthe (BSS), a Bangalore-based microfinance institution that made joint-liability loans, principally to women organised in small borrowing groups. The 75 sample villages span five districts of rural Karnataka, in southern India, and BSS began lending in 43 of them from 2007 onward, entering villages one at a time. We retain the $29$ of these $43$ villages for which the household census records each household's caste, which leaves $6{,}733$ households. The action $y_i$ equals one if the household \emph{participates}, as recorded in the lender's administrative records. Take-up was gradual and far from universal, with $17.4\%$ of households in our estimation sample participating.

Crucially, \citet{banerjee2013} fielded a baseline survey in 2006, \emph{before} BSS entered any village, so the network is measured prior to the adoption decisions it later shapes, therefore is pre-determined. The survey paired a household census with a social-network module covering twelve dimensions of interaction, such as visits, borrowing and lending, and advice. In line with \citet{lambotte2025peer}, we use the visiting network, which links two households if either reports visiting the other or being visited by it. It is sparse, with a mean degree of $5.75$ and $9.3\%$ of households isolated. The individual characteristics $X_i$ comprise the number of rooms and beds in the home; indicators for a non-Hindu head of household and for scheduled-caste or scheduled-tribe status; and indicators for owning a latrine, for being a designated village ``leader'' household, and for renting rather than owning the home. Summary statistics are reported in Appendix~Table~\ref{tab:summary_stats}.

\begin{figure}[!ht]
  \centering
  \includegraphics[width=1\linewidth]{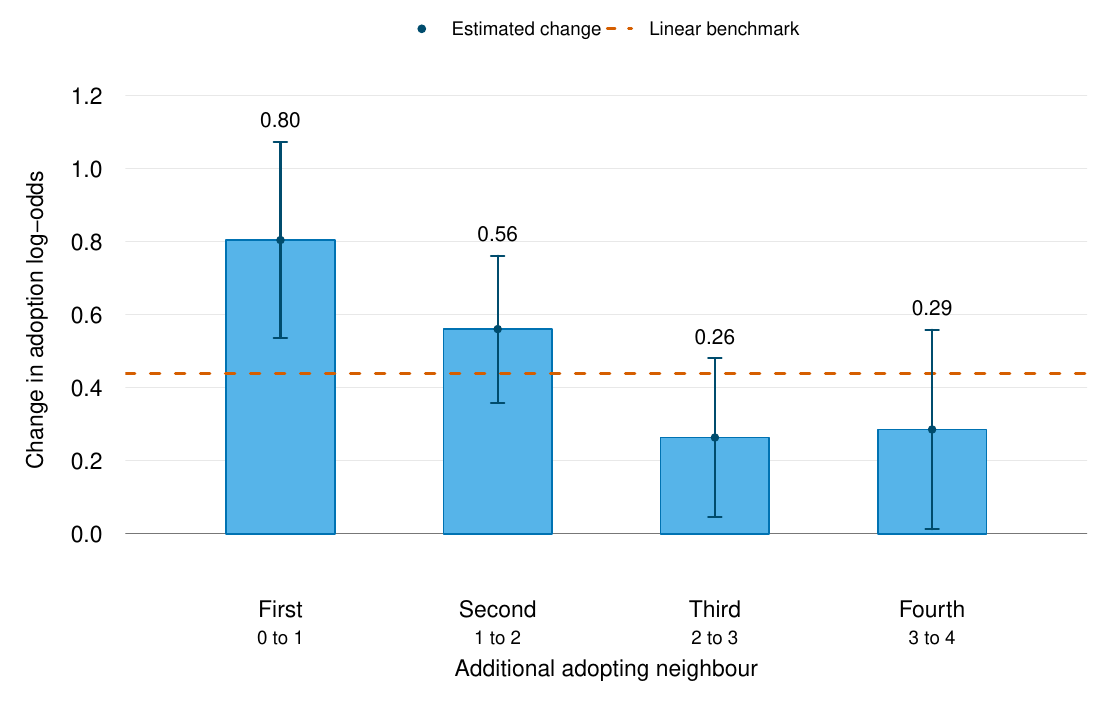}
  \caption{Microfinance adoption and additional adopting neighbours.}
  \label{fig:first-adopter}
  \begin{minipage}{0.95\linewidth}
    \footnotesize
    \textit{Notes:} $4{,}123$ households with at least four neighbours. Each bar is the estimated change in the log-odds of adoption from one more adopting neighbour, from a logit of adoption on dummies for the number of adopting neighbours and on their share, with village and degree fixed effects and household controls. The dashed line is the constant change implied when the number and share enter linearly, as under a linear norm. The changes decline over the first four adopting neighbours, and a village-clustered Wald test rejects their equality ($p=0.011$). Whiskers are village-clustered $90\%$ confidence intervals.
  \end{minipage}
\end{figure}

Two features of microfinance diffusion make the linear assumption ex-ante implausible. \citet{banerjee2013} find that, once informed, households barely respond to more neighbours adopting, pointing to a maximum-like norm. The group-liability structure of microfinance, by contrast, requires a minimum group size, suggesting a minimum-like norm. The opposite pull of these two mechanisms implies that the curvature of the social norm is an empirical question instead of an \emph{a priori} assumption. Figure~\ref{fig:first-adopter} illustrates that the data favour a non-linear norm. The effect of each additional adopting neighbour on adoption declines, whereas a linear norm implies a constant effect.

We first estimate the model on the pooled sample of all 29 villages. Since villages are not inter-connected and differ in their average household characteristics, we then partition them into four segments along two dimensions, caste homogeneity and income, the latter proxied by asset ownership (beds and latrines). Our main specification includes village fixed effects and the isolated-household dummy. Appendix Table~\ref{tab:emp_full_app} reports estimates without the dummy and relates them to \citet{lambotte2025peer}, whose control function for network endogeneity appears to absorb the same isolated-household heterogeneity.

\subsection{Pooled sample}\label{ss:pooled}

\begin{figure}[!ht]\centering
\includegraphics[width=\textwidth]{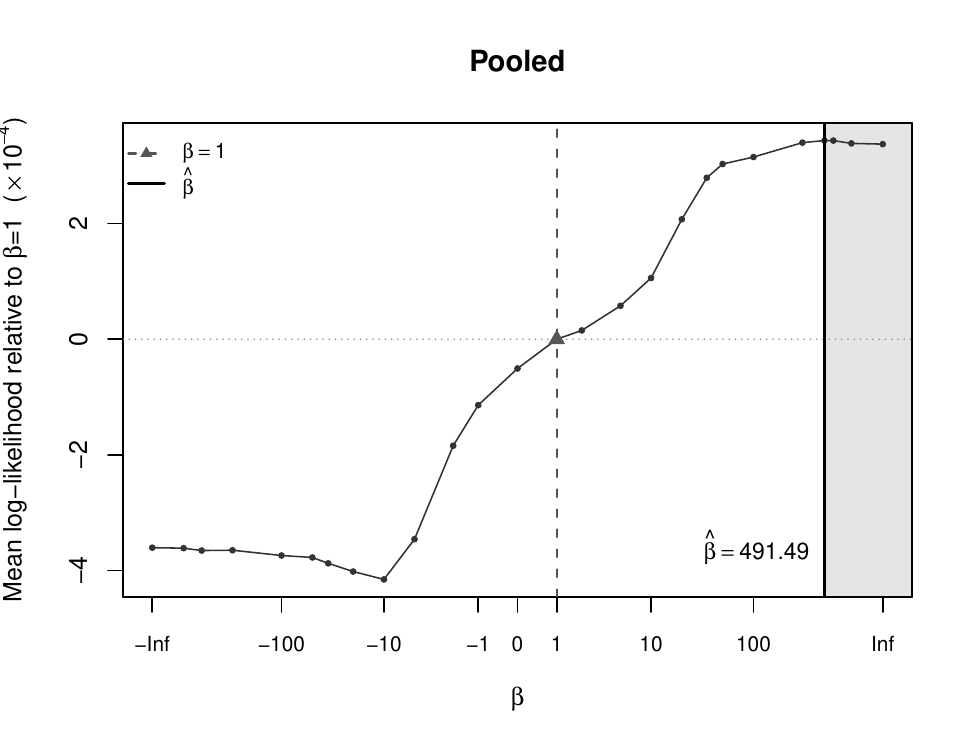}
\caption{Profiled log-likelihood in $\beta$ for pooled sample.}
\label{fig:lik_pooled}
\begin{minipage}{0.9\textwidth}
\footnotesize
\emph{Notes:} The y-axis is the Log-Likelihood of the unrestricted CES model relative to the linear case. The profiled likelihood rises towards the maximum aggregator and then plateaus, with refined estimate $\hat\beta=491.5$, while the linear norm ($\beta=1$) is dominated. Shaded region denotes the flat tail-region where $\beta$ is numerically indistinguishable.
\end{minipage}
\end{figure}

Table~\ref{tab:emp_full} reports the pooled estimates with the full estimates presented in Table~\ref{tab:emp_full_app}. The data reject the linear norm in favour of a strongly max-leaning one, with an estimated $\beta$ of $491.5$ and the likelihood-ratio test rejects imposing linearity ($\mathrm{LR}=4.63$, $p=0.03$). However, as Figure~\ref{fig:lik_pooled} illustrates, the profiled likelihood rises monotonically towards the maximum aggregator but plateaus (within set tolerance), indicating tail like behaviour where $\beta$ is numerically indistinguishable and we report the optimal value in that regime.  This is in line with the pattern in Figure~\ref{fig:first-adopter} where the first adopting neighbour does most of the work.

\begin{table}[!ht]
\begin{singlespacing}\centering\small
\caption{Microfinance Adoption for pooled sample: Point Estimates for CES Elasticity and Peer Effects (CES vs.\ Linear Norm) and the LR Test of $\beta=1$.}
\label{tab:emp_full}
\setlength{\tabcolsep}{6pt}
\begin{tabular}{c|ccc|cc|cc}
\toprule
 & \multicolumn{3}{c|}{\textbf{CES ($\beta$ free)}} & \multicolumn{2}{c|}{\textbf{Linear ($\beta=1$)}} & \multicolumn{2}{c}{\textbf{LR: $\beta=1$}} \\
\cmidrule(lr){2-4}\cmidrule(lr){5-6}\cmidrule(lr){7-8}
$N$ & $\beta$ & $\theta_1$ & $\theta_2$ & $\theta_1$ & $\theta_2$ & stat. & $p$ \\
\midrule
6733 & 491.5$^{\dagger}$ & 0.053$^{**}$ & 2.064$^{***}$ & 0.180$^{***}$ & 1.538$^{***}$ & 4.63 & 0.031$^{**}$ \\
 & & (0.023) & (0.445) & (0.032) & (0.501) & & \\
\bottomrule
\end{tabular}
\\[0.3em]\footnotesize\parbox{0.9\textwidth}{\emph{Notes:} All 29 villages pooled with village fixed effects and an isolated-household dummy. $^{\dagger}$ marks a flat-tail fit, where $\beta$ is numerically indistinguishable and the likelihood is flat. The reported value is the refined estimate in the plateau. Standard errors in parentheses, stars are Wald tests against zero. The LR statistic tests $\beta=1$ against a $\chi^2_1$ reference. $^{*}p<0.10$, $^{**}p<0.05$, $^{***}p<0.01$.}
\end{singlespacing}
\end{table}

Furthermore, imposing linearity distorts the two peer channels as anticipated by our simulations. The restricted fit overstates strategic complementarity more than threefold ($0.18$ vs.\ $0.05$) and understates conformity
($1.54$ vs.\ $2.06$). A researcher assuming the linear norm would thus attribute adoption far more to spillovers, and less to conformity, than the data support.

Moreover, the dummy on isolated households is significantly different from zero and negative under both the linear and the non-linear norm ($-0.894$ and $-1.033$, respectively). Appendix Table~\ref{tab:emp_full_app} re-estimates the pooled model with and without it, and the exclusion changes the results drastically with conformity turning negative ($\hat\theta_2=-0.725$ under the CES norm and $-0.182$ under the linear norm) and the profiled $\beta$ runs to the minimum aggregator. Without a separate intercept, isolated households' lower baseline adoption is mechanically loaded onto the conformity term, which in turn distorts the estimated norm and its curvature $\beta$. This mirrors the pattern in \citet{lambotte2025peer},\footnote{They include exogenous peer effects $\mathbf{GX}$, which we do not as these effects enter only through feedback and provide identifying information, similar to \citet{boucher2024peer}. Their inclusion also reduces the information available to identify $\beta$ precisely.} whose homophily-based network-endogeneity control plays the same role as our isolated-household dummy. Comparing their Models~3 and~6 in Table~2, removing the control turns the conformity coefficient from positive to negative while complementarity does not change much, similar to our no-dummy results, for both the linear and the non-linear specifications. Their control function is therefore likely to capture, at least in part, the difference between connected and isolated households rather than network endogeneity.\footnote{Their control enters as a quartic polynomial in a node position $\mu$ estimated from a homophily-based link-formation model with node fixed effects (\texttt{homophily.fe} in \textsf{CDatanet}). A household with no links has an unidentified fixed effect that the estimator drives to the lower boundary ($\mu\to-\infty$, numerically a floor), so after the normalisation to $[-1,1]$ isolated households are pinned at the bottom of the $\mu$ range.}

\subsection{Heterogeneity across segments}\label{ss:segments}

Two dimensions of heterogeneity that are especially salient in Indian villages can shape peer influence and adoption-- social stratification, captured by caste, and
wealth. Caste in our data is a household's Scheduled Caste or Tribe (SC/ST) status, and caste-based networks act as channels along which information, norms and risk-sharing flow in rural India. This makes caste homogeneity the natural axis for conformity
\citep{munshirosenzweig2006,banerjee2013}. Wealth, in contrast, shapes both the demand for and the returns to microcredit, and hence take-up \citep{banerjee2015miracle,banerjeekarlanzinman2015}, while also conditioning the strategic, complementarity side of adoption through credit constraints and the social-interaction returns to borrowing \citep{feigenberg2013social}. Both dimensions also shape the norm and whose behaviour matters for adoption decisions.

We therefore split villages along these two axes at their village-level medians, giving four segments (heterogeneous/homogeneous $\times$ poor/rich). Since there are multiple wealth proxies in the data, we pick two of them, bed and latrine ownership, based on how well they segment the villages and report results for both.\footnote{An alternative would be to construct a composite wealth index from all the income-related covariates.} The segment composition and the proxies are described in Appendix Tables~\ref{tab:cluster_composition} and~\ref{tab:cluster_variant_choice}. Here we discuss results for the case with the isolated-dummy included while Appendix Tables~\ref{tab:emp_caste_beds_dummy_off_full_app} and~\ref{tab:emp_caste_latrine_dummy_off_full_app} present the other specification.

\begin{table}[H]
\begin{singlespacing}\centering\small
\caption{Norm elasticity and test of $\beta=1$ by segment for the Caste $\times$ Beds and Caste $\times$ Latrine partitions.}
\label{tab:emp_segments_beta_lr}
\setlength{\tabcolsep}{5pt}
\begin{tabular}{l|cccc|cccc}
\toprule
 & \multicolumn{4}{c|}{\textbf{Caste $\times$ Beds}} & \multicolumn{4}{c}{\textbf{Caste $\times$ Latrine}} \\
\cmidrule(lr){2-5}\cmidrule(lr){6-9}
Segment & $N$ & $\hat\beta$ & LR & $p$ & $N$ & $\hat\beta$ & LR & $p$ \\
\midrule
Het/P & 1,616 & -83.1$^{\dagger}$ & 1.81 & 0.179 & 1,691 & -0.97 & 0.67 & 0.412 \\
Het/R & 1,387 & 491.5$^{\dagger}$ & 3.39 & 0.066$^{*}$ & 1,312 & -282.8$^{\dagger}$ & 3.16 & 0.075$^{*}$ \\
Hom/P & 1,728 & -296.7$^{\dagger}$ & 1.56 & 0.212 & 1,718 & -4.05 & 5.23 & 0.022$^{**}$ \\
Hom/R & 2,002 & 787.7$^{\dagger}$ & 0.75 & 0.387 & 2,012 & 2.00 & 0.35 & 0.556 \\
\bottomrule
\end{tabular}
\\[0.3em]\footnotesize\parbox{0.92\textwidth}{\emph{Notes:} Segments are Het/Hom (heterogeneous vs homogeneous on the caste-minority share) $\times$ P/R (poor vs rich), split at village-level medians of each partition's wealth proxy (bed or latrine ownership). $^{\dagger}$ marks a flat-tail fit ($\beta$ numerically indistinguishable; the value is the refined estimate in the plateau). LR tests $\beta=1$ against a $\chi^2_1$ reference, with $^{*}p<0.10$, $^{**}p<0.05$, $^{***}p<0.01$.}
\end{singlespacing}
\end{table}

Table~\ref{tab:emp_segments_beta_lr} reports the $\beta$ estimates by segment for both partitions, with the LR statistic and $p$-value for $H_0:\beta=1$. Only 3 out of 8 segments reject the $\beta=1$ restriction, likely due to the smaller sample size. The norm is markedly heterogeneous, where for the Caste $\times$ Beds case every segment sits in the flat-tail regime, whereas under Caste $\times$ Latrine three of the four segments yield interior $\hat\beta$ (Figures~\ref{fig:lik_caste_beds} and~\ref{fig:lik_caste_latrine} show the likelihood profiles).\footnote{In several segments the estimated peer parameters are large enough that our sufficient condition for equilibrium uniqueness (the best-response elasticity bound) is likely violated, under both the CES and the linear ($\beta=1$) specifications, thus multiplicity of equilibria likely exists.} The norm regime lines up with wealth with poorer segments leaning towards the minimum ($\hat\beta<1$) in both partitions and the richer ones towards the maximum, the one exception being the heterogeneous--rich latrine segment, which is min-like.

\begin{figure}[!ht]\centering
\includegraphics[width=\textwidth]{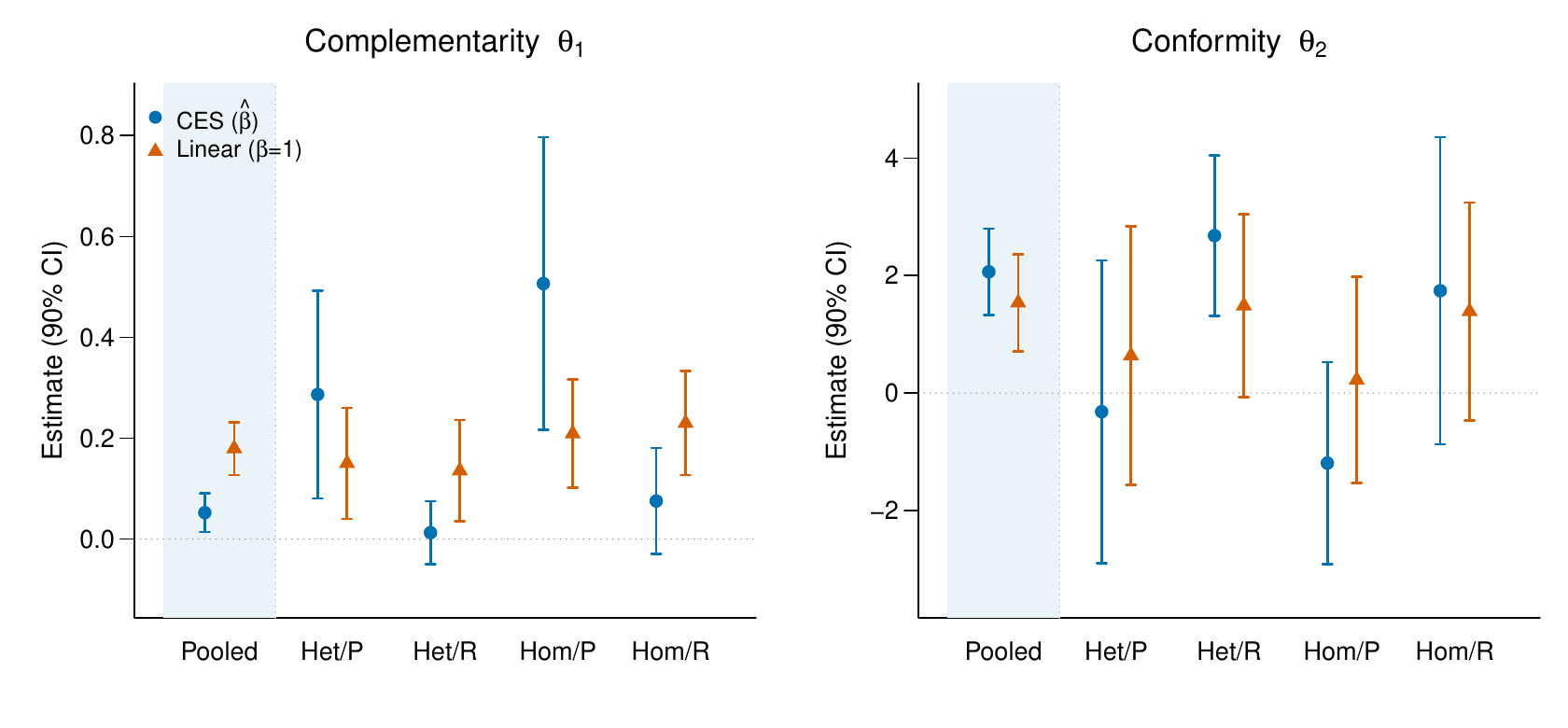}
\caption{Peer parameters by segment, Caste $\times$ Beds: CES vs.\ linear, with 90\%
CIs and the pooled estimates for reference.}
\label{fig:theta_caste_beds}
\end{figure}

\begin{figure}[!ht]\centering
\includegraphics[width=\textwidth]{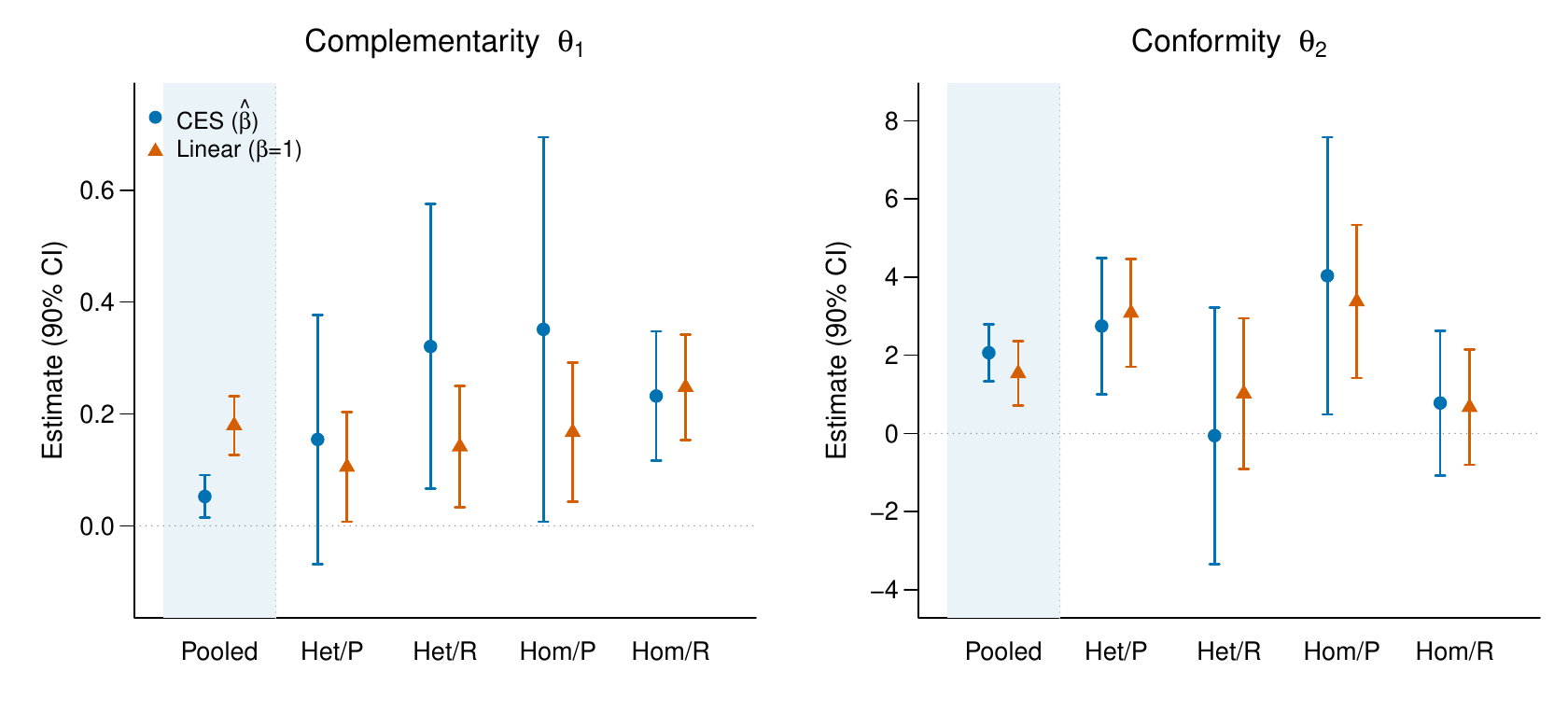}
\caption{Peer parameters by segment, Caste $\times$ Latrine: CES vs.\ linear, with 90\%
CIs and the pooled estimates for reference.}
\label{fig:theta_caste_latrine}
\end{figure}

Figures~\ref{fig:theta_caste_beds}--\ref{fig:theta_caste_latrine} plot the peer parameters against the pooled estimates and the linear restriction. As in the pooled sample, imposing linearity distorts the peer-effect estimates in several segments, even flipping sign in some segments, although the segment estimates are noisier given the smaller sample size. Relative to the pooled fit, the segments are highly heterogeneous with complementarity generally stronger, and conformity is far more dispersed and even negative. Under the beds split, poorer villages show stronger complementarity and weaker conformity while for the latrine split the conformity ordering reverses, with the poorer segments showing the strongest conformity. Among poorer villages, caste-homogeneous villages exhibit stronger conformity than heterogeneous ones holding wealth fixed, although this pattern does not survive the bed-ownership split. The results illustrate that there is indeed heterogeneity in how peer influence shapes adoption and how we define its dimensions matters. These are patterns that the pooled sample would mask both in terms of the norm as well as the peer effect parameters. Appendix Tables~\ref{tab:emp_caste_beds_full_app} and~\ref{tab:emp_caste_latrine_full_app} report the full estimates.

\section{Optimal Targeting}\label{S_seed}

The previous section illustrated the impact of norm misspecification on the peer effect estimates. An important next issue is how this misspecification biases counterfactual analysis and the resulting policy recommendations. A natural policy question in the case of microfinance adoption is whom a planner should target in the network, and by how much, with financial inclusion an explicit government objective in many developing countries, including India (for example, Pradhan Mantri Jan Dhan Yojana).\footnote{A separate but important question is whether to target welfare improvements instead of adoption in discrete choice models, for example \citet{bhattacharya2024, volpe2025discrete} and others. We focus on adoption rather than welfare for two reasons. First, raising take-up is the explicit objective of microfinance and financial-inclusion programmes \citep{banerjee2013,banerjeekarlanzinman2015,banerjee2015miracle}, and more broadly of the network-targeting literature on product and technology adoption that we build on \citep{beaman2021network,banerjee2019gossip,bursztyn2014peer}. Second, welfare analysis in these models is especially difficult, requiring a stance on the private value of microfinance, as well as on whether a planner should consider conformity as welfare relevant. Moreover, \citet{bhattacharya2024} show that choice data, while sufficient to predict counterfactual demand, may not identify welfare, since distinct mechanisms behind the same interaction coefficient imply different welfare effects.} The answer depends on which households are central, and hence on the shape of the norm. The optimal allocation is a network-centrality problem, and misspecifying the norm as linear distorts which households count as central for propagation of peer influence, so the planner targets on the wrong centrality measure.

We set the problem of the planner as follows. The planner maximises expected adoption by allocating subsidies, such as conditional cash transfers or take-up incentives conditional on adoption, to individuals in the network subject to a fixed budget. A subsidy $\tau_i\geq0$ raises agent $i$'s incentive from $\alpha_i$ to $\alpha_i+\tau_i$ and we assume that these subsidies are public knowledge. For an agent with at least one peer, the post-policy equilibrium satisfies
\begin{equation}\label{eq:targeting_equilibrium}
 p_i=\Gamma_i(\mathbf p,\boldsymbol\tau;\beta)
 =F_\eta\!\left(
 \alpha_i+\tau_i+(\theta_1d_i+\theta_2)\widetilde y^e_{-i}(\mathbf p;\beta)
 -\frac{\theta_2}{2}\right).
\end{equation}
Denote the new conditional choice probabilities for a vector of subsidies by $\mathbf p(\boldsymbol\tau;\beta)$. For a fixed and finite budget $B$, the planner's problem is to,
\begin{equation}\label{eq:targeting_problem}
 \max_{\boldsymbol\tau\geq\mathbf 0}\;
 \frac{1}{N}\sum_{i=1}^N p_i(\boldsymbol\tau;\beta)
 \qquad\text{subject to}\qquad
 \sum_{i=1}^N\tau_i=B \qquad \mathbf{p}(\boldsymbol{\tau},\beta) = \Gamma(\mathbf{p},\boldsymbol{\tau};\beta).
\end{equation}
The Jacobian of the post-policy best-response mapping $\Gamma(\cdot;\beta)$ in \eqref{eq:targeting_equilibrium}, evaluated at the equilibrium, has typical element
\[
J_{ij}=\frac{\partial \Gamma_i}{\partial p_j}
=p_i(1-p_i)\,(\theta_1 d_i+\theta_2)\,
\frac{\partial\widetilde y^e_{-i}(\mathbf p;\beta)}{\partial p_j},
\qquad J_{ij}=0 \text{ for isolated } i.
\]
Here
$\partial\widetilde y^e_{-i}/\partial p_j$ is the marginal effect of $j$'s adoption on
$i$'s norm. This term defines an effective influence network through which peer effects propagate (similarly defined in \citet{Tchuente2026}). Its form however, is
governed by $\beta$ and in the linear case this factor is simply $\tilde{g}_{ij}$. While a closed-form solution for optimal adoption does not exist, Proposition~\ref{prop:optimal_targeting} below presents the solution for a generalised norm with elasticity $\beta$.
\begin{prop}[Optimal Targeting]\label{prop:optimal_targeting}
Suppose Assumptions~\ref{asum:A1}--\ref{asum:ident} hold with $\alpha_i$ replaced by the post-policy incentives $\alpha_i+\tau^*_i$. At any optimal allocation $\boldsymbol\tau^*$ where the norm is continuously differentiable, there exists $\lambda$ such that
\[
\frac{p_i(1-p_i)}{N}
\left[(I-J)^{-\top}\iota\right]_i \qquad
\begin{cases}
=\lambda, & \tau_i^*>0,\\[2pt]
\leq\lambda, & \tau_i^*=0,
\end{cases}
\]
where all quantities are evaluated at the post-policy equilibrium.
\end{prop}
\begin{proof}
    Proof in Appendix Section \ref{SA.4}.
\end{proof}
\begin{remark}
Here, $\lambda$ is the Lagrange multiplier on the budget constraint and the left-hand side is the marginal return to subsidising agent $i$. The planner funds households until these marginal returns are equalised with any household left unsubsidised having a weakly lower return. $[(I-J)^{-\top}\iota]_i$, with $\iota$ the vector of ones, is the total increase in adoption, across all households, from a unit increase in $i$'s adoption probability after network feedback is accounted for. It is $i$'s Katz--Bonacich centrality in the transposed effective network $J^\top$. Multiplying it by $p_i(1-p_i)/N$, the response of $i$'s adoption probability to a unit subsidy, gives the marginal return. For $\beta=1$, this is similar to the standard centrality-based targeting of \citet{ballester2006,galeotti2020targeting} but for a non-linear norm the same centrality is computed on the influence network defined by $\partial\widetilde y^e_{-i}/\partial p_j$. Note that the objective is not globally concave in $\boldsymbol\tau$ since the logit is not globally concave. Therefore, optimal allocation need not be unique, and the condition above is necessary but not sufficient. In practice we compute the optimal policy $\boldsymbol\tau^*$ numerically.\footnote{We solve the planner's problem by spectral projected gradient \citep{birgin2000nonmonotone}, projecting onto the budget set $\{\boldsymbol\tau\geq\mathbf 0:\sum_i\tau_i=B\}$, and stop when the duality gap falls below $2\times10^{-6}$. Because the objective is not concave, we start from three allocations (uniform, the budget spread over the $1{,}000$ households with the highest marginal returns at $\boldsymbol\tau=\mathbf 0$, and a rescaled larger-budget solution) and report the stationary allocation with the highest adoption. Uniqueness is checked by iterating from the $\mathbf{p} = \mathbf 0$ and $\mathbf{p} = \mathbf 1$ end points.}
\end{remark}

\begin{figure}[!ht]
  \centering
  \includegraphics[width=\linewidth]{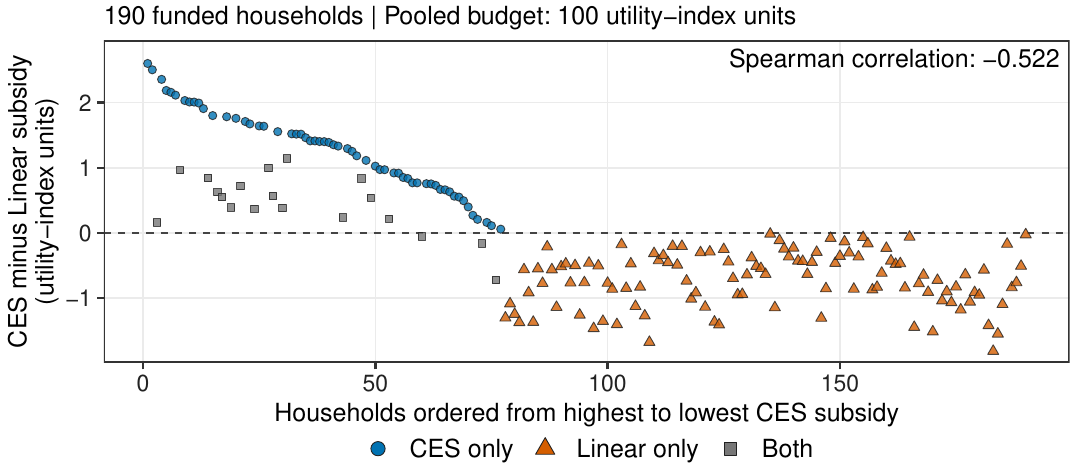}
  \caption{Optimal subsidy allocation under CES and linear norms in the pooled sample}
  \label{fig:pooled-subsidy-comparison}
  \begin{minipage}{\linewidth}
    \footnotesize
    \textit{Notes:} Each point is one of the $190$ households funded under at least one norm, ordered by CES subsidy and the vertical axis is the CES minus linear subsidy. Each model uses its own estimates and equilibrium to allocate the same $100$ utility-index-unit budget. The allocations are strongly negatively rank-correlated (Spearman $-0.52$), and across all $6{,}733$ households the budget raises predicted adoption by $1.9$ points under the linear norm ($17.4\%\to19.3\%$) and $10.9$ under CES ($17.4\%\to28.2\%$).
  \end{minipage}
\end{figure}

We take the estimates from Section~\ref{S_emp} for the pooled sample and compare the optimal policy for the estimated $\beta$ to that of the restricted linear model under their respective peer effect estimates with a common budget of $B=100$ utility-index units.\footnote{The budget is small and holding peers' adoption probabilities at their baseline values, the optimal allocation raises average adoption by only about a third of a percentage point under either norm.} Figure \ref{fig:pooled-subsidy-comparison} presents the comparison. Of the 190 households funded under at least one norm (77 under CES and 132 in linear), only 19 households overlapping. Figure~\ref{fig:pooled-village-subsidy-comparison} presents the difference for an illustrative village. Who is targeted and by how much is fundamentally different under the two norms, with a Spearman rank correlation of $-0.52$. Because the pooled norm is strongly max-leaning, the CES planner concentrates the budget on a small set of high-centrality hubs in the effective network, whereas the linear planner spreads it far more evenly. For the same budget this gap in targeting translates into an increase in predicted adoption of $10.9$ percentage points for the CES case while only $1.9$ under the linear norm, reflecting the gap due to both the norm and the peer effect estimates.

Table~\ref{tab:pooled_targeting_characteristics} shows that the two norms disagree far less about \emph{how connected} their targets are than about \emph{who} those targets are. Both policies fund well-connected households, with mean degree near $15$ against $5.5$ in the population that is not targeted, and the households the two norms agree on are the most connected and wealthiest of all, although the size of the subsidy still differs. The decisive difference is whether the household is a leader or not. $62\%$ of the CES-only recipients are village leaders, against only $12\%$ of the linear-only recipients which is close to their proportion in the population. Leaders have a similar positive own coefficient under both norms ($0.56$ and $0.57$), with higher baseline adoption ($0.239$ versus $0.165$) and degree ($7.5$ versus $5.5$). Under a max-like norm a household's reference point is its highest-adopting neighbour, so a leader, who adopts more and has more links, generates especially strong spillovers when it is the maximal peer for others. 

This parallels \citet{banerjee2013}, where information about the programme was seeded through leaders (teachers, shopkeepers, and self-help-group heads) and they found that the centrality of these first contacts predicted participation. Although their mechanism is one of information diffusion, our estimated norm in the simultaneous belief-based game recovers the same pattern by identifying those leaders from the data. In the linear norm, leadership carries no special value as a neighbour's influence is independent of its likelihood of adoption and the budget instead flows to a broader, somewhat wealthier set of high-degree households. Misspecifying the norm as linear therefore does not just rescale a common ranking, it ignores the very households the max-leaning norm makes pivotal.

\begin{table}[!ht]
\begin{singlespacing}\centering\small
\caption{Household characteristics by subsidy-recipient group for pooled sample.}
\label{tab:pooled_targeting_characteristics}
\setlength{\tabcolsep}{5pt}
\begin{tabular}{lrrrrrrrr}
\toprule
Recipient group & $N$ & Degree & Rooms & Beds & Non-Hindu & SC/ST & Latrine & Leader\\
\midrule
CES only    & 58    & 15.9 & 2.22 & 0.69 & 12.1\% & 56.9\% & 17.2\% & 62.1\% \\
Linear only & 113   & 14.7 & 2.74 & 0.78 & 3.5\%  & 38.9\% & 33.6\% & 12.4\% \\
Both        & 19    & 21.2 & 3.21 & 1.05 & 10.5\% & 52.6\% & 36.8\% & 36.8\% \\
Neither     & 6,543 & 5.5  & 2.44 & 0.76 & 4.9\%  & 32.6\% & 26.1\% & 11.5\% \\
\bottomrule
\end{tabular}

\par\smallskip
\begin{minipage}{0.98\linewidth}\footnotesize
\emph{Notes:} Entries are group averages. A household is a recipient if its subsidy exceeds $10^{-6}$. Degree is the number of network links, and Rooms and Beds are the household's respective numbers of rooms and beds. SC/ST is the Scheduled-Caste or Scheduled-Tribe flag and Non-Hindu the non-Hindu household-head flag, while Latrine and Leader indicate whether the household has a latrine and is a village leader.
\end{minipage}
\end{singlespacing}
\end{table}

A similar exercise for illustrative segments (heterogeneous and rich) under the two wealth proxies is reported in Appendix~\ref{SD} in Figure~\ref{fig:het-r-subsidy-ranks} and Table~\ref{tab:het_r_targeting_characteristics} with a budget of $B=20$ for each segment. For the Caste-Beds segment the norm is again max-leaning ($\hat\beta\approx491$) and optimal subsidy is concentrated on $15$ high-degree households with $40\%$ of them being village leaders mirroring the pooled sample optimal policy. For the Caste-Latrine segment the estimated norm is instead minimum-like ($\hat\beta\approx-283$), and the targeting is very different. The CES planner now spreads the budget across $93$ households and raises adoption by only $0.9$ points, similar to the linear norm. Its recipients are low-degree ($6.1$ versus $13.6$ for the linear) but on average richer ($3.4$ rooms and $1.2$ beds, versus $2.3$ rooms and $0.6$ beds), and, in contrast to the maximum-like regime, they are not village leaders ($3.3\%$ versus $26.1\%$ under the linear norm). Under the min-like norm, it is more important to target the ones with the lowest adoption probability. This illustrates that a different estimated norm delivers very different policy conclusions and misspecification would induce bias.

\section{Concluding Remarks}\label{S_con}

With the growing availability of network data and the policy relevance of economic questions that require modelling peer effects, it is crucial to understand the implications of the underlying model primitives and assumptions. Any form of misspecification can lead to biased counterfactual predictions and, consequently, sub-optimal policy recommendations. This paper studies one particular form of misspecification in discrete choice models of peer effects, specifically the social norm or how peer influence is aggregated. Ex-ante assumptions on the social norm are typically ad hoc, and while recent work has relaxed them for continuous outcomes \citep{boucher2024peer,Tchuente2026}, many empirically relevant settings, for example microfinance adoption, have
discrete outcomes. We study norm
misspecification with both conformity and complementarity effects extending the CES norm of \citet{boucher2024peer} to a setting that raises challenges absent in the continuous case.

We model the game under incomplete information, where agents form beliefs about others' actions, and give a full structural treatment of the model. We establish existence and uniqueness of equilibrium, and identification of the structural parameters. Because the CES norm and the discrete choice setting makes the best-response system highly non-linear, none of these follows directly from the linear or continuous case. Building on \citet{aguirregabiria2007sequential} and \citet{Aguirregabiria2021}, we propose a consistent and asymptotically normal estimator of the structural parameters and equilibrium conditional choice probabilities. In Monte Carlo simulations, its peer-effect estimates approach their true values as the sample grows, whereas imposing a linear norm leaves them biased.

In the microfinance data of \citet{banerjee2013}, the pooled estimates reject the linear norm in favour of a max-like one, and imposing linearity overstates complementarity more than threefold while understating conformity. The norm also varies across caste and wealth segments, heterogeneity that pooling hides. The misspecification carries over to policy. The optimal subsidies under the estimated and the linear norm fund almost disjoint sets of households, and under the max-like norm the planner favours village leaders, echoing their role in \citet{banerjee2013}. More broadly, wherever peer effects shape take-up, from financial inclusion to other adoption programmes that governments and NGOs pursue, getting the norm wrong would not only bias the estimated peer effects but also misdirect the subsidy, so getting the shape of the norm right is important.

\bibliography{mybibliography}

@article{boucher2024peer,
author = {Boucher, Vincent and Rendall, Michelle and Ushchev, Philip and Zenou, Yves},
title = {Toward a General Theory of Peer Effects},
journal = {Econometrica},
volume = {92},
number = {2},
pages = {543-565},
doi = {https://doi.org/10.3982/ECTA21048},
url = {https://onlinelibrary.wiley.com/doi/abs/10.3982/ECTA21048},
eprint = {https://onlinelibrary.wiley.com/doi/pdf/10.3982/ECTA21048},
year = {2024}
}

@article{lambotte2025peer,
author = {Lambotte, Mathieu},
title = {Peer Effects in Binary Outcomes: Strategic Complementarity and Taste for Conformity With Endogenous Networks},
journal = {Journal of Applied Econometrics},
volume = {40},
number = {6},
pages = {608-626},
doi = {https://doi.org/10.1002/jae.3128},
url = {https://onlinelibrary.wiley.com/doi/abs/10.1002/jae.3128},
eprint = {https://onlinelibrary.wiley.com/doi/pdf/10.1002/jae.3128},
year = {2025}
}

@article{liu2014,
title = {Endogenous peer effects: local aggregate or local average?},
journal = {Journal of Economic Behavior \& Organization},
volume = {103},
pages = {39-59},
year = {2014},
issn = {0167-2681},
doi = {https://doi.org/10.1016/j.jebo.2014.03.025},
url = {https://www.sciencedirect.com/science/article/pii/S0167268114001036},
author = {Xiaodong Liu and Eleonora Patacchini and Yves Zenou},
}

@article{Kasahara2012,
author = {Kasahara, Hiroyuki and Shimotsu, Katsumi},
title = {Sequential Estimation of Structural Models With a Fixed Point Constraint},
journal = {Econometrica},
volume = {80},
number = {5},
pages = {2303-2319},
doi = {https://doi.org/10.3982/ECTA8291},
url = {https://onlinelibrary.wiley.com/doi/abs/10.3982/ECTA8291},
eprint = {https://onlinelibrary.wiley.com/doi/pdf/10.3982/ECTA8291},
year = {2012}
}

@article{aguirregabiria2007sequential,
  title={Sequential estimation of dynamic discrete games},
  author={Aguirregabiria, Victor and Mira, Pedro},
  journal={Econometrica},
  volume={75},
  number={1},
  pages={1--53},
  year={2007},
  publisher={Wiley Online Library}
}

@article{Lin2024,
title = {Binary choice with misclassification and social interactions, with an application to peer effects in attitude},
journal = {Journal of Econometrics},
volume = {238},
number = {1},
pages = {105551},
year = {2024},
issn = {0304-4076},
doi = {https://doi.org/10.1016/j.jeconom.2023.105551},
url = {https://www.sciencedirect.com/science/article/pii/S0304407623002671},
author = {Zhongjian Lin and Yingyao Hu},
}

@article{Aguirregabiria2021,
author = {Aguirregabiria, Victor and Marcoux, Mathieu},
title = {Imposing equilibrium restrictions in the estimation of dynamic discrete games},
journal = {Quantitative Economics},
volume = {12},
number = {4},
pages = {1223-1271},
doi = {https://doi.org/10.3982/QE1735},
url = {https://onlinelibrary.wiley.com/doi/abs/10.3982/QE1735},
eprint = {https://onlinelibrary.wiley.com/doi/pdf/10.3982/QE1735},
year = {2021}
}

@article{Cruz2006,
author = {La Cruz, William and Mart{\'i}nez, Jos{\'e} Mario and Raydan, Marcos},
year = {2006},
month = {07},
pages = {1429-1448},
title = {Spectral Residual Method without Gradient Information for Solving Large-Scale Nonlinear Systems of Equations},
volume = {75},
journal = {Mathematics of Computation},
number = {255},
doi = {10.1090/S0025-5718-06-01840-0}
}

@article{Menzel2016,
author = {Menzel, Konrad},
year = {2016},
month = {07},
number = {1}, 
pages = {306--337},
title = {Inference for Games with Many Players},
volume = {83},
journal = {The Review of Economic Studies},
doi = {10.1093/restud/rdv038}
}

@article{BrockDurlauf2001,
 ISSN = {00346527, 1467937X},
 URL = {http://www.jstor.org/stable/2695928},
 author = {William A. Brock and Steven N. Durlauf},
 journal = {The Review of Economic Studies},
 number = {2},
 pages = {235--260},
 publisher = {[Oxford University Press, Review of Economic Studies, Ltd.]},
 title = {Discrete Choice with Social Interactions},
 urldate = {2025-08-15},
 volume = {68},
 year = {2001}
}

@article{lee2014binarychoice,
	author = {Lee, Lung-fei and Li, Ji and Lin, Xu},
	doi = {10.1162/REST_a_00401},
	eprint = {https://direct.mit.edu/rest/article-pdf/96/3/402/1917628/rest\_a\_00401.pdf},
	issn = {0034-6535},
	journal = {The Review of Economics and Statistics},
	month = {07},
	number = {3},
	pages = {402-417},
	title = {Binary Choice Models with Social Network under Heterogeneous Rational Expectations},
	url = {https://doi.org/10.1162/REST\_a\_00401},
	volume = {96},
	year = {2014}}

@article{bramoulle2020peer,
  title={Peer effects in networks: A survey},
  author={Bramoull{\'e}, Yann and Djebbari, Habiba and Fortin, Bernard},
  journal={Annual Review of Economics},
  volume={12},
  number={1},
  pages={603--629},
  year={2020},
  publisher={Annual Reviews}
}

@TechReport{joshua2022,
  author={Shemesh, Joshua and Zapatero, Fernando and Zenou, Yves},
  title={{Heterogenous Peer Effects: How Community Connectivity Affects Car Purchases}},
  year=2022,
  month=5,
  institution={C.E.P.R. Discussion Papers},
  type={CEPR Discussion Papers},
  url={https://ideas.repec.org/p/cpr/ceprdp/17306.html},
  number={17306},
  doi={},
}

@article{yakovlev2018,
Author = {Yakovlev, Evgeny},
Title = {Demand for Alcohol Consumption in {Russia} and Its Implication for Mortality},
Journal = {American Economic Journal: Applied Economics},
Volume = {10},
Number = {1},
Year = {2018},
Month = {1},
Pages = {106–49},
DOI = {10.1257/app.20130170},
URL = {https://www.aeaweb.org/articles?id=10.1257/app.20130170}}

@article{bollinger2012peer,
  title={Peer effects in the diffusion of solar photovoltaic panels},
  author={Bollinger, Bryan and Gillingham, Kenneth},
  journal={Marketing Science},
  volume={31},
  number={6},
  pages={900--912},
  year={2012},
  publisher={INFORMS}
}

@article{calvo2009peer,
  title={Peer effects and social networks in education},
  author={Calv{\'o}-Armengol, Antoni and Patacchini, Eleonora and Zenou, Yves},
  journal={Review of Economic Studies},
  volume={76},
  number={4},
  pages={1239--1267},
  year={2009},
  publisher={Wiley-Blackwell}
}

@article{manski1993identification,
  title={Identification of endogenous social effects: The reflection problem},
  author={Manski, Charles F},
  journal={Review of Economic Studies},
  volume={60},
  number={3},
  pages={531--542},
  year={1993},
  publisher={Wiley-Blackwell}
}

@incollection{boucher2016,
    author = {Boucher, Vincent and Fortin, Bernard},
    isbn = {9780199948277},
    title = {Some Challenges in the Empirics of the Effects of Networks},
    booktitle = {The Oxford Handbook of the Economics of Networks},
    publisher = {Oxford University Press},
    year = {2016},
    month = {04},
    doi = {10.1093/oxfordhb/9780199948277.013.22},
    url = {https://doi.org/10.1093/oxfordhb/9780199948277.013.22},
    eprint = {https://academic.oup.com/book/0/chapter/212020532/chapter-pdf/44271983/oxfordhb-9780199948277-e-22.pdf},
}

@article{
banerjee2013,
author = {Abhijit Banerjee  and Arun G. Chandrasekhar  and Esther Duflo  and Matthew O. Jackson },
title = {The Diffusion of Microfinance},
journal = {Science},
volume = {341},
number = {6144},
pages = {1236498},
year = {2013},
doi = {10.1126/science.1236498},
URL = {https://www.science.org/doi/abs/10.1126/science.1236498},
eprint = {https://www.science.org/doi/pdf/10.1126/science.1236498},
}

@article{bajari2010identification,
 ISSN = {00129682, 14680262},
 URL = {http://www.jstor.org/stable/40928965},
 author = {Patrick Bajari and Han Hong and Stephen P. Ryan},
 journal = {Econometrica},
 number = {5},
 pages = {1529--1568},
 publisher = {[Wiley, Econometric Society]},
 title = {Identification And Estimation Of A Discrete Game Of Complete Information},
 urldate = {2025-05-03},
 volume = {78},
 year = {2010}
}

@incollection{BrockDurlauf2001b,
  author  = {Brock, William A. and Durlauf, Steven N.},
  title   = {Interactions-based Models},
  booktitle = {Handbook of Econometrics},
  year    = {2001},
  volume  = {5},
  pages   = {3297--3380},
  publisher = {Elsevier}
}

@incollection{blume2010,
  author={Blume, Lawrence E. and Brock, William A. and Durlauf, Steven N. and Ioannides, Yannis M.},
  title={Identification of Social Interactions},
  booktitle={Handbook of Social Economics}, editor={Benhabib, Jess and Bisin, Alberto and Jackson, Matthew O.},
  volume={1}, pages={853--964}, publisher={North-Holland}, year={2011}}

@article{Carrell2010,
journal={The Quarterly Journal of Economics},
author={Scott E. Carrell and Marianne E. Page and James E. West},
title={Sex and Science: How Professor Gender Perpetuates the Gender Gap},
year={2010},
month={},
pages={1101-1144},
volume={125},
number={3},
doi = {10.1162/qjec.2010.125.3.1101},
url={https://ideas.repec.org/a/oup/qjecon/v125y2010i3p1101-1144..html},
}

@article{Hahn2020Friendship,
  author       = {Hahn, Youjin and Islam, Asadul and Patacchini, Eleonora and Zenou, Yves},
  title        = {Friendship and Female Education: Evidence from a Field Experiment in {Bangladeshi} Primary Schools},
  journal      = {The Economic Journal},
  volume       = {130},
  number       = {627},
  pages        = {740--764},
  year         = {2020},
  month        = {04},
  doi          = {10.1093/ej/uez064}
}

@TechReport{Tatsi2015,
type={VfS Annual Conference 2015 (Muenster): Economic Development - Theory and Policy},
institution={Verein f{\"u}r Socialpolitik / German Economic Association},
author={Tatsi, Eirini},
title={Endogenous Social Interactions: Which Peers Matter?},
year={2015},
number={113168},
url={https://ideas.repec.org/p/zbw/vfsc15/113168.html},
}

@article{chambers2022,
	author = {Chambers, Christopher P and Cuhadaroglu, Tugce and Masatlioglu, Yusufcan},
	doi = {10.1093/jeea/jvac028},
	eprint = {https://academic.oup.com/jeea/article-pdf/21/1/135/49159753/jvac028.pdf},
	issn = {1542-4766},
	journal = {Journal of the European Economic Association},
	month = {05},
	number = {1},
	pages = {135-166},
	title = {Behavioral Influence},
	url = {https://doi.org/10.1093/jeea/jvac028},
	volume = {21},
	year = {2023}}

@article{guerra2022,
    author = {Guerra, José-Alberto and Mohnen, Myra},
    title = {Multinomial Choice with Social Interactions: Occupations in {Victorian London}},
    journal = {The Review of Economics and Statistics},
    volume = {104},
    number = {4},
    pages = {736-747},
    year = {2022},
    month = {07},
    issn = {0034-6535},
    doi = {10.1162/rest_a_00979},
    url = {https://doi.org/10.1162/rest\_a\_00979},
    eprint = {https://direct.mit.edu/rest/article-pdf/104/4/736/2033338/rest\_a\_00979.pdf},
}

@article{dePaula2018,
author = {de Paula, {\'A}ureo and Richards-Shubik, Seth and Tamer, Elie},
title = {Identifying Preferences in Networks With Bounded Degree},
journal = {Econometrica},
volume = {86},
number = {1},
pages = {263-288},
doi = {https://doi.org/10.3982/ECTA13564},
url = {https://onlinelibrary.wiley.com/doi/abs/10.3982/ECTA13564},
eprint = {https://onlinelibrary.wiley.com/doi/pdf/10.3982/ECTA13564},
year = {2018}
}

@article{Bowden1973,
 ISSN = {00129682, 14680262},
 URL = {http://www.jstor.org/stable/1914036},
 author = {Roger Bowden},
 journal = {Econometrica},
 number = {6},
 pages = {1069--1074},
 publisher = {[Wiley, Econometric Society]},
 title = {The Theory of Parametric Identification},
 urldate = {2025-08-18},
 volume = {41},
 year = {1973}
}

@article{Rothenberg1971,
 ISSN = {00129682, 14680262},
 URL = {http://www.jstor.org/stable/1913267},
 author = {Thomas J. Rothenberg},
 journal = {Econometrica},
 number = {3},
 pages = {577--591},
 publisher = {[Wiley, The Econometric Society]},
 title = {Identification in Parametric Models},
 urldate = {2025-08-18},
 volume = {39},
 year = {1971}
}

@book{Topkis1998,
 ISBN = {9780691032443},
 URL = {http://www.jstor.org/stable/j.ctt7s83q},
 author = {Donald Topkis},
 publisher = {Princeton University Press},
 title = {Supermodularity and Complementarity},
 urldate = {2025-08-19},
 year = {1998}
}

@article{BrockDurlauf2007,
  author  = {Brock, William A. and Durlauf, Steven N.},
  title   = {Identification of Binary Choice Models with Social Interactions},
  journal = {Journal of Econometrics},
  year    = {2007},
  volume  = {140},
  number  = {1},
  pages   = {52--75},
}

@TechReport{bajari2015dynamic,
  author={Patrick Bajari and Victor Chernozhukov and Han Hong and Denis Nekipelov},
  title={{Identification and Efficient Semiparametric Estimation of a Dynamic Discrete Game}},
  year=2015,
  month=Apr,
  institution={National Bureau of Economic Research, Inc},
  type={NBER Working Papers},
  url={https://ideas.repec.org/p/nbr/nberwo/21125.html},
  number={21125},
  doi={},
}

@article{tao2014,
author = {Tao, Ji and Lee, Lung-fei},
title = {A social interaction model with an extreme order statistic},
journal = {The Econometrics Journal},
volume = {17},
number = {3},
pages = {197-240},
doi = {https://doi.org/10.1111/ectj.12031},
url = {https://onlinelibrary.wiley.com/doi/abs/10.1111/ectj.12031},
eprint = {https://onlinelibrary.wiley.com/doi/pdf/10.1111/ectj.12031},
year = {2014}
}

@article{diaz2021,
title = {Leaders in juvenile crime},
journal = {Journal of Economic Behavior \& Organization},
volume = {192},
pages = {638-667},
year = {2021},
issn = {0167-2681},
doi = {https://doi.org/10.1016/j.jebo.2021.10.034},
url = {https://www.sciencedirect.com/science/article/pii/S0167268121004649},
author = {Carlos Díaz and Eleonora Patacchini and Thierry Verdier and Yves Zenou},
}

@techreport{islam2021,
  title        = {Centrality-Based Spillover Effects},
  author       = {Islam, Asadul and Vlassopoulos, Michael and Zenou, Yves and Zhang, Xin},
  institution  = {Centre for Economic Policy Research (CEPR)},
  type         = {Discussion Paper},
  number       = {DP16321},
  year         = {2021},
  month        = jul,
  note         = {Available at SSRN},
}

@article{bietenbeck2020,
    author = {Bietenbeck, Jan},
    title = {The Long-Term Impacts of Low-Achieving Childhood Peers: Evidence from {Project STAR}},
    journal = {Journal of the European Economic Association},
    volume = {18},
    number = {1},
    pages = {392-426},
    year = {2020},
    month = {01},
    issn = {1542-4766},
    doi = {10.1093/jeea/jvy049},
    url = {https://doi.org/10.1093/jeea/jvy049},
    eprint = {https://academic.oup.com/jeea/article-pdf/18/1/392/33050701/jvy049.pdf},
}

@unpublished{hoxby2005,
  title        = {Taking Race Out of the Equation: School Reassignment and the Structure of Peer Effects},
  author       = {Hoxby, Caroline and Weingarth, Gretchen},
  year         = {2005},
  note         = {Unpublished manuscript, Harvard University},
}

@article{Pesendorfer2008,
  author    = {Martin Pesendorfer and Paul Schmidt-Dengler},
  title     = {Asymptotic Least Squares Estimators for Dynamic Games},
  journal   = {The Review of Economic Studies},
  year      = {2008},
  volume    = {75},
  number    = {3},
  pages     = {901--928},
}

@article{Pesendorfer2010,
  author    = {Martin Pesendorfer and Paul Schmidt-Dengler},
  title     = {Sequential Estimation of Dynamic Discrete Games: A Comment},
  journal   = {Econometrica},
  year      = {2010},
  volume    = {78},
  number    = {2},
  pages     = {833--842}
}

@misc{Tchuente2026,
      title={Who Matters to Whom? Identifying Peer Effects with Propagation Geometry}, 
      author={Guy Tchuente},
      year={2026},
      note={arXiv:2602.23594}, 
      eprint={2602.23594},
      archivePrefix={arXiv},
      primaryClass={econ.EM},
      url={https://arxiv.org/abs/2602.23594}, 
}

@techreport{volpe2025discrete,
  title={Discrete Choice with Generalized Social Interactions},
  author={Volpe, Oscar},
  year={2025},
  institution={Working paper, Harvard University}
}

@article{Cai2015,
Author = {Cai, Jing and De Janvry, Alain and Sadoulet, Elisabeth},
Title = {Social Networks and the Decision to Insure},
Journal = {American Economic Journal: Applied Economics},
Volume = {7},
Number = {2},
Year = {2015},
Month = {April},
Pages = {81–108},
DOI = {10.1257/app.20130442},
URL = {https://www.aeaweb.org/articles?id=10.1257/app.20130442}}

@article{banerjee2015miracle,
  author={Banerjee, Abhijit and Duflo, Esther and Glennerster, Rachel and Kinnan, Cynthia}, title={The Miracle of Microfinance? Evidence from a Randomized Evaluation},
  journal={American Economic Journal: Applied Economics}, volume={7}, number={1}, pages={22--53}, year={2015}}

@article{banerjeekarlanzinman2015,
  author={Banerjee, Abhijit and Karlan, Dean and Zinman, Jonathan}, title={Six Randomized Evaluations of Microcredit: Introduction and Further Steps},
  journal={American Economic Journal: Applied Economics}, volume={7}, number={1}, pages={1--21}, year={2015}}

@article{banerjee2019gossip,
  author={Banerjee, Abhijit and Chandrasekhar, Arun G. and Duflo, Esther and Jackson, Matthew O.}, title={Using Gossips to Spread Information: Theory and Evidence from Two Randomized Controlled Trials},
  journal={Review of Economic Studies}, volume={86}, number={6}, pages={2453--2490}, year={2019}}

@article{munshirosenzweig2006,
  author={Munshi, Kaivan and Rosenzweig, Mark}, title={Traditional Institutions Meet the Modern World: Caste, Gender, and Schooling Choice in a Globalizing Economy},
  journal={American Economic Review}, volume={96}, number={4}, pages={1225--1252}, year={2006}}

@article{beaman2021network,
  author={Beaman, Lori and BenYishay, Ariel and Magruder, Jeremy and Mobarak, Ahmed Mushfiq}, title={Can Network Theory-Based Targeting Increase Technology Adoption?},
  journal={American Economic Review}, volume={111}, number={6}, pages={1918--1943}, year={2021}}

@article{galeotti2020targeting,
  author={Galeotti, Andrea and Golub, Benjamin and Goyal, Sanjeev}, title={Targeting Interventions in Networks},
  journal={Econometrica}, volume={88}, number={6}, pages={2445--2471}, year={2020}}

@article{GeanakoplosPearceStacchetti1989,
  author  = {Geanakoplos, John and Pearce, David and Stacchetti, Ennio},
  title   = {Psychological Games and Sequential Rationality},
  journal = {Games and Economic Behavior},
  volume  = {1},
  number  = {1},
  pages   = {60--79},
  year    = {1989},
  doi     = {10.1016/0899-8256(89)90005-5}
}

@article{BattigalliDufwenberg2009,
  author  = {Battigalli, Pierpaolo and Dufwenberg, Martin},
  title   = {Dynamic Psychological Games},
  journal = {Journal of Economic Theory},
  volume  = {144},
  number  = {1},
  pages   = {1--35},
  year    = {2009},
  doi     = {10.1016/j.jet.2008.01.004}
}

@article{BursztynGonzalezYanagizawaDrott2020,
  author  = {Bursztyn, Leonardo and Gonz\'{a}lez, Alessandra L. and Yanagizawa-Drott, David},
  title   = {Misperceived Social Norms: Women Working Outside the Home in {Saudi Arabia}},
  journal = {American Economic Review},
  volume  = {110},
  number  = {10},
  pages   = {2997--3029},
  year    = {2020},
  doi     = {10.1257/aer.20180975}
}

@article{chunglu2002,
  author  = {Chung, Fan and Lu, Linyuan},
  title   = {Connected Components in Random Graphs with Given Expected Degree Sequences},
  journal = {Annals of Combinatorics},
  volume  = {6},
  number  = {2},
  pages   = {125--145},
  year    = {2002},
  doi     = {10.1007/PL00012580}
}

@article{feigenberg2013social,
  author  = {Feigenberg, Benjamin and Field, Erica and Pande, Rohini},
  title   = {The Economic Returns to Social Interaction: Experimental Evidence from Microfinance},
  journal = {Review of Economic Studies},
  year    = {2013}, volume = {80}, number = {4}, pages = {1459--1483}
}

@article{ballester2006,
  author={Ballester, Coralio and Calv\'o-Armengol, Antoni and Zenou, Yves},
  title={Who's Who in Networks. Wanted: The Key Player},
  journal={Econometrica}, year={2006}, volume={74}, number={5}, pages={1403--1417}}

@article{mas2009peers,
  author={Mas, Alexandre and Moretti, Enrico},
  title={Peers at Work}, journal={American Economic Review},
  year={2009}, volume={99}, number={1}, pages={112--145}}

@article{boucherbramoulle2026binary,
  author  = {Boucher, Vincent and Bramoull\'{e}, Yann},
  title   = {Binary Outcomes and Linear Interactions},
  journal = {Journal of Business \& Economic Statistics},
  year    = {2026},
  volume  = {44},
  number  = {3},
  pages   = {865--875},
  doi     = {10.1080/07350015.2025.2565281}
}

@article{koszegi2006reference,
  author  = {K{\H{o}}szegi, Botond and Rabin, Matthew},
  title   = {A Model of Reference-Dependent Preferences},
  journal = {Quarterly Journal of Economics},
  year    = {2006}, volume = {121}, number = {4}, pages = {1133--1165}
}

@article{charness2019opportunistic,
  title   = {Opportunistic conformism},
  author  = {Charness, Gary and Naef, Michael and Sontuoso, Alessandro},
  journal = {Journal of Economic Theory},
  volume  = {180},
  pages   = {100--134},
  year    = {2019},
  doi     = {10.1016/j.jet.2018.12.003},
  publisher = {Elsevier}
}

@article{BattigalliDufwenberg2022,
  title   = {Belief-Dependent Motivations and Psychological Game Theory},
  author  = {Battigalli, Pierpaolo and Dufwenberg, Martin},
  journal = {Journal of Economic Literature},
  volume  = {60},
  number  = {3},
  pages   = {833--882},
  year    = {2022},
  doi     = {10.1257/jel.20201378}
}

@article{norros2006conditionally,
  title     = {On a conditionally {P}oissonian graph process},
  author    = {Norros, Ilkka and Reittu, Hannu},
  journal   = {Advances in Applied Probability},
  volume    = {38},
  number    = {1},
  pages     = {59--75},
  year      = {2006},
  publisher = {Cambridge University Press}
}

@article{bhattacharya2024,
  author  = {Bhattacharya, Debopam and Dupas, Pascaline and Kanaya, Shin},
  title   = {Demand and Welfare Analysis in Discrete Choice Models with Social Interactions},
  journal = {Review of Economic Studies},
  year    = {2024},
  volume  = {91},
  number  = {2},
  pages   = {748--784}
}

@article{bursztyn2014peer,
  author  = {Bursztyn, Leonardo and Ederer, Florian and Ferman, Bruno and Yuchtman, Noam},
  title   = {Understanding Mechanisms Underlying Peer Effects: Evidence from a Field Experiment on Financial Decisions},
  journal = {Econometrica},
  year    = {2014}, volume = {82}, number = {4}, pages = {1273--1301}
}

@article{meng2025,
title = {Targeting network intervention with social norm},
journal = {Journal of Economic Theory},
volume = {230},
pages = {106092},
year = {2025},
issn = {0022-0531},
doi = {https://doi.org/10.1016/j.jet.2025.106092},
url = {https://www.sciencedirect.com/science/article/pii/S0022053125001383},
author = {Dawen Meng},
}

@misc{houndetoungan2025quantile,
  author={Houndetoungan, Aristide}, title={Quantile Peer Effect Models},
  year={2025}, note={arXiv:2506.12920}}

@misc{herstad2026heterogeneous,
  author={Herstad, Eyo I. and Houndetoungan, Aristide and Shin, Myungkou},
  title={Identification of Heterogeneous Peer Effects}, year={2026},
  note={arXiv:2410.14317}}

@misc{lambotte2025heterogeneity,
  author={Lambotte, Mathieu}, title={Heterogeneity in Peer Effects for Binary Outcomes},
  year={2025}, note={arXiv:2511.15891}}

@article{houndetoungan2026count,
  author={Houndetoungan, Aristide},
  title={Count Data Models with Heterogeneous Peer Effects under Rational Expectations},
  journal={Journal of Applied Econometrics},
  year={2026}, volume={41}, number={3}, pages={295--309},
  doi={10.1002/jae.70045}}

@misc{houndetoungan2026asymmetries,
  author={Houndetoungan, Aristide and Lambotte, Mathieu},
  title={Asymmetries in Peer Effects}, year={2026}, note={arXiv:2608.09219}}

@article{birgin2000nonmonotone, author={Birgin, Ernesto G. and Mart{\'\i}nez, Jos{\'e} Mario and Raydan, Marcos}, title={Nonmonotone Spectral Projected Gradient Methods on Convex Sets}, journal={SIAM Journal on Optimization}, volume={10}, number={4}, pages={1196--1211}, year={2000}}

@article{nakajima2007,
  author  = {Nakajima, Ryo},
  title   = {Measuring Peer Effects on Youth Smoking Behaviour},
  journal = {The Review of Economic Studies},
  volume  = {74},
  number  = {3},
  pages   = {897--935},
  year    = {2007},
  doi     = {10.1111/j.1467-937X.2007.00448.x}
}

\newpage
\pagenumbering{roman}
\setcounter{page}{1}
\counterwithin{figure}{section}
\counterwithin{table}{section}
\appendix

\section*{Appendix}\label{SA}

\renewcommand{\thesection}{\Alph{section}}
\renewcommand{\theHassumption}{appendix.\thesection.\arabic{assumption}}
\renewcommand{\theHdefinition}{appendix.\thesection.\arabic{definition}}

\section{Proofs of Main Results}

\subsection{Existence}\label{SA.1}

In this section we prove the Existence of a Rational Expectations equilibrium of the game using Tarski's Fixed Point Theorem. We first state the Theorem.

\begin{definition}[Tarski's Fixed Point Theorem]
Let \( L \) be a complete lattice and let \( f: L \to L \) be an order-preserving (i.e., monotone) function. Then the set of fixed points of \( f \) is a non-empty complete lattice. In particular, \( f \) has a least and a greatest fixed point.
\end{definition}
Moreover, define the component-wise order as below:
\begin{definition}[Component-wise Order]\label{def:comp}
Let \( \mathbf{x} = (x_1, \dots, x_n) \in \mathbb{R}^N \) and \( \mathbf{y} = (y_1, \dots, y_n) \in \mathbb{R}^N \). We define the component-wise (partial) order as follows:
\begin{itemize}
    \item \( \mathbf{x} \geq \mathbf{y} \) if and only if \( x_i \geq y_i \) for all \( i = 1, \dots, N \),
    \item \( \mathbf{x} > \mathbf{y} \) if and only if \( \mathbf{x} \geq \mathbf{y} \) and there exists at least one index \( j \) such that \( x_j > y_j \).
\end{itemize}
\end{definition}

Note that L $:= [0,1]^N$ is a complete lattice under the component-wise order. For any subset \( S \subseteq [0,1]^N \), define the infimum and supremum component-wise:
\[
\inf S = \left( \inf_{x \in S} x_1, \dots, \inf_{x \in S} x_n \right), \quad 
\sup S = \left( \sup_{x \in S} x_1, \dots, \sup_{x \in S} x_n \right)
\]
Since each \( [0,1] \) is a complete lattice under the usual order, each component infimum and supremum exists and lies in \( [0,1] \). Hence, \( [0,1]^N \) is a complete lattice under the component-wise partial order. We need to show that $\Gamma(p,\boldsymbol \theta)$ is a monotone function in p then by Tarski's Fixed Point Theorem, the set of fixed points of $\Gamma$ would be non-empty and in itself form a complete lattice under pair-wise comparison i.e. there exist a least and greatest fixed point.

We now present the proof of Theorem~\ref{thm:existence}.

\bigskip

\begin{customproof}{Theorem \ref{thm:existence}}

We will use Tarski's fixed point theorem to prove existence.

\textbf{Self-map.} Since \( F_{\eta_i}\in(0,1) \) for every \( i \), we have
\( \Gamma(\mathbf p;\boldsymbol \theta)\in(0,1)^N\subset[0,1]^N \), so \( \Gamma \) maps
\( [0,1]^N \) into itself.

\textbf{Complete Lattice.} The domain \( [0,1]^N \), equipped with the component-wise partial order (Definition \ref{def:comp}), is a complete lattice. For any subset \( S \subseteq [0,1]^N \), the infimum and supremum are defined component wise and lie in \( [0,1]^N \).

\textbf{Monotonicity.} We show that \( \Gamma(\cdot; \boldsymbol \theta) \) is monotone. Let \( \mathbf{p}, \mathbf{p}' \in [0,1]^N \) with \( \mathbf{p} \leq \mathbf{p}' \) component wise. Consider:
\[
\tilde{y}^e_{-i}(\mathbf{p}) = \left( \sum_{j=1}^N \tilde{g}_{ij} p_j^\beta \right)^{1/\beta}, \quad \text{with } \tilde{g}_{ij} \geq 0.
\]
Since \( p_j \leq p_j' \) and all \( \tilde{g}_{ij} \geq 0 \), it follows that:
\[
\tilde{y}^e_{-i}(\mathbf{p}) \leq \tilde{y}^e_{-i}(\mathbf{p}').
\]
Now, using Assumption~\ref{asum:A5}, we have \( \theta_1 d_i + \theta_2 > 0 \). Therefore,
\[
z_i(\mathbf{p}) = \alpha_i + (\theta_1 d_i + \theta_2) \tilde{y}^e_{-i}(\mathbf{p},\beta) - \frac{\theta_2}{2}
\leq
z_i(\mathbf{p}') = \alpha_i + (\theta_1 d_i + \theta_2) \tilde{y}^e_{-i}(\mathbf{p}',\beta) - \frac{\theta_2}{2}.
\]
Since \( F_{\eta_i} \) is strictly increasing (as the logistic CDF), we have
\[
\Gamma_i(\mathbf{p}; \boldsymbol \theta) = F_{\eta_i}(z_i(\mathbf{p})) \leq F_{\eta_i}(z_i(\mathbf{p}')) = \Gamma_i(\mathbf{p}'; \boldsymbol \theta).
\]
Thus, \( \Gamma(\cdot; \boldsymbol \theta) \) is monotone on \( [0,1]^N \). Since \( \Gamma \) is a monotone function from a complete lattice into itself, by Tarski’s Fixed Point Theorem, the set of fixed points of \( \Gamma \) is non-empty and forms a complete lattice. Therefore, there exist minimal ($\mathbf{\underline{p}}^*$) and maximal ($\mathbf{\overline{p}}^*$) fixed points with
\[
\mathbf{\underline{p}}^* = \Gamma(\mathbf{\underline{p}}^*; \boldsymbol \theta) \ \text{and} \ \mathbf{\overline{p}}^{*} = \Gamma(\mathbf{\overline{p}}^{*}; \boldsymbol \theta)
\]
\end{customproof}

This result yields implications beyond mere existence. It establishes that the set of fixed points of $\Gamma$ is a complete lattice in the induced order, with a least and a greatest element. However, it need not be a sublattice of $[0,1]^N$, so the component-wise maximum of two fixed points need not be a fixed point.

\subsection{Uniqueness}\label{SA.2}

We first define the logarithmic $L_\infty$ distance, also known as the Thompson metric, used in the uniqueness proof.
\begin{definition}[Logarithmic $L_\infty$ distance]\label{def:log_metric}
For any $\mathbf p,\mathbf q\in\mathbb R_{++}^N$, define
\[
d_{\log,\infty}(\mathbf p,\mathbf q)
:=
\left\|
\log\mathbf p-\log\mathbf q
\right\|_\infty
=
\max_{1\leq i\leq N}
\left|
\log\left(\frac{p_i}{q_i}\right)
\right|.
\]
\end{definition}
The logarithm is applied component-wise. The Thompson metric measures the largest proportional difference between corresponding components of two strictly positive vectors. We also re-state Assumption 1 of \citet{boucher2024peer} which holds here too.
\begin{assumption*}[Assumption 1 of \citet{boucher2024peer}]\label{asum:A1_boucher2024}
For all $i = 1,2,\ldots,N$, agent $i$'s social norm $\tilde{y}^e_i: \mathbb{R}^N \rightarrow \mathbb{R}$ is monotone increasing, continuous, linearly homogeneous, and normalised such that $\tilde{y}^e_i(\mathbf{1}) = 1$.
\end{assumption*}

The condition in Assumption~\ref{asum:A6} is \emph{radial} (multiplicative) instead of the usual additive restrictions used in the linear case. An additive condition contracts absolute differences $\lvert p_i-p_i'\rvert$ in beliefs, whereas our radial condition contracts relative differences $\lvert\log(p_i/p_i')\rvert$ (the Thompson metric). Because the CES norm is scale-free, the relative scale is the natural one. \citet[appendix]{Tchuente2026} gives an additive, Lipschitz-based uniqueness proof for the CES norm, but that argument does not generally deliver a contraction when $\beta<1$, for which the norm need not be globally non-expansive in levels. For $0\le\beta<1$, $\partial\tilde y^e_{-i}/\partial p_j \propto p_j^{\beta-1}$ diverges as a belief approaches zero, so a Lipschitz bound holds only on beliefs bounded away from zero, with a constant that depends on $\beta$ and that range. For $\beta<0$ a Lipschitz bound holds on the whole domain, but its constant, up to $d_i^{1/|\beta|}$, exceeds one and diverges as $\beta\to0^-$, so peer effects must be correspondingly smaller. Our radial condition needs no such restriction. Monotonicity and degree-one homogeneity make the CES norm non-expansive in the Thompson metric for every $\beta$, and the best-response elasticity vanishes as the norm approaches zero, so uniqueness holds on the entire domain under an explicit condition.

We now present the proof of Theorem~\ref{thm:uniqueness}.

\bigskip

\begin{customproof}{Theorem \ref{thm:uniqueness}}

By Theorem \ref{thm:existence}, $\Gamma(\cdot)$ has a least fixed point $\underline{\mathbf p}^{*}$ and a greatest fixed point $\overline{\mathbf p}^{*}$ satisfying $ \underline{\mathbf p}^{*} \leq \overline{\mathbf p}^{*}.$
Because the logistic CDF takes values strictly between zero and one, both fixed points are strictly positive. Applying the CES transformation then implies that the corresponding social norms $\tilde{\mathbf{y}}^e(\cdot)$ at the extreme fixed points are also strictly positive in the absence of isolated agents.

Let $D=d_{\log,\infty}(\overline{\mathbf p}^{*},\underline{\mathbf p}^{*})$ denote the
Thompson distance between the extreme fixed points. Since both are fixed points,
\begin{equation}\label{eq:unique_1}
D=\left\|\log\Gamma(\overline{\mathbf p}^{*};\boldsymbol \theta)-\log\Gamma(\underline{\mathbf p}^{*};\boldsymbol \theta)\right\|_\infty.
\end{equation}
We will show that Assumption~\ref{asum:A6} forces $D=0$.

Let $\underline y_i=\tilde y^e_{-i}(\underline{\mathbf p}^{*};\beta)$ and $\overline y_i=\tilde y^e_{-i}(\overline{\mathbf p}^{*};\beta)$ for the norms at the two extreme fixed points. By Assumption~\ref{asum:A6} these lie in the interval $[a,b]$ on which the best-response elasticity, $\varepsilon_i^{BR}$, is bounded. From
\eqref{eq:best_response_elasticity}, the elasticity of $\Gamma_i(\cdot)$ with respect to the norm of $i$ is,
\[
\varepsilon_i^{BR}(\tilde y^e_{-i})
=(\theta_1d_i+\theta_2)\,\tilde y^e_{-i}\,
\bigl[1-F_\eta\bigl(\alpha_i-\tfrac{\theta_2}{2}+(\theta_1d_i+\theta_2)\tilde y^e_{-i}\bigr)\bigr].
\]
By the mean value theorem, there exists a 
$y_i^\dagger\in[\underline y_i,\overline y_i]\subseteq[a,b]$ such that
\[
\left|\log\Gamma_i(\overline{\mathbf p}^{*};\boldsymbol \theta)-\log\Gamma_i(\underline{\mathbf p}^{*};\boldsymbol \theta)\right|
=\varepsilon_i^{BR}(y_i^\dagger)\,
\left|\log\tilde y^e_{-i}(\overline{\mathbf p}^{*};\beta)-\log\tilde y^e_{-i}(\underline{\mathbf p}^{*};\beta)\right|.
\]
By Assumption~\ref{asum:A6} we get that
\begin{align}
\left|
\log\Gamma_i(\overline{\mathbf p}^{*};\boldsymbol \theta)
-
\log\Gamma_i(\underline{\mathbf p}^{*};\boldsymbol \theta)
\right|
&\leq
\overline\varepsilon^{BR}
\left|
\log\tilde y^e_{-i}(\overline{\mathbf p}^{*};\beta)
-
\log\tilde y^e_{-i}(\underline{\mathbf p}^{*};\beta)
\right|,
\label{eq:unique_2}
\end{align}
where $\overline\varepsilon^{BR}=\max_{1\le i\le N}\ \sup_{\tilde y^e_{-i}\in[a,b]}\ \varepsilon_i^{BR}$. Taking the maximum over $i$ and using equation~\eqref{eq:unique_1} gives
\begin{align}
D
\leq
\overline\varepsilon^{BR}
\left\|
\log\tilde{\mathbf y}^{e}(\overline{\mathbf p}^{*};\beta)
-
\log\tilde{\mathbf y}^{e}(\underline{\mathbf p}^{*};\beta)
\right\|_\infty .
\label{eq:unique_3}
\end{align}

For every $i$, the definition of $D$ and the ordering of the extremal fixed
points implies
\begin{align*}
 0
\leq
\log\left(
\frac{\overline p_i^{*}}{\underline p_i^{*}}
\right)
\leq D  \quad \Longleftrightarrow \quad \underline{\mathbf p}^{*}
\leq
\overline{\mathbf p}^{*}
\leq
\exp(D)\,\underline{\mathbf p}^{*}.
\end{align*}

Applying the CES transformation and using its monotonicity and homogeneous of degree one property gives us,
\[
\tilde{\mathbf y}^{e}(\underline{\mathbf p}^{*};\beta)
\leq
\tilde{\mathbf y}^{e}(\overline{\mathbf p}^{*};\beta)
\leq
\exp(D)\,
\tilde{\mathbf y}^{e}(\underline{\mathbf p}^{*};\beta).
\]
Finally, taking component-wise logarithms therefore yields
\begin{align}
\left\|
\log\tilde{\mathbf y}^{e}(\overline{\mathbf p}^{*};\beta)
-
\log\tilde{\mathbf y}^{e}(\underline{\mathbf p}^{*};\beta)
\right\|_\infty
\leq D.
\label{eq:unique_4}
\end{align}

Combining \eqref{eq:unique_3} and \eqref{eq:unique_4} gives $ D\leq\overline\varepsilon^{BR}D$. Since Assumption \ref{asum:A6} requires $\overline\varepsilon^{BR}<1$, this implies $D=0$ and hence
\[
\overline{\mathbf p}^{*}
=
\underline{\mathbf p}^{*}.
\]
Since the maximal and minimal fixed points coincide, this implies that the set of fixed points of $\Gamma$ is a singleton.
\end{customproof}

\bigskip

Relatedly, we also prove a property concerning the Jacobian of the Best-Response mapping which is used in the proofs in Sections~\ref{SA.3} and \ref{SA.4}. For its proof, we require the following definitions:

\begin{definition}[Spectral Radius]\label{def:rho}
For $A\in\mathbb R^{N\times N}$ with (complex) eigenvalues $\lambda_1,\dots,\lambda_N$, the
\emph{spectral radius} is
\[
\rho(A):=\max_{1\le k\le N}|\lambda_k|,
\]
the largest modulus of its eigenvalues.
\end{definition}

\begin{definition}[Spectral Radius: Invertibility, Row-Sum Bound, and Similarity]\label{def:specrad}
Let $A\in\mathbb R^{N\times N}$.
\begin{itemize}
\item \emph{Invertibility (Neumann series):} if $\rho(A)<1$, then $I-A$ is invertible and
$(I-A)^{-1}=\sum_{k=0}^{\infty}A^{k}$.
\item \emph{Row-sum bound:} $\rho(A)\le\|A\|_\infty=\max_{1\le i\le N}\sum_{j=1}^{N}|A_{ij}|$;
if $A\ge\mathbf 0$ entrywise, this equals $\max_i\sum_j A_{ij}$.
\item \emph{Similarity invariance:} for any invertible diagonal $D$,
$\rho(D^{-1}AD)=\rho(A)$.
\end{itemize}
\end{definition}

\begin{definition}[Euler's Theorem for Homogeneous Functions]\label{def:euler}
If $g:\mathbb R^{N}_{++}\to\mathbb R$ is differentiable and homogeneous of degree one, then
$\sum_{j=1}^{N} x_j\,\dfrac{\partial g}{\partial x_j}(\mathbf x)=g(\mathbf x)$ for every
$\mathbf x\in\mathbb R^{N}_{++}$.
\end{definition}

\begin{definition}[Partitioned Inverse]\label{def:partinv}
Let $A$ be an invertible matrix, partitioned as
\[
A=\begin{pmatrix}A_{SS}&A_{ST}\\ A_{TS}&A_{TT}\end{pmatrix},
\]
with $A_{TT}$ invertible. Then $A_{SS}-A_{ST}A_{TT}^{-1}A_{TS}$ is invertible, and
\[
\bigl[A^{-1}\bigr]_{SS}=\bigl(A_{SS}-A_{ST}A_{TT}^{-1}A_{TS}\bigr)^{-1}.
\]
\end{definition}

\bigskip

\begin{lemma}[Best-Response Jacobian]\label{lem:jacobian}
Let $\mathbf p^*=\Gamma(\mathbf p^*;\boldsymbol \theta)$ be the equilibrium and $J=\partial\Gamma(\mathbf p^*;\theta)/\partial\mathbf p^\top$ the Jacobian of the best-response map at $\mathbf p^*$. Then under Assumptions~\ref{asum:A1}, \ref{asum:A5} and~\ref{asum:A6}, for every set $S$ of individuals,
\[
I-J,\qquad I-J_{SS}\qquad\text{and}\qquad\bigl[(I-J)^{-1}\bigr]_{SS}
\]
are invertible, where the subscript $SS$ denotes the block with rows and columns in $S$.
\end{lemma}
\begin{proof}
By Assumption~\ref{asum:A1}, $F_\eta$ is logistic, so its density at the equilibrium is $p_i^*(1-p_i^*)$. Differentiating $\Gamma_i$ therefore gives zero rows of $J$ for isolated individuals and
\[
J_{ij}=p_i^*(1-p_i^*)(\theta_1d_i+\theta_2)
\frac{\partial\tilde y^e_{-i}(\mathbf p^*;\beta)}{\partial p_j},
\qquad d_i>0.
\]
By Assumption~\ref{asum:A5}, $\theta_1d_i+\theta_2>0$, and the norm is increasing in each $p_j$ (Assumption~1 of \citet{boucher2024peer}, restated above), so $J\geq\mathbf 0$. The similar matrix
\[
\widetilde J=\operatorname{diag}(\mathbf p^*)^{-1}J\operatorname{diag}(\mathbf p^*)
\]
is therefore nonnegative and has the same spectral radius as $J$ (Definition~\ref{def:specrad}). The norm is homogeneous of degree one, so, by Euler's theorem (Definition~\ref{def:euler}), the row sums of $\widetilde J$ are
\[
\begin{aligned}
\sum_j\widetilde J_{ij}
&=(1-p_i^*)(\theta_1d_i+\theta_2)\sum_j
\frac{\partial\tilde y^e_{-i}(\mathbf p^*;\beta)}{\partial p_j}\,p_j^*\\
&=(1-p_i^*)(\theta_1d_i+\theta_2)\,\tilde y^e_{-i}(\mathbf p^*;\beta)
=\varepsilon_i^{BR}\leq\overline\varepsilon^{BR}<1
\end{aligned}
\]
for $d_i>0$, by \eqref{eq:best_response_elasticity} and Assumption~\ref{asum:A6}, since the equilibrium norm lies in $[a,b]$. By the row-sum bound (Definition~\ref{def:specrad}),
\[
\rho(J)=\rho(\widetilde J)<1,
\]
so $I-J$ is invertible (Definition~\ref{def:specrad}). For a set $S$, $J_{SS}$ is similar to $\widetilde J_{SS}$, which is nonnegative with row sums no larger than those of $\widetilde J$, so $\rho(J_{SS})<1$ and $I-J_{SS}$ is invertible. Let $T$ be the complement of $S$. Since $I-J$ and $I-J_{TT}$ are invertible, Definition~\ref{def:partinv} with $A=I-J$ gives
\[
\bigl[(I-J)^{-1}\bigr]_{SS}
=\bigl(I-J_{SS}-J_{ST}(I-J_{TT})^{-1}J_{TS}\bigr)^{-1},
\]
so the block is invertible. If $T$ is empty, the block is $(I-J)^{-1}$ itself.
\end{proof}

\subsection{Identification}\label{SA.3}

Before we prove Theorem~\ref{thm:ident}, we state intermediate definitions and results used for it. First we define the following:

\begin{definition}[Power-Mean Inequality]\label{def:powermean}
For positive numbers $p_1,\dots,p_n$, the power mean $\bigl(\frac1n\sum_{j=1}^np_j^t\bigr)^{1/t}$, read as the geometric mean at $t=0$, is non-decreasing in $t$, and strictly increasing unless $p_1=\dots=p_n$.
\end{definition}

\begin{definition}[Real Analytic Functions]\label{def:real_analytic}
A function $f$ on an open set $U\subseteq\mathbb R^n$ is \emph{real analytic} if, near every point of $U$, it equals a convergent power series. We use two standard facts.
\begin{itemize}
\item Sums, products, quotients and compositions of real analytic functions, including exponentials and logarithms, are real analytic where defined.
\item \emph{Identity theorem:} if $U$ is connected and $f=0$ on a non-empty open subset of $U$, then $f=0$ on all of $U$.
\end{itemize}
\end{definition}

We also formally define variation in one's neighbourhood that identifies the parameters.
\begin{definition}[Neighbour Variation]\label{def:nv}
Let $\mathbf x$ be the payoff-relevant covariate, and $\mathbf X^{-x}$ the other covariates. For a set $S$ of individuals, let $\mathbf x_{-S}=(x_k)_{k\notin S}$.
\begin{itemize}
\item \emph{Neighbour variation} holds for player $i$ through a neighbour $j$ if, for fixed $(\mathbf G,\mathbf M,\mathbf X^{-x},\mathbf x_{-j})$, the support of $x_j$ contains at least two points. It holds at degree $d$ if it holds at some $i$ with $d_i=d$.
\item \emph{Continuous neighbour variation} holds for agent $i$ jointly across all its neighbours if, for fixed $(\mathbf G,\mathbf M,\mathbf X^{-x},\mathbf x_{-N_i})$, the support of $(x_j)_{j\in N_i}$ contains a non-empty open set.
\end{itemize}
\end{definition}

Lemmas~\ref{lem:nv} and~\ref{lem:covariates} link neighbour variation in the payoff-relevant covariate to the equilibrium choice probabilities. Lemma~\ref{lem:nv} shows that when a change in $x_j$ raises $j$'s intrinsic utility, no equilibrium probability falls and $p_j^*$ rises strictly. Lemma~\ref{lem:covariates} shows how much variation in the neighbours' probabilities this creates, holding $i$'s own characteristics fixed. In part (a), continuous variation in the neighbours' covariates moves their probabilities over a whole open set and since the best-response Jacobian is invertible (Lemma~\ref{lem:jacobian}), the neighbours' probabilities can be varied independently over small ranges. In part (b), a covariate taking three values at one neighbour of an individual with two neighbours changes the gap between the two neighbours' probabilities. The neighbour whose covariate rises gains proportionally more than the other, who responds only through its norm and, by Assumption~\ref{asum:A6}, less than proportionally, hence the spread takes at least two values. Finally, Lemma~\ref{lem:powermean} shows that either kind of variation is enough to tell different values of $\beta$ apart.

\bigskip

\begin{lemma}[Neighbour Variation]\label{lem:nv}
Under Assumptions~\ref{asum:A1}--\ref{asum:ident}, suppose two supported configurations, $L$ and $H$, differ only in $j$'s payoff-relevant covariate, whose values $x_j^L$ and $x_j^H$ satisfy $\gamma_xx_j^H>\gamma_xx_j^L$. Then their equilibria satisfy
\[
\mathbf p^{*,H}\geq\mathbf p^{*,L},\qquad p_j^{*,H}>p_j^{*,L}.
\]
\end{lemma}
\begin{proof}
Let $\Gamma^L$ and $\Gamma^H$ be the best-response maps of the two configurations. They differ only in $j$'s index, which is higher under $\Gamma^H$ by $\gamma_x(x_j^H-x_j^L)>0$. By Assumption~\ref{asum:A5}, $\theta_1d_k+\theta_2>0$ for every $k$ with $d_k>0$, the norm is increasing in each neighbour probability (Assumption~1 of \citet{boucher2024peer}, restated in Appendix~\ref{SA.2}) and $F_\eta$ is strictly increasing by Assumption~\ref{asum:A1}. Hence both maps are increasing in $\mathbf p$ in the component-wise order (Definition~\ref{def:comp}), and
\[
\Gamma^H(\mathbf p^{*,L})\geq\Gamma^L(\mathbf p^{*,L})=\mathbf p^{*,L},
\qquad
\Gamma_j^H(\mathbf p^{*,L})>p_j^{*,L}.
\]
Iterating $\Gamma^H$ from $\mathbf p^{*,L}$ therefore gives an increasing sequence, bounded by one. Its limit is a fixed point of $\Gamma^H$ by continuity and equals $\mathbf p^{*,H}$, since the equilibrium is unique under Assumptions~\ref{asum:A1}--\ref{asum:A6} (Theorem~\ref{thm:uniqueness}). Hence
\[
\mathbf p^{*,H}\geq\Gamma^H(\mathbf p^{*,L}),
\]
which gives both inequalities.
\end{proof}

\bigskip

\begin{lemma}[Covariates and Neighbour Probabilities]\label{lem:covariates}
Under Assumptions~\ref{asum:A1}--\ref{asum:ident},
\begin{enumerate}[(a)]
\item If continuous neighbour variation holds at $i$, then $i$'s neighbour probabilities $(p_j^*)_{j\in N_i}$ take every value in some non-empty open set $\prod_{j\in N_i}I_j'$ of open intervals.
\item If $d_i=2$, with neighbours $j$ and $l$, and, for fixed $(\mathbf G,\mathbf M,\mathbf X^{-x},\mathbf x_{-j})$, the support of $j$'s payoff-relevant covariate $x_j$ contains at least three points, then the spread $|\log(p_j^*/p_l^*)|$ takes at least two values.
\end{enumerate}
\end{lemma}
\begin{proof}
(a) Let $\mathcal O=\prod_{j\in N_i}I_j$ be the non-empty open set in Definition~\ref{def:nv}, so that, for fixed $(\mathbf G,\mathbf M,\mathbf X^{-x},\mathbf x_{-N_i})$, $(x_j)_{j\in N_i}$ takes every value in $\mathcal O$. Changing $j$'s payoff-relevant covariate $x_j$ shifts only $j$'s index, by $\gamma_xx_j$. At each point of $\mathcal O$, the equilibrium is unique (Theorem~\ref{thm:uniqueness}) and $I-J$ is invertible (Lemma~\ref{lem:jacobian}). Since $F_\eta$ is logistic (Assumption~\ref{asum:A1}), the implicit function theorem shows that $(p_j^*)_{j\in N_i}$ is continuously differentiable in $(x_j)_{j\in N_i}$ on $\mathcal O$, with Jacobian
\[
\frac{\partial(p_j^*)_{j\in N_i}}{\partial(x_j)_{j\in N_i}^\top}
=\gamma_x\bigl[(I-J)^{-1}\bigr]_{N_iN_i}
\operatorname{diag}\{p_j^*(1-p_j^*)\}_{j\in N_i}.
\]
This matrix is invertible, since $[(I-J)^{-1}]_{N_iN_i}$ is invertible by Lemma~\ref{lem:jacobian} and $\gamma_x\neq0$ by Assumption~\ref{asum:ident}. By the inverse function theorem, the image of $\mathcal O$ contains a non-empty open set $\prod_{j\in N_i}I_j'$ of open intervals.

(b) There are three supported configurations that differ only in $x_j$, and, since $\gamma_x\neq0$ (Assumption~\ref{asum:ident}), $\gamma_xx_j$ takes a different value at each. It suffices to show that $\log(p_l^*/p_j^*)$ is decreasing in $\gamma_xx_j$.

Take two of these configurations, $L$ and $H$, with $\gamma_xx_j$ higher at $H$. By Lemma~\ref{lem:nv}, $\mathbf p^{*,H}\geq\mathbf p^{*,L}$ with $p_j^{*,H}>p_j^{*,L}$. The norm is increasing and homogeneous of degree one (Assumption~1 of \citet{boucher2024peer}), and, by Definition~\ref{def:log_metric},
\[
\mathbf p^{*,L}\leq\mathbf p^{*,H}
\leq\exp\bigl(d_{\log,\infty}(\mathbf p^{*,H},\mathbf p^{*,L})\bigr)\,\mathbf p^{*,L},
\]
so, for every $k$ with $d_k>0$,
\[
0\leq\log\tilde y^e_{-k}(\mathbf p^{*,H};\beta)-\log\tilde y^e_{-k}(\mathbf p^{*,L};\beta)
\leq d_{\log,\infty}(\mathbf p^{*,H},\mathbf p^{*,L}).
\]
For $k\neq j$ with $d_k>0$, $k$'s index changes only through its norm, and both norms lie in the interval $[a,b]$ of Assumption~\ref{asum:A6}, which is common to all supported configurations. By the mean value theorem,
\[
\log\frac{p_k^{*,H}}{p_k^{*,L}}
\leq\overline\varepsilon^{BR}\,d_{\log,\infty}(\mathbf p^{*,H},\mathbf p^{*,L}),
\qquad k\neq j,
\]
and the left side is zero for isolated $k$. If the largest increase $d_{\log,\infty}(\mathbf p^{*,H},\mathbf p^{*,L})$ occurred at some $k\neq j$, the last display would give $d_{\log,\infty}(\mathbf p^{*,H},\mathbf p^{*,L})\leq\overline\varepsilon^{BR}d_{\log,\infty}(\mathbf p^{*,H},\mathbf p^{*,L})$, so the distance would be zero, contradicting $p_j^{*,H}>p_j^{*,L}$. It therefore occurs at $j$:
\[
d_{\log,\infty}(\mathbf p^{*,H},\mathbf p^{*,L})=\log\frac{p_j^{*,H}}{p_j^{*,L}}.
\] For $k=l$, this gives
\[
\log\frac{p_l^{*,H}}{p_l^{*,L}}
\leq\overline\varepsilon^{BR}\log\frac{p_j^{*,H}}{p_j^{*,L}}
<\log\frac{p_j^{*,H}}{p_j^{*,L}},
\]
so $\log(p_l^*/p_j^*)$ is lower at $H$ than at $L$.
Applied to each pair of the three configurations, this shows that $\log(p_l^*/p_j^*)$ takes three different values. At most two of them can have the same absolute value, so the spread $|\log(p_j^*/p_l^*)|$ takes at least two values.
\end{proof}

\bigskip

\begin{lemma}[Power Means]\label{lem:powermean}
Let $d_i\geq2$, let the neighbour probabilities $(p_j)_{j\in N_i}$ range freely over $(0,1)$, and let $\beta,\beta'$ be finite and $\kappa_0,\kappa_1$ constants.
\begin{enumerate}[(a)]
\item If $\tilde y^e_{-i}(\mathbf p;\beta)=\kappa_0+\kappa_1\tilde y^e_{-i}(\mathbf p;\beta')$ for every $(p_j)_{j\in N_i}$ in some non-empty open set $\mathcal V\subseteq(0,1)^{d_i}$, then $\kappa_0=0$, $\kappa_1=1$ and $\beta=\beta'$.
\item If $d_i=2$, with neighbours $j$ and $l$, and $\tilde y^e_{-i}(\mathbf p;\beta)=\kappa_1\tilde y^e_{-i}(\mathbf p;\beta')$ at two different spreads $|\log(p_j/p_l)|$, then $\kappa_1=1$ and $\beta=\beta'$.
\end{enumerate}
\end{lemma}
\begin{proof}
(a) For finite $\beta$, the norm is built from sums, exponentials and logarithms of $p_j$, hence it is real analytic on the connected open set $(0,1)^{d_i}$ (Definition~\ref{def:real_analytic}). Therefore, the difference $\tilde y^e_{-i}(\mathbf p;\beta)-\kappa_0-\kappa_1\tilde y^e_{-i}(\mathbf p;\beta')$
is also real analytic. By the hypothesis of (a), it is zero on $\mathcal V$, and by the identity theorem, it is zero on all of $(0,1)^{d_i}$. Thus,
\[
\tilde y^e_{-i}(\mathbf p;\beta)=\kappa_0+\kappa_1\tilde y^e_{-i}(\mathbf p;\beta')
\qquad\text{for every }(p_j)_{j\in N_i}\in(0,1)^{d_i}.
\]
Replace $\mathbf p$ by $\lambda\mathbf p$, with $\lambda\in(0,1)$. The norm is homogeneous of degree one (Assumption~1 of \citet{boucher2024peer}), so
\[
\lambda\tilde y^e_{-i}(\mathbf p;\beta)
=\kappa_0+\lambda\kappa_1\tilde y^e_{-i}(\mathbf p;\beta')
=\kappa_0+\lambda\bigl(\tilde y^e_{-i}(\mathbf p;\beta)-\kappa_0\bigr),
\]
so $\kappa_0=\lambda\kappa_0$ and $\kappa_0=0$. If all $p_j$ equal some $q$, both norms equal $q$, so $\kappa_1=1$. If the $p_j$ are not all equal, the norm is strictly increasing in its exponent (Definition~\ref{def:powermean}), so $\tilde y^e_{-i}(\mathbf p;\beta)=\tilde y^e_{-i}(\mathbf p;\beta')$ gives $\beta=\beta'$.

(b) For finite $\beta\neq0$, factoring out $(p_jp_l)^{\beta/2}$ gives
\[
\frac{p_j^\beta+p_l^\beta}{2}
=(p_jp_l)^{\beta/2}\,\frac{(p_j/p_l)^{\beta/2}+(p_l/p_j)^{\beta/2}}{2}
=(p_jp_l)^{\beta/2}\cosh\Bigl(\tfrac\beta2\log\frac{p_j}{p_l}\Bigr),
\]
and taking the power $1/\beta$ and then logarithms gives, since $\cosh$ is even,
\[
\log\tilde y^e_{-i}(\mathbf p;\beta)
=\tfrac12\log(p_jp_l)
+\frac1\beta\log\cosh\Bigl(\tfrac\beta2\Bigl|\log\frac{p_j}{p_l}\Bigr|\Bigr).
\]
At $\beta=0$ the last term is read as zero, its limit, which gives the geometric mean. The first term does not depend on $\beta$, so the norm depends on $\beta$ only through the spread $|\log(p_j/p_l)|$. Both norms are positive, so $\kappa_1>0$, and taking logarithms of the identity gives, at both values,
\[
\log\kappa_1
=\frac1\beta\log\cosh\Bigl(\tfrac\beta2\Bigl|\log\frac{p_j}{p_l}\Bigr|\Bigr)
-\frac1{\beta'}\log\cosh\Bigl(\tfrac{\beta'}2\Bigl|\log\frac{p_j}{p_l}\Bigr|\Bigr).
\]
The derivative of the right side with respect to the spread is
\[
\tfrac12\Bigl[\tanh\Bigl(\tfrac\beta2\Bigl|\log\frac{p_j}{p_l}\Bigr|\Bigr)
-\tanh\Bigl(\tfrac{\beta'}2\Bigl|\log\frac{p_j}{p_l}\Bigr|\Bigr)\Bigr].
\]
If $\beta\neq\beta'$, it has the sign of $\beta-\beta'$ at every positive spread, so the right side is strictly monotone in the spread and cannot take the same value at two different spreads. Hence $\beta=\beta'$, and then $\kappa_1=1$.
\end{proof}

We are now ready to prove Theorem \ref{thm:ident}.

\bigskip

\begin{customproof}{Theorem~\ref{thm:ident}}

Suppose two admissible parameter vectors
\[
(\xi,\gamma,\delta,\theta_1,\theta_2,\beta),\qquad
(\xi',\gamma',\delta',\theta_1',\theta_2',\beta')
\]
generate the same conditional choice probabilities $\mathbf p^*$ at every supported configuration, the first being the true parameter vector. Since $F_\eta$ is strictly increasing (Assumption~\ref{asum:A1}), their indices coincide. For $d_i>0$, this gives
\begin{equation}\label{eq:id_oe}
\begin{aligned}
0={}&M_i(\xi-\xi')+X_i(\gamma-\gamma')
 -\frac{\theta_2-\theta_2'}{2}\\
&+(\theta_1d_i+\theta_2)\tilde y^e_{-i}(\mathbf p^*;\beta)
 -(\theta_1'd_i+\theta_2')\tilde y^e_{-i}(\mathbf p^*;\beta'),
\end{aligned}
\end{equation}
and for $d_i=0$ it gives
\begin{equation}\label{eq:id_oe_iso}
0=M_i(\xi-\xi')+X_i(\gamma-\gamma')+\delta-\delta'.
\end{equation}

\medskip\noindent
\emph{Neighbour variation.}
Suppose neighbour variation holds at degree $d$, with individual $i$ and neighbour $j$. Then two supported configurations, $L$ and $H$, differ only in $j$'s payoff-relevant covariate, with $\gamma_xx_j^H>\gamma_xx_j^L$. Since $x_j$ does not enter $i$'s own index, differencing \eqref{eq:id_oe} at $i$ between them gives
\begin{equation}\label{eq:id_nv}
\begin{aligned}
&(\theta_1d+\theta_2)\bigl[
\tilde y^e_{-i}(\mathbf p^{*,H};\beta)
-\tilde y^e_{-i}(\mathbf p^{*,L};\beta)\bigr]\\
&\qquad=(\theta_1'd+\theta_2')\bigl[
\tilde y^e_{-i}(\mathbf p^{*,H};\beta')
-\tilde y^e_{-i}(\mathbf p^{*,L};\beta')\bigr].
\end{aligned}
\end{equation}
By Lemma~\ref{lem:nv}, $\mathbf p^{*,H}\geq\mathbf p^{*,L}$ with $p_j^{*,H}>p_j^{*,L}$. In the no-peer model this is immediate, since only $p_j^*$ changes. The norm is strictly increasing in each neighbour probability, so both brackets in \eqref{eq:id_nv} are strictly positive. We now prove each part separately.

\medskip\noindent
\emph{Part (i).}
Under non-zero peer effects and continuous neighbour variation at an individual $i$ with $d_i\geq2$, $i$'s choices over the open set of neighbour probabilities from Lemma~\ref{lem:covariates}(a) identify $\beta$ and $\theta_1d_i+\theta_2$. Neighbour variation at a second degree then separates $\theta_1$ from $\theta_2$, and full rank identifies the remaining coefficients.

\emph{Curvature.}
By (i)(a) and Lemma~\ref{lem:covariates}(a), with $i$'s degree and own characteristics held fixed, the neighbour probabilities $(p_j^*)_{j\in N_i}$ take every value in a nonempty open set $\mathcal V\subseteq(0,1)^{d_i}$, and \eqref{eq:id_oe} holds at each of these configurations. Since $\theta_1d_i+\theta_2>0$ by Assumption~\ref{asum:A5}, dividing \eqref{eq:id_oe} by it gives
\[
\tilde y^e_{-i}(\mathbf p;\beta)
=-\frac{M_i(\xi-\xi')+X_i(\gamma-\gamma')-(\theta_2-\theta_2')/2}
{\theta_1d_i+\theta_2}
+\frac{\theta_1'd_i+\theta_2'}{\theta_1d_i+\theta_2}\,
\tilde y^e_{-i}(\mathbf p;\beta')
\]
for every $(p_j)_{j\in N_i}\in\mathcal V$, where the two fractions are constants. By Lemma~\ref{lem:powermean}(a), the first fraction is zero, the second equals one, and $\beta=\beta'$. Hence
\[
\theta_1d_i+\theta_2=\theta_1'd_i+\theta_2'.
\]

\emph{Remaining coefficients.}
Apply \eqref{eq:id_nv} at the degree $d\neq d_i$ in (i)(b). Since $\beta=\beta'$, its two brackets coincide and are positive, so
\[
\theta_1d+\theta_2=\theta_1'd+\theta_2'.
\]
Together with the curvature step, and because $d\neq d_i$, this gives $\theta_1=\theta_1'$ and $\theta_2=\theta_2'$. The peer terms and the conformity constants in \eqref{eq:id_oe} then cancel, so, together with \eqref{eq:id_oe_iso}, every individual satisfies
\begin{equation}\label{eq:id_own_only}
M_i(\xi-\xi')+X_i(\gamma-\gamma')+\mathbf 1\{d_i=0\}(\delta-\delta')=0,
\end{equation}
and the full-rank condition in Assumption~\ref{asum:ident} gives $\xi=\xi'$, $\gamma=\gamma'$ and $\delta=\delta'$. This proves part (i).

\medskip\noindent
\emph{Part (ii).}
Without continuous variation, we use individuals with one and two neighbours. At degree one, the norm equals the neighbour's probability for every $\beta$, so neighbour variation at $d=1$ identifies $\theta_1+\theta_2$, and full rank among these individuals identifies $\gamma$, and $\xi$ up to $\theta_2/2$. At degree two, the different spreads from Lemma~\ref{lem:covariates}(b) identify $\beta$ and $2\theta_1+\theta_2$, which separates $\theta_1$ from $\theta_2$, and full rank then identifies the remaining coefficients.

\emph{Degree-one.}
Apply \eqref{eq:id_nv} at $d=1$, with the individual $i$ and sole neighbour $j$ in (ii)(a). For every $\beta$, $\tilde y^e_{-i}(\mathbf p;\beta)=p_j$, so \eqref{eq:id_nv} reads
\[
\bigl[(\theta_1+\theta_2)-(\theta_1'+\theta_2')\bigr]
\bigl(p_j^{*,H}-p_j^{*,L}\bigr)=0.
\]
By Lemma~\ref{lem:nv}, the probability difference is strictly positive. Therefore
\begin{equation}\label{eq:id_degree_one}
\theta_1+\theta_2=\theta_1'+\theta_2'.
\end{equation}

\emph{Own coefficients up to the conformity constant.}
By \eqref{eq:id_degree_one}, equation \eqref{eq:id_oe} at individuals with $d_i=1$ reduces to
\[
M_i(\xi-\xi')+X_i(\gamma-\gamma')-\frac{\theta_2-\theta_2'}{2}=0.
\]
Each row of $\mathbf M$ has a single one, so $M_i\mathbf 1=1$. Since, by (ii)(b), $[\,\mathbf M\ \ \mathbf X\,]$ restricted to these individuals has full column rank, we obtain
\begin{equation}\label{eq:id_own}
\gamma=\gamma',\qquad
\xi-\frac{\theta_2}{2}\mathbf 1
=\xi'-\frac{\theta_2'}{2}\mathbf 1.
\end{equation}
Hence $M_i\xi+X_i\gamma-\theta_2/2=M_i\xi'+X_i\gamma'-\theta_2'/2$ for every connected individual, and \eqref{eq:id_oe} reduces to
\begin{equation}\label{eq:id_peer_only}
(\theta_1d_i+\theta_2)\tilde y^e_{-i}(\mathbf p^*;\beta)
=(\theta_1'd_i+\theta_2')\tilde y^e_{-i}(\mathbf p^*;\beta'),
\qquad d_i>0.
\end{equation}

\emph{Curvature from the spread.}
By Lemma~\ref{lem:covariates}(b), the individual $i$ in (ii)(c), with neighbours $j$ and $l$, has different spreads $|\log(p_j^*/p_l^*)|$ at two supported configurations. Since $2\theta_1+\theta_2>0$ by Assumption~\ref{asum:A5}, \eqref{eq:id_peer_only} gives, at both,
\[
\tilde y^e_{-i}(\mathbf p^*;\beta)
=\frac{2\theta_1'+\theta_2'}{2\theta_1+\theta_2}\,
\tilde y^e_{-i}(\mathbf p^*;\beta').
\]
By Lemma~\ref{lem:powermean}(b), $\beta=\beta'$ and $2\theta_1+\theta_2=2\theta_1'+\theta_2'$.

\emph{Remaining coefficients.}
Together with \eqref{eq:id_degree_one}, the curvature step gives $\theta_1=\theta_1'$ and $\theta_2=\theta_2'$. As in part (i), \eqref{eq:id_own_only} then holds for every individual, and the full-rank condition in Assumption~\ref{asum:ident} gives $\xi=\xi'$, $\gamma=\gamma'$ and $\delta=\delta'$. This proves part (ii).

\medskip\noindent
\emph{Part (iii).}
When $\theta_1=\theta_2=0$, $i$'s choice does not respond to a change in a neighbour's covariate, so an observationally equivalent alternative cannot have peer effects either. Full rank then identifies $(\xi,\gamma,\delta)$, while $\beta$ does not enter the model.

Let $\theta_1=\theta_2=0$, and apply \eqref{eq:id_nv} at the degree $d>0$ in part (iii). Its left side is zero and its right bracket is positive, so
\[
\theta_1'd+\theta_2'=0.
\]
By Assumption~\ref{asum:A5}, peer effects satisfy $\theta_1'd+\theta_2'>0$ at every supported $d>0$, so $\theta_1'=\theta_2'=0$. Equations \eqref{eq:id_oe} and \eqref{eq:id_oe_iso} then reduce to \eqref{eq:id_own_only} for every individual. Since $[\,\mathbf M\ \ \mathbf 1_{d=0}\ \ \mathbf X\,]$ has full column rank by Assumption~\ref{asum:ident}, $\xi=\xi'$, $\gamma=\gamma'$ and $\delta=\delta'$. The norm does not enter any index, so every admissible finite $\beta'$ is observationally equivalent. This proves part (iii).
\end{customproof}
\subsection{Optimal targeting}\label{SA.4}

We first state the KKT conditions that we use in the proof for optimal targeting.

\begin{definition}[Karush--Kuhn--Tucker Conditions]\label{def:kkt}
Let $V:\mathbb R^{N}\to\mathbb R$ be differentiable. At any local maximiser
$\boldsymbol\tau^{*}$ of $V$ on $\{\boldsymbol\tau\ge\mathbf 0:\sum_j\tau_j=B\}$ there exists
$\lambda\in\mathbb R$ with $\partial V/\partial\tau_i=\lambda$ when $\tau_i^{*}>0$ and
$\partial V/\partial\tau_i\le\lambda$ when $\tau_i^{*}=0$.
\end{definition}

We now provide the results on optimal targeting for the generalised norm for the problem discussed in Section \ref{S_seed}.

\bigskip

\begin{customproof}{Proposition~\ref{prop:optimal_targeting}}

Consider the Lagrangian
\[
\mathcal L
=\frac1N\sum_j p_j(\boldsymbol\tau;\beta)
-\lambda\left(\sum_j\tau_j-B\right),
\qquad \boldsymbol\tau\geq\mathbf0.
\]
By the Karush--Kuhn--Tucker conditions (Definition~\ref{def:kkt}), at a differentiable optimum there is $\lambda$ with
\[
\frac1N\sum_j\frac{\partial p_j}{\partial\tau_i}
\begin{cases}
=\lambda, & \tau_i^*>0,\\[2pt]
\leq\lambda, & \tau_i^*=0.
\end{cases}
\]
It remains to evaluate the marginal effects $\partial p_j/\partial\tau_i$ at the
post-policy equilibrium.

For finite $\beta$ the CES norm is continuously differentiable at the strictly positive
equilibrium probabilities, and the logistic density is $p_i(1-p_i)$. Differentiating each
connected agent's equilibrium condition \eqref{eq:targeting_equilibrium} in $\boldsymbol\tau$
gives
\[
(I-J)\,d\mathbf p=\operatorname{diag}\{p_i(1-p_i)\}\,d\boldsymbol\tau,
\qquad
J_{ij}:=p_i(1-p_i)(\theta_1 d_i+\theta_2)\,\frac{\partial\widetilde y^e_{-i}(\mathbf p;\beta)}{\partial p_j},
\]
with $J$ the best-response Jacobian in $\mathbf p$ (zero rows for isolated agents). By Lemma~\ref{lem:jacobian}, applied with intrinsic utilities $\alpha_i+\tau_i$, $I-J$ is invertible.

The implicit function theorem then gives
$\partial\mathbf p/\partial\boldsymbol\tau^\top=(I-J)^{-1}\operatorname{diag}\{p_i(1-p_i)\}$,
so the marginal effect on average adoption of a transfer to agent $i$ is
\[
\frac1N\sum_j\frac{\partial p_j}{\partial\tau_i}
=\frac{p_i(1-p_i)}{N}\big[(I-J)^{-\top}\iota\big]_i .
\]
Substituting into the first-order conditions gives the stated result.
\end{customproof}

\newpage

\section{Estimation}\label{SB.1}

We make the following regularity assumption in line with \citet{Lin2024}, required for the asymptotic properties of the NPL estimator and for the proof of Theorem~1 of \citet{Lin2024}.
\begin{assumption}[A8]\label{assump:estimation}
    \begin{enumerate}
        \item $\boldsymbol \Theta$ is compact, $\boldsymbol \theta_0$ is an interior point of $\boldsymbol \Theta$ and $\mathcal{P}$ is a compact and convex subset of $(0,1)^N$.
        \item $(\boldsymbol \theta_0, \mathbf{p}^*)$ is an isolated population NPL fixed point, i.e. it is unique or there is an open ball around it that does not contain any other fixed point in $\Lambda_{0N}$.
        \item The operator $\Gamma_0(\mathbf{p}) - \mathbf{p}$ has a non-singular Jacobian matrix at $\mathbf{p}^*$.
        \item $V_1(\boldsymbol \theta_0)$ and $V_2(\boldsymbol \theta_0)$ as defined in equation \eqref{eq:variances} are non-singular matrices with $V_1(\boldsymbol \theta_0)$ additionally being negative definite.
    \item The family $\{q_i(\boldsymbol \theta, \mathbf{p}) : \boldsymbol \theta \in \boldsymbol \Theta\}$ is a Vapnik-Chervonenkis (VC) class of functions.
    \item $\mathbb{E} \left[ \left| q_i(\boldsymbol \theta, \mathbf{p}, y) \right| \, \middle| \, \cdot \right]$ and 
    $\left| \mathbb{E} \left[ q_i(\boldsymbol \theta, \mathbf{p}, 1) - q_i(\boldsymbol \theta, \mathbf{p}, 0) \right] \right|$ are bounded by a constant;
    \item The class $\{q_i(\boldsymbol \theta, \mathbf{p}, y)\}$ is Donsker with respect to the distribution of $(Z(\mathbf{p}))_i$ for $y = 1$ or $y = 0$, with a square-integrable function.
        \end{enumerate}
\end{assumption}
\begin{remark}
    Parts (1)-(4) are standard regularity conditions that ensure the NPL estimator converges to the true parameter and its asymptotic variance is well defined. Parts (5)-(7) allow the use of uniform Law of Large Numbers and Central Limit Theorem results for network games from \citet{Menzel2016}. Together, these assumptions guarantee consistency and asymptotic normality of the NPL estimator in our setting. Since $\beta$ ranges over the real line, we take its parameter space to be the extended line $[-\infty,\infty]$, whose endpoints are the max- and min-aggregator limits $\beta\to\pm\infty$, to which the CES norm extends continuously. This set is compact, so Part~(1) holds. In estimation, $\beta$ is profiled over a grid that includes these endpoints, as described in Section~\ref{S_est}.
\end{remark}

\citet{aguirregabiria2007sequential} provide an algorithm (NPL algorithm) which is a recursive version of the two step PML. Starting with an initial random guess for the probabilities, the \textit{K-stage} estimator is defined as
\begin{equation}
\hat{\boldsymbol \theta}_K = \arg\max_{\boldsymbol \theta \in \boldsymbol \Theta} \mathcal{Q}_N(\boldsymbol \theta, \mathbf{\hat{p}}_{K-1})
\end{equation}
and the probabilities \( \{\mathbf{\hat{p}}_K : K \geq 1\} \) are obtained recursively as
\begin{equation}
\mathbf{\hat{p}}_K = \Gamma(\mathbf{\hat{p}}_{K-1},\hat{\boldsymbol \theta}_K).
\end{equation}
The algorithm terminates when the difference between two consecutive probabilities is less than some tolerance. However, as shown in later works such as \citet{Kasahara2012} and \citet{Aguirregabiria2021}, the NPL algorithm has finite-sample convergence issues especially in the case of multiple fixed points. While \citet{Kasahara2012} propose a relaxation of the NPL algorithm by replacing the fixed-point mapping with a Cobb--Douglas-type combination of the mapping and the current probabilities which performs better, it requires knowing the population Jacobian and probabilities else can have similar convergence issues. Other methods include solving the non-linear system of equations subject to the fixed point constraint which requires the calculation of the Jacobian at every step using Newton's method which can be computationally costly.

Thus, for the estimation of parameters and probabilities using the NPL estimator, we employ the spectral algorithm proposed by \citet{Aguirregabiria2021}. This approach addresses the convergence issues present in the original NPL algorithm of \citet{aguirregabiria2007sequential} and eliminates the need for arbitrary selection of the tuning parameter $\alpha$ as required in \citet{Kasahara2012}, while maintaining computational efficiency. The spectral method (see \citet{Cruz2006} for technical details) approximates the Jacobian matrix using a scalar value. This scalar is typically computed as a weighted average of the minimum and maximum eigenvalues of the Jacobian of the mapping function $\hat{\phi}(\hat{\mathbf{p}}) = \hat{\Gamma}(\hat{\mathbf{p}}) - \hat{\mathbf{p}}$. These can also be easily estimated using in-built solvers in R where $\alpha_k$ at every step k is given by:
\begin{align}\label{eq:alpha_k}
    \alpha_k = \frac{\Delta \hat{\mathbf{p}}'_k \Delta \hat{\mathbf{p}}_k}{\Delta \hat{\mathbf{p}}'_k \Delta \hat{\phi}(\hat{\mathbf{p}}_k) },
\end{align}
where $\Delta \hat{\mathbf{p}}_k = \hat{\mathbf{p}}_{k} - \hat{\mathbf{p}}_{k-1}$ and in every iteration, the conditional choice probabilities are updated using:
\begin{align}\label{eq:update_rule}
    \hat{\mathbf{p}}_{k+1} = \hat{\mathbf{p}}_k - \alpha_k \hat{\phi}(\hat{\mathbf{p}}_k).
\end{align}

\newpage

\section{Simulation Set-up and Discussion}\label{SB.2}

We present the set-up of the Monte Carlo simulations in this section. Networks are generated from the Chung--Lu model \citep{chunglu2002}. Each node $i$ draws a fitness $w_i \sim \mathrm{Gamma}(\text{shape}=k,\ \text{scale}=\mu/k)$ with $\mathbb{E}[w_i]=\mu$, and nodes $i$ and $j$ are linked independently with probability $1-\exp\!\big(-w_i w_j/\textstyle\sum_l w_l\big)$, giving expected degree $\mu$. This is the Norros--Reittu \citep{norros2006conditionally} form. The degree then has variance $\mu+\mu^2/k$, so smaller $k$ produces more degree heterogeneity and more isolated nodes. We set $\mu=6$ and $k=1.5$ to match the mean degree ($5.75$) and share of isolated households ($9.3\%$) of the microfinance network in Section~\ref{S_emp} (Table~\ref{tab:summary_stats}). For each $N\in\{500,1000,2000,4000\}$ we fix the Monte Carlo number to $500$. We draw two covariates  independently from $\mathcal{N}(0,2^2)$ and $\mathcal{N}(0,1^2)$ with coefficients $\gamma=(-0.2,0.1)$. The peer parameters are $\theta_1=0.15$ and $\theta_2=1$, with the isolated-household dummy set to $\delta=-0.7$. Since the strength of identification depends on adoption rates, we try two values of the common intercept, ($-1.70$ and $-1.30$). We vary $\beta\in\{-8.3,\,3.9,\,22.3\}$, which covers interior values, and also consider the extreme values $\beta\in\{-386,\,371.8\}$, which fall in the flat-tail regime.

For a given $N$ the network is fixed across replications. In each replication we solve the equilibrium fixed point for the true choice probabilities, draw logistic preference shocks to realise the actions $\mathbf y$, and estimate both the CES and the restricted linear model with the algorithm of Section~\ref{S_est}. All simulation tables are in Appendix~\ref{SD}. For each design they report the median and mean of $\hat\beta$, the share of tail solutions, the mean and standard deviation of the peer effect estimates under both norms, the LR test of $\beta=1$, the bias and MSE under both norms, and the $90\%$ Wald coverage under the CES norm.

\textbf{Discussion}: The adoption rate matters for identification. In the low-adoption design (intercept $-1.70$), the CES estimate of $\theta_2$ is $0.16$ for $\beta=-8.3$ and $0.33$ for $\beta=3.9$ at $N=500$, and still $0.70$ and $0.78$ at $N=4000$, whereas in the high-adoption design it is within $0.07$ of its true value at $N=4000$ in every design. At $N=4000$, the median $\hat\beta$ is $-3.1$ and $-5.8$ for $\beta=-8.3$, $3.9$ and $4.0$ for $\beta=3.9$, and $21.5$ and $22.7$ for $\beta=22.3$, in the low- and high-adoption design respectively. The share of tail solutions falls from $37$--$69\%$ at $N=500$ to $10$--$60\%$ at $N=4000$, and because of these draws $\hat\beta$ has a standard deviation across replications of about $70$ to $165$, which distorts its mean. The coverage of $\hat\theta_1$ rises with $N$ to about $82\%$ at $\beta=22.3$, stays at $85$ to $90\%$ at $\beta=3.9$, and falls to about $80\%$ at $\beta=-8.3$. When $\beta=3.9$, imposing linearity matters less: at $N=4000$ the linear model over-estimates complementarity by about a quarter ($0.19$ against $0.15$), and the LR test rejects linearity in $27$ to $53\%$ of draws. When $\beta=22.3$, by contrast, the test rejects in about half of the draws even at $N=500$.

For the extreme values (Tables~\ref{tab:sim_extremes_CL}--\ref{tab:sim_coverage_ext_CL_b}), at $N=4000$ the median $\hat\beta$ has the right sign but lies only between $-6.6$ and $-2.4$ for $\beta=-386$ and between $54$ and $70$ for $\beta=371.8$. For $\beta=371.8$, the CES estimates of $\theta_1$ and $\theta_2$ stay close to their true values, while the linear model misses both by $65\%$ or more, and the LR test rejects $\beta=1$ in almost every draw. For $\beta=-386$, both models under-estimate $\theta_1$, the linear model by about twice as much, and the test rejects in only $12$ to $27\%$ of draws.

\newpage

\section{Belief-Aggregated Perceived Social Norms}
\label{SC}

In this section, we discuss Assumption~\ref{asum:A3}, under which agents evaluate the social norm at their beliefs about peers' actions, so that these beliefs enter utility directly. The standard alternative is expected utility, in which agent $i$ takes the expectation of the utility in \eqref{eq:pay-offs} over the realised actions of its peers. Since the shocks are independent across agents (Assumption~\ref{asum:A1}), each peer $j$ then adopts independently with probability $p_j$.

Under expected utility, agent $i$ compares the expected utilities of the two actions. For a given realisation of peers' actions, differencing \eqref{eq:pay-offs} across $y_i\in\{0,1\}$ gives
\begin{align*}
U_i(1,y_{-i},\mathbf G)-U_i(0,y_{-i},\mathbf G)
&=\alpha_i+\theta_1 d_i\,\tilde y_{-i}(\beta)-\frac12\theta_2\Big[\big(1-\tilde y_{-i}(\beta)\big)^2-\tilde y_{-i}(\beta)^2\Big]+\epsilon_i(1)-\epsilon_i(0)\\
&=\alpha_i+\theta_1 d_i\,\tilde y_{-i}(\beta)+\theta_2\Big(\tilde y_{-i}(\beta)-\frac12\Big)-\eta_i ,
\end{align*}
since $(1-\tilde y_{-i})^2-\tilde y_{-i}^2=1-2\tilde y_{-i}$. The squared norm enters both utilities with the same coefficient and cancels, exactly as in \eqref{eq:br}, so the difference is linear in $\tilde y_{-i}(\beta)$. Peers' actions are independent of $\eta_i$ by Assumption~\ref{asum:A1}, so taking expectations over them shows that agent $i$ adopts if and only if
\[
\alpha_i+\theta_1 d_i\,\mathbb E\big[\tilde y_{-i}(\beta)\big]+\theta_2\Big(\mathbb E\big[\tilde y_{-i}(\beta)\big]-\frac12\Big)\ \ge\ \eta_i ,
\]
which is \eqref{eq:br} with $\tilde y^e_{-i}(\beta)$ replaced by $\mathbb E[\tilde y_{-i}(\beta)]$. Only the mean of the realised norm matters. Its second moment $\mathbb E[\tilde y_{-i}(\beta)^2]$, which would carry the variance, drops out.

At $\beta=1$ the two coincide, since the norm is the weighted average of peers' actions and, by linearity of expectations, $\mathbb E[\tilde y_{-i}(1)]=\sum_j\tilde g_{ij}p_j=\tilde y^e_{-i}(1)$, the benchmark of \citet{BrockDurlauf2001}. For $\beta\neq1$ they differ for any agent with at least two peers. The CES norm is convex in peers' actions for $\beta>1$ and concave for $\beta<1$, and $\tilde y^e_{-i}(\beta)$ evaluates it at the expected actions $p_j$, so Jensen's inequality gives
\[
\tilde y^e_{-i}(\beta)\le\mathbb E\big[\tilde y_{-i}(\beta)\big]\ \text{ for }\beta>1,
\qquad
\tilde y^e_{-i}(\beta)\ge\mathbb E\big[\tilde y_{-i}(\beta)\big]\ \text{ for }\beta<1 .
\]

We adopt the norm at beliefs for three reasons. First, it is tractable. The expected realised norm sums the CES norm over all $2^{d_i}$ action profiles of $i$'s peers and has no closed form, whereas $\tilde y^e_{-i}(\beta)$ is closed-form and differentiable in $\mathbf p$ and $\beta$, which the equilibrium, identification and estimation results use. Second, it keeps $\beta$ informative with binary actions. Since $y_j^\beta=y_j$, the realised norm is a power of the share of adopting peers when $\beta>0$, and when $\beta<0$ it is one if every peer adopts and zero otherwise. Every $\beta<0$ then gives the same expected realised norm, $\prod_{j\in N_i}p_j$, so under expected utility negative values of $\beta$ cannot be told apart. Beliefs lie in $(0,1)$, so $p_j^\beta$ varies with $\beta$ over its whole range and different values of $\beta$ imply different norms. Third, it has a behavioural foundation. As discussed in the remark following Assumption~\ref{asum:A3}, utility that depends directly on beliefs places the model among psychological games \citep{GeanakoplosPearceStacchetti1989,BattigalliDufwenberg2009}, whose equilibrium requires beliefs to be correct. Here $\beta$ is the rule by which an agent aggregates its beliefs about individual peers into a perceived norm. A misspecified $\beta$ therefore distorts the norm even when the beliefs are correct, unlike in \citet{BursztynGonzalezYanagizawaDrott2020}, where the beliefs themselves are wrong.

\newpage

\section{Additional Tables and Figures}\label{SD}

\subsection{Simulation Results}

These tables report the simulation results discussed in Subsection~\ref{Sub_sim}.

\begin{table}[H]
\begin{singlespacing}\centering\small
\caption{Simulation Results with Intercept $=-1.70$: Point Estimates for CES Elasticity and Peer Effects (CES vs.\ Linear Norm), LR Test Statistic and Rejection Rates.}
\label{tab:sim_main_CL_a}
\renewcommand{\arraystretch}{0.95}\setlength{\tabcolsep}{3pt}
\begin{tabular*}{\linewidth}{@{\extracolsep{\fill}}cccccccccc@{}}
\toprule
& \multicolumn{3}{c}{$\hat\beta$} & \multicolumn{2}{c}{CES} & \multicolumn{2}{c}{Linear} & \multicolumn{2}{c}{LR test} \\
\cmidrule(lr){2-4}\cmidrule(lr){5-6}\cmidrule(lr){7-8}\cmidrule(lr){9-10}
$N$ & Median & Mean & \% tail & $\hat\theta_1$ & $\hat\theta_2$ & $\hat\theta_1$ & $\hat\theta_2$ & Statistic & Rejection \\
\midrule
\multicolumn{10}{c}{\textbf{$\beta = -8.3$\quad(adoption $=12.1\%$)}} \\
\midrule
500 & 1.19 & \makecell[t]{14.23 \\ (73.58)} & 69\% & \makecell[t]{0.04 \\ (0.38)} & \makecell[t]{0.16 \\ (1.89)} & \makecell[t]{0.02 \\ (0.24)} & \makecell[t]{0.15 \\ (1.80)} & 0.72 & 2\% \\
1000 & -1.17 & \makecell[t]{15.06 \\ (94.43)} & 67\% & \makecell[t]{0.12 \\ (0.26)} & \makecell[t]{0.40 \\ (1.76)} & \makecell[t]{0.06 \\ (0.15)} & \makecell[t]{0.33 \\ (1.67)} & 0.94 & 5\% \\
2000 & -1.62 & \makecell[t]{14.05 \\ (99.34)} & 61\% & \makecell[t]{0.12 \\ (0.17)} & \makecell[t]{0.58 \\ (1.66)} & \makecell[t]{0.07 \\ (0.10)} & \makecell[t]{0.53 \\ (1.53)} & 1.02 & 8\% \\
4000 & -3.05 & \makecell[t]{2.63 \\ (101.26)} & 60\% & \makecell[t]{0.13 \\ (0.13)} & \makecell[t]{0.70 \\ (1.44)} & \makecell[t]{0.07 \\ (0.07)} & \makecell[t]{0.58 \\ (1.31)} & 1.39 & 15\% \\
\addlinespace[6pt]
\midrule
\multicolumn{10}{c}{\textbf{$\beta = 3.9$\quad(adoption $=14.1\%$)}} \\
\midrule
500 & 0.99 & \makecell[t]{16.17 \\ (95.35)} & 58\% & \makecell[t]{0.20 \\ (0.30)} & \makecell[t]{0.33 \\ (1.78)} & \makecell[t]{0.16 \\ (0.17)} & \makecell[t]{0.21 \\ (1.78)} & 0.86 & 5\% \\
1000 & 1.65 & \makecell[t]{18.87 \\ (103.61)} & 43\% & \makecell[t]{0.22 \\ (0.19)} & \makecell[t]{0.56 \\ (1.59)} & \makecell[t]{0.19 \\ (0.09)} & \makecell[t]{0.52 \\ (1.52)} & 1.16 & 11\% \\
2000 & 3.01 & \makecell[t]{21.16 \\ (102.28)} & 39\% & \makecell[t]{0.18 \\ (0.13)} & \makecell[t]{0.73 \\ (1.41)} & \makecell[t]{0.18 \\ (0.06)} & \makecell[t]{0.67 \\ (1.34)} & 1.22 & 12\% \\
4000 & 3.93 & \makecell[t]{19.31 \\ (107.83)} & 28\% & \makecell[t]{0.17 \\ (0.10)} & \makecell[t]{0.78 \\ (1.08)} & \makecell[t]{0.19 \\ (0.04)} & \makecell[t]{0.85 \\ (1.03)} & 2.17 & 27\% \\
\addlinespace[6pt]
\midrule
\multicolumn{10}{c}{\textbf{$\beta = 22.3$\quad(adoption $=26.3\%$)}} \\
\midrule
500 & 13.41 & \makecell[t]{46.02 \\ (90.91)} & 43\% & \makecell[t]{0.21 \\ (0.12)} & \makecell[t]{0.95 \\ (0.92)} & \makecell[t]{0.32 \\ (0.05)} & \makecell[t]{0.57 \\ (0.98)} & 3.84 & 57\% \\
1000 & 17.01 & \makecell[t]{73.33 \\ (124.32)} & 38\% & \makecell[t]{0.18 \\ (0.06)} & \makecell[t]{1.02 \\ (0.60)} & \makecell[t]{0.30 \\ (0.03)} & \makecell[t]{0.59 \\ (0.67)} & 7.51 & 83\% \\
2000 & 15.74 & \makecell[t]{76.04 \\ (137.95)} & 27\% & \makecell[t]{0.16 \\ (0.04)} & \makecell[t]{1.04 \\ (0.48)} & \makecell[t]{0.29 \\ (0.02)} & \makecell[t]{0.66 \\ (0.53)} & 11.85 & 94\% \\
4000 & 21.46 & \makecell[t]{91.64 \\ (143.76)} & 28\% & \makecell[t]{0.16 \\ (0.02)} & \makecell[t]{0.99 \\ (0.29)} & \makecell[t]{0.29 \\ (0.01)} & \makecell[t]{0.56 \\ (0.35)} & 26.66 & 100\% \\
\bottomrule
\end{tabular*}

\par\smallskip\begin{minipage}{\linewidth}\footnotesize
\emph{Notes:} Peer estimates are means over 500 replications, with standard deviations across replications in parentheses. We report the median and mean $\hat\beta$, since extreme tail draws distort the mean. Likelihood Ratio (LR) test columns report the mean test statistic and the rejection rate for $H_0:\beta=1$ at 10\% using $\chi^2_1$. \% tail is the share flagged as tail solutions. Adoption is the mean share of adopting households in the simulated data.
\end{minipage}
\end{singlespacing}
\end{table}

\begin{table}[H]
\begin{singlespacing}\centering\small
\caption{Simulation Results with Intercept $=-1.30$: Point Estimates for CES Elasticity and Peer Effects (CES vs.\ Linear Norm), LR Test Statistic and Rejection Rates.}
\label{tab:sim_main_CL_b}
\renewcommand{\arraystretch}{0.95}\setlength{\tabcolsep}{3pt}
\begin{tabular*}{\linewidth}{@{\extracolsep{\fill}}cccccccccc@{}}
\toprule
& \multicolumn{3}{c}{$\hat\beta$} & \multicolumn{2}{c}{CES} & \multicolumn{2}{c}{Linear} & \multicolumn{2}{c}{LR test} \\
\cmidrule(lr){2-4}\cmidrule(lr){5-6}\cmidrule(lr){7-8}\cmidrule(lr){9-10}
$N$ & Median & Mean & \% tail & $\hat\theta_1$ & $\hat\theta_2$ & $\hat\theta_1$ & $\hat\theta_2$ & Statistic & Rejection \\
\midrule
\multicolumn{10}{c}{\textbf{$\beta = -8.3$\quad(adoption $=18.3\%$)}} \\
\midrule
500 & 0.00 & \makecell[t]{18.44 \\ (97.91)} & 64\% & \makecell[t]{0.10 \\ (0.22)} & \makecell[t]{0.37 \\ (1.78)} & \makecell[t]{0.05 \\ (0.13)} & \makecell[t]{0.26 \\ (1.67)} & 0.90 & 5\% \\
1000 & -2.56 & \makecell[t]{1.00 \\ (101.13)} & 56\% & \makecell[t]{0.14 \\ (0.15)} & \makecell[t]{0.71 \\ (1.48)} & \makecell[t]{0.08 \\ (0.08)} & \makecell[t]{0.56 \\ (1.42)} & 1.34 & 15\% \\
2000 & -3.93 & \makecell[t]{1.43 \\ (113.71)} & 54\% & \makecell[t]{0.13 \\ (0.11)} & \makecell[t]{0.87 \\ (1.29)} & \makecell[t]{0.07 \\ (0.05)} & \makecell[t]{0.80 \\ (1.18)} & 1.60 & 19\% \\
4000 & -5.83 & \makecell[t]{-21.69 \\ (127.77)} & 48\% & \makecell[t]{0.14 \\ (0.09)} & \makecell[t]{0.93 \\ (1.00)} & \makecell[t]{0.08 \\ (0.04)} & \makecell[t]{0.83 \\ (0.97)} & 2.47 & 32\% \\
\addlinespace[6pt]
\midrule
\multicolumn{10}{c}{\textbf{$\beta = 3.9$\quad(adoption $=24.6\%$)}} \\
\midrule
500 & 2.36 & \makecell[t]{24.84 \\ (120.66)} & 40\% & \makecell[t]{0.19 \\ (0.14)} & \makecell[t]{0.59 \\ (1.48)} & \makecell[t]{0.18 \\ (0.06)} & \makecell[t]{0.59 \\ (1.34)} & 1.27 & 13\% \\
1000 & 3.74 & \makecell[t]{24.82 \\ (95.79)} & 30\% & \makecell[t]{0.17 \\ (0.09)} & \makecell[t]{0.90 \\ (0.98)} & \makecell[t]{0.19 \\ (0.03)} & \makecell[t]{0.88 \\ (0.93)} & 1.89 & 26\% \\
2000 & 3.97 & \makecell[t]{31.71 \\ (113.28)} & 19\% & \makecell[t]{0.15 \\ (0.06)} & \makecell[t]{0.97 \\ (0.79)} & \makecell[t]{0.18 \\ (0.03)} & \makecell[t]{0.93 \\ (0.79)} & 2.00 & 27\% \\
4000 & 4.01 & \makecell[t]{30.33 \\ (95.83)} & 10\% & \makecell[t]{0.15 \\ (0.04)} & \makecell[t]{0.97 \\ (0.48)} & \makecell[t]{0.19 \\ (0.02)} & \makecell[t]{0.98 \\ (0.52)} & 4.11 & 53\% \\
\addlinespace[6pt]
\midrule
\multicolumn{10}{c}{\textbf{$\beta = 22.3$\quad(adoption $=37.7\%$)}} \\
\midrule
500 & 12.51 & \makecell[t]{58.49 \\ (107.67)} & 37\% & \makecell[t]{0.20 \\ (0.09)} & \makecell[t]{0.95 \\ (0.89)} & \makecell[t]{0.25 \\ (0.04)} & \makecell[t]{0.55 \\ (0.89)} & 3.23 & 50\% \\
1000 & 16.47 & \makecell[t]{81.64 \\ (129.76)} & 37\% & \makecell[t]{0.17 \\ (0.06)} & \makecell[t]{0.98 \\ (0.61)} & \makecell[t]{0.25 \\ (0.03)} & \makecell[t]{0.51 \\ (0.58)} & 5.84 & 75\% \\
2000 & 19.38 & \makecell[t]{91.21 \\ (145.64)} & 32\% & \makecell[t]{0.16 \\ (0.03)} & \makecell[t]{1.03 \\ (0.48)} & \makecell[t]{0.24 \\ (0.02)} & \makecell[t]{0.54 \\ (0.44)} & 9.51 & 91\% \\
4000 & 22.70 & \makecell[t]{105.69 \\ (164.55)} & 32\% & \makecell[t]{0.15 \\ (0.02)} & \makecell[t]{0.99 \\ (0.30)} & \makecell[t]{0.24 \\ (0.01)} & \makecell[t]{0.51 \\ (0.30)} & 19.20 & 98\% \\
\bottomrule
\end{tabular*}

\par\smallskip\begin{minipage}{\linewidth}\footnotesize
\emph{Notes:} Peer estimates are means over 500 replications, with standard deviations across replications in parentheses. We report the median and mean $\hat\beta$, since extreme tail draws distort the mean. Likelihood Ratio (LR) test columns report the mean test statistic and the rejection rate for $H_0:\beta=1$ at 10\% using $\chi^2_1$. \% tail is the share flagged as tail solutions. Adoption is the mean share of adopting households in the simulated data.
\end{minipage}
\end{singlespacing}
\end{table}

\begin{table}[H]
\begin{singlespacing}\centering\small
\caption{Simulation Results with Intercept $=-1.70$: Bias and mean squared error of the peer-effect estimates under the CES and linear norms.}
\label{tab:sim_biasmse_main_CL_a}
\renewcommand{\arraystretch}{0.95}\setlength{\tabcolsep}{3pt}
\begin{tabular*}{\linewidth}{@{\extracolsep{\fill}}ccccccccc@{}}
\toprule
& \multicolumn{4}{c}{$\theta_1$} & \multicolumn{4}{c}{$\theta_2$} \\
\cmidrule(lr){2-5}\cmidrule(lr){6-9}
& \multicolumn{2}{c}{CES} & \multicolumn{2}{c}{Linear} & \multicolumn{2}{c}{CES} & \multicolumn{2}{c}{Linear} \\
\cmidrule(lr){2-3}\cmidrule(lr){4-5}\cmidrule(lr){6-7}\cmidrule(lr){8-9}
$N$ & Bias & MSE & Bias & MSE & Bias & MSE & Bias & MSE \\
\midrule
\multicolumn{9}{c}{\textbf{$\beta = -8.3$\quad(adoption $=12.1\%$)}} \\
\midrule
500 & -0.11 & 0.15 & -0.13 & 0.08 & -0.84 & 4.26 & -0.85 & 3.96 \\
1000 & -0.03 & 0.07 & -0.09 & 0.03 & -0.60 & 3.44 & -0.67 & 3.22 \\
2000 & -0.03 & 0.03 & -0.08 & 0.02 & -0.42 & 2.93 & -0.47 & 2.56 \\
4000 & -0.02 & 0.02 & -0.08 & 0.01 & -0.30 & 2.16 & -0.42 & 1.89 \\
\midrule
\multicolumn{9}{c}{\textbf{$\beta = 3.9$\quad(adoption $=14.1\%$)}} \\
\midrule
500 & 0.05 & 0.09 & 0.01 & 0.03 & -0.67 & 3.62 & -0.79 & 3.79 \\
1000 & 0.07 & 0.04 & 0.04 & 0.01 & -0.44 & 2.71 & -0.48 & 2.53 \\
2000 & 0.03 & 0.02 & 0.03 & 0.01 & -0.27 & 2.06 & -0.33 & 1.90 \\
4000 & 0.02 & 0.01 & 0.04 & 0.00 & -0.22 & 1.22 & -0.15 & 1.08 \\
\midrule
\multicolumn{9}{c}{\textbf{$\beta = 22.3$\quad(adoption $=26.3\%$)}} \\
\midrule
500 & 0.06 & 0.02 & 0.17 & 0.03 & -0.05 & 0.86 & -0.43 & 1.15 \\
1000 & 0.03 & 0.00 & 0.15 & 0.02 & 0.02 & 0.36 & -0.41 & 0.62 \\
2000 & 0.01 & 0.00 & 0.14 & 0.02 & 0.04 & 0.23 & -0.34 & 0.40 \\
4000 & 0.01 & 0.00 & 0.14 & 0.02 & -0.01 & 0.09 & -0.44 & 0.32 \\
\bottomrule
\end{tabular*}

\par\smallskip\begin{minipage}{\linewidth}\footnotesize
\emph{Notes:} Bias $=\mathrm{mean}(\hat\theta-\theta)$ and MSE $=\mathrm{mean}((\hat\theta-\theta)^2)$ over 500 replications. True $\theta_1=0.15$ and $\theta_2=1$. Adoption is the mean share of adopting households in the simulated data.
\end{minipage}
\end{singlespacing}
\end{table}

\begin{table}[H]
\begin{singlespacing}\centering\small
\caption{Simulation Results with Intercept $=-1.30$: Bias and mean squared error of the peer-effect estimates under the CES and linear norms.}
\label{tab:sim_biasmse_main_CL_b}
\renewcommand{\arraystretch}{0.95}\setlength{\tabcolsep}{3pt}
\begin{tabular*}{\linewidth}{@{\extracolsep{\fill}}ccccccccc@{}}
\toprule
& \multicolumn{4}{c}{$\theta_1$} & \multicolumn{4}{c}{$\theta_2$} \\
\cmidrule(lr){2-5}\cmidrule(lr){6-9}
& \multicolumn{2}{c}{CES} & \multicolumn{2}{c}{Linear} & \multicolumn{2}{c}{CES} & \multicolumn{2}{c}{Linear} \\
\cmidrule(lr){2-3}\cmidrule(lr){4-5}\cmidrule(lr){6-7}\cmidrule(lr){8-9}
$N$ & Bias & MSE & Bias & MSE & Bias & MSE & Bias & MSE \\
\midrule
\multicolumn{9}{c}{\textbf{$\beta = -8.3$\quad(adoption $=18.3\%$)}} \\
\midrule
500 & -0.05 & 0.05 & -0.10 & 0.03 & -0.63 & 3.54 & -0.74 & 3.31 \\
1000 & -0.01 & 0.02 & -0.07 & 0.01 & -0.29 & 2.27 & -0.44 & 2.21 \\
2000 & -0.02 & 0.01 & -0.08 & 0.01 & -0.13 & 1.67 & -0.20 & 1.43 \\
4000 & -0.01 & 0.01 & -0.07 & 0.01 & -0.07 & 1.01 & -0.17 & 0.97 \\
\midrule
\multicolumn{9}{c}{\textbf{$\beta = 3.9$\quad(adoption $=24.6\%$)}} \\
\midrule
500 & 0.04 & 0.02 & 0.03 & 0.00 & -0.41 & 2.35 & -0.41 & 1.97 \\
1000 & 0.02 & 0.01 & 0.04 & 0.00 & -0.10 & 0.97 & -0.12 & 0.88 \\
2000 & 0.00 & 0.00 & 0.03 & 0.00 & -0.03 & 0.63 & -0.07 & 0.63 \\
4000 & -0.00 & 0.00 & 0.04 & 0.00 & -0.03 & 0.23 & -0.02 & 0.27 \\
\midrule
\multicolumn{9}{c}{\textbf{$\beta = 22.3$\quad(adoption $=37.7\%$)}} \\
\midrule
500 & 0.05 & 0.01 & 0.10 & 0.01 & -0.05 & 0.80 & -0.45 & 0.99 \\
1000 & 0.02 & 0.00 & 0.10 & 0.01 & -0.02 & 0.38 & -0.49 & 0.58 \\
2000 & 0.01 & 0.00 & 0.09 & 0.01 & 0.03 & 0.23 & -0.46 & 0.40 \\
4000 & 0.00 & 0.00 & 0.09 & 0.01 & -0.01 & 0.09 & -0.49 & 0.33 \\
\bottomrule
\end{tabular*}

\par\smallskip\begin{minipage}{\linewidth}\footnotesize
\emph{Notes:} Bias $=\mathrm{mean}(\hat\theta-\theta)$ and MSE $=\mathrm{mean}((\hat\theta-\theta)^2)$ over 500 replications. True $\theta_1=0.15$ and $\theta_2=1$. Adoption is the mean share of adopting households in the simulated data.
\end{minipage}
\end{singlespacing}
\end{table}

\begin{table}[H]
\begin{singlespacing}\centering\small
\caption{Simulation Results with Intercept $=-1.70$: 90\% Wald Coverage of Peer Effect Estimates under CES Norm.}
\label{tab:sim_coverage_main_CL_a}
\renewcommand{\arraystretch}{0.95}\setlength{\tabcolsep}{3pt}
\begin{tabular*}{\linewidth}{@{\extracolsep{\fill}}cccccccc@{}}
\toprule
& & \multicolumn{3}{c}{$\theta_1$} & \multicolumn{3}{c}{$\theta_2$} \\
\cmidrule(lr){3-5}\cmidrule(lr){6-8}
$N$ & \% tail & All & Interior & Tail & All & Interior & Tail \\
\midrule
\multicolumn{8}{c}{\textbf{$\beta = -8.3$\quad(adoption $=12.1\%$)}} \\
\midrule
500 & 69\% & 92\% & 95\% & 90\% & 99\% & 100\% & 98\% \\
1000 & 67\% & 88\% & 88\% & 88\% & 97\% & 99\% & 96\% \\
2000 & 61\% & 83\% & 85\% & 81\% & 96\% & 99\% & 93\% \\
4000 & 60\% & 72\% & 80\% & 66\% & 88\% & 92\% & 86\% \\
\midrule
\multicolumn{8}{c}{\textbf{$\beta = 3.9$\quad(adoption $=14.1\%$)}} \\
\midrule
500 & 58\% & 90\% & 90\% & 90\% & 97\% & 100\% & 96\% \\
1000 & 43\% & 83\% & 86\% & 80\% & 97\% & 99\% & 93\% \\
2000 & 39\% & 81\% & 89\% & 69\% & 94\% & 98\% & 87\% \\
4000 & 28\% & 73\% & 88\% & 35\% & 89\% & 94\% & 74\% \\
\midrule
\multicolumn{8}{c}{\textbf{$\beta = 22.3$\quad(adoption $=26.3\%$)}} \\
\midrule
500 & 43\% & 79\% & 70\% & 91\% & 94\% & 96\% & 91\% \\
1000 & 38\% & 80\% & 77\% & 86\% & 96\% & 96\% & 97\% \\
2000 & 27\% & 75\% & 76\% & 74\% & 91\% & 91\% & 90\% \\
4000 & 28\% & 74\% & 82\% & 56\% & 91\% & 91\% & 92\% \\
\bottomrule
\end{tabular*}

\par\smallskip\begin{minipage}{\linewidth}\footnotesize
\emph{Notes:} Coverage over 500 replications, split by the estimator's tail flag. \% tail is the share flagged as tail solutions where the profile likelihood is flat in $\beta$. Adoption is the mean share of adopting households in the simulated data.
\end{minipage}
\end{singlespacing}
\end{table}

\begin{table}[H]
\begin{singlespacing}\centering\small
\caption{Simulation Results with Intercept $=-1.30$: 90\% Wald Coverage of Peer Effect Estimates under CES Norm.}
\label{tab:sim_coverage_main_CL_b}
\renewcommand{\arraystretch}{0.95}\setlength{\tabcolsep}{3pt}
\begin{tabular*}{\linewidth}{@{\extracolsep{\fill}}cccccccc@{}}
\toprule
& & \multicolumn{3}{c}{$\theta_1$} & \multicolumn{3}{c}{$\theta_2$} \\
\cmidrule(lr){3-5}\cmidrule(lr){6-8}
$N$ & \% tail & All & Interior & Tail & All & Interior & Tail \\
\midrule
\multicolumn{8}{c}{\textbf{$\beta = -8.3$\quad(adoption $=18.3\%$)}} \\
\midrule
500 & 64\% & 85\% & 92\% & 81\% & 96\% & 99\% & 94\% \\
1000 & 56\% & 81\% & 86\% & 77\% & 93\% & 97\% & 90\% \\
2000 & 54\% & 71\% & 79\% & 65\% & 89\% & 93\% & 85\% \\
4000 & 48\% & 72\% & 78\% & 66\% & 89\% & 94\% & 83\% \\
\midrule
\multicolumn{8}{c}{\textbf{$\beta = 3.9$\quad(adoption $=24.6\%$)}} \\
\midrule
500 & 40\% & 81\% & 88\% & 70\% & 91\% & 96\% & 84\% \\
1000 & 30\% & 71\% & 85\% & 40\% & 93\% & 97\% & 85\% \\
2000 & 19\% & 74\% & 86\% & 18\% & 92\% & 96\% & 76\% \\
4000 & 10\% & 79\% & 86\% & 15\% & 88\% & 91\% & 62\% \\
\midrule
\multicolumn{8}{c}{\textbf{$\beta = 22.3$\quad(adoption $=37.7\%$)}} \\
\midrule
500 & 37\% & 78\% & 72\% & 89\% & 94\% & 93\% & 96\% \\
1000 & 37\% & 81\% & 79\% & 84\% & 93\% & 92\% & 95\% \\
2000 & 32\% & 78\% & 82\% & 70\% & 90\% & 91\% & 89\% \\
4000 & 32\% & 72\% & 81\% & 53\% & 91\% & 90\% & 93\% \\
\bottomrule
\end{tabular*}

\par\smallskip\begin{minipage}{\linewidth}\footnotesize
\emph{Notes:} Coverage over 500 replications, split by the estimator's tail flag. \% tail is the share flagged as tail solutions where the profile likelihood is flat in $\beta$. Adoption is the mean share of adopting households in the simulated data.
\end{minipage}
\end{singlespacing}
\end{table}

\begin{table}[H]
\begin{singlespacing}\centering\footnotesize
\caption{Simulation Results for Extreme $\beta$: Point Estimates for CES Elasticity and Peer Effects (CES vs.\ Linear Norm), LR Test Statistic and Rejection Rates.}
\label{tab:sim_extremes_CL}
\renewcommand{\arraystretch}{0.9}\setlength{\tabcolsep}{3pt}
\begin{tabular*}{\linewidth}{@{\extracolsep{\fill}}ccccccccc@{}}
\toprule
& \multicolumn{2}{c}{$\hat\beta$} & \multicolumn{2}{c}{CES} & \multicolumn{2}{c}{Linear} & \multicolumn{2}{c}{LR test} \\
\cmidrule(lr){2-3}\cmidrule(lr){4-5}\cmidrule(lr){6-7}\cmidrule(lr){8-9}
$N$ & Median & Mean & $\hat\theta_1$ & $\hat\theta_2$ & $\hat\theta_1$ & $\hat\theta_2$ & Statistic & Rejection \\
\midrule
\multicolumn{9}{c}{\textbf{Intercept $=-1.70$}} \\
\midrule
\multicolumn{9}{c}{\textbf{$\beta = -386.0$\quad(adoption $=11.8\%$)}} \\
\midrule
500 & 1.37 & \makecell[t]{16.29 \\ (80.29)} & \makecell[t]{-0.01 \\ (0.40)} & \makecell[t]{0.12 \\ (1.89)} & \makecell[t]{-0.02 \\ (0.26)} & \makecell[t]{0.16 \\ (1.82)} & 0.71 & 2\% \\
1000 & -0.95 & \makecell[t]{12.76 \\ (90.56)} & \makecell[t]{0.08 \\ (0.27)} & \makecell[t]{0.47 \\ (1.74)} & \makecell[t]{0.03 \\ (0.16)} & \makecell[t]{0.37 \\ (1.68)} & 0.89 & 5\% \\
2000 & -1.27 & \makecell[t]{15.11 \\ (95.19)} & \makecell[t]{0.07 \\ (0.18)} & \makecell[t]{0.60 \\ (1.69)} & \makecell[t]{0.03 \\ (0.11)} & \makecell[t]{0.51 \\ (1.57)} & 0.93 & 5\% \\
4000 & -2.41 & \makecell[t]{7.14 \\ (116.13)} & \makecell[t]{0.09 \\ (0.13)} & \makecell[t]{0.67 \\ (1.49)} & \makecell[t]{0.04 \\ (0.08)} & \makecell[t]{0.56 \\ (1.34)} & 1.23 & 12\% \\
\addlinespace[3pt]\midrule
\multicolumn{9}{c}{\textbf{$\beta = 371.8$\quad(adoption $=31.6\%$)}} \\
\midrule
500 & 15.90 & \makecell[t]{61.81 \\ (107.10)} & \makecell[t]{0.23 \\ (0.10)} & \makecell[t]{0.91 \\ (0.86)} & \makecell[t]{0.33 \\ (0.04)} & \makecell[t]{0.37 \\ (0.87)} & 4.35 & 62\% \\
1000 & 24.09 & \makecell[t]{105.22 \\ (153.18)} & \makecell[t]{0.19 \\ (0.05)} & \makecell[t]{1.00 \\ (0.58)} & \makecell[t]{0.31 \\ (0.03)} & \makecell[t]{0.37 \\ (0.58)} & 8.52 & 88\% \\
2000 & 38.54 & \makecell[t]{133.98 \\ (171.59)} & \makecell[t]{0.17 \\ (0.03)} & \makecell[t]{1.01 \\ (0.42)} & \makecell[t]{0.30 \\ (0.02)} & \makecell[t]{0.37 \\ (0.41)} & 14.82 & 99\% \\
4000 & 54.31 & \makecell[t]{174.38 \\ (192.00)} & \makecell[t]{0.17 \\ (0.02)} & \makecell[t]{0.97 \\ (0.28)} & \makecell[t]{0.30 \\ (0.01)} & \makecell[t]{0.32 \\ (0.30)} & 30.44 & 100\% \\
\addlinespace[3pt]\midrule
\multicolumn{9}{c}{\textbf{Intercept $=-1.30$}} \\
\midrule
\multicolumn{9}{c}{\textbf{$\beta = -386.0$\quad(adoption $=17.6\%$)}} \\
\midrule
500 & 0.47 & \makecell[t]{20.93 \\ (97.68)} & \makecell[t]{0.06 \\ (0.24)} & \makecell[t]{0.36 \\ (1.80)} & \makecell[t]{0.02 \\ (0.15)} & \makecell[t]{0.25 \\ (1.72)} & 0.85 & 5\% \\
1000 & -1.45 & \makecell[t]{9.06 \\ (95.48)} & \makecell[t]{0.09 \\ (0.15)} & \makecell[t]{0.69 \\ (1.54)} & \makecell[t]{0.04 \\ (0.09)} & \makecell[t]{0.56 \\ (1.46)} & 1.13 & 9\% \\
2000 & -2.33 & \makecell[t]{3.78 \\ (117.73)} & \makecell[t]{0.08 \\ (0.11)} & \makecell[t]{0.83 \\ (1.39)} & \makecell[t]{0.04 \\ (0.06)} & \makecell[t]{0.74 \\ (1.26)} & 1.38 & 16\% \\
4000 & -6.55 & \makecell[t]{-17.64 \\ (118.75)} & \makecell[t]{0.10 \\ (0.08)} & \makecell[t]{0.97 \\ (1.07)} & \makecell[t]{0.05 \\ (0.04)} & \makecell[t]{0.81 \\ (1.02)} & 2.05 & 27\% \\
\addlinespace[3pt]\midrule
\multicolumn{9}{c}{\textbf{$\beta = 371.8$\quad(adoption $=41.9\%$)}} \\
\midrule
500 & 20.44 & \makecell[t]{84.23 \\ (126.48)} & \makecell[t]{0.21 \\ (0.08)} & \makecell[t]{0.93 \\ (0.90)} & \makecell[t]{0.27 \\ (0.04)} & \makecell[t]{0.41 \\ (0.83)} & 3.52 & 52\% \\
1000 & 34.56 & \makecell[t]{120.03 \\ (156.14)} & \makecell[t]{0.18 \\ (0.05)} & \makecell[t]{0.97 \\ (0.59)} & \makecell[t]{0.26 \\ (0.03)} & \makecell[t]{0.35 \\ (0.54)} & 6.39 & 80\% \\
2000 & 46.45 & \makecell[t]{143.58 \\ (171.19)} & \makecell[t]{0.17 \\ (0.03)} & \makecell[t]{1.01 \\ (0.47)} & \makecell[t]{0.26 \\ (0.02)} & \makecell[t]{0.36 \\ (0.40)} & 11.11 & 95\% \\
4000 & 70.22 & \makecell[t]{176.18 \\ (186.75)} & \makecell[t]{0.16 \\ (0.02)} & \makecell[t]{0.99 \\ (0.30)} & \makecell[t]{0.26 \\ (0.01)} & \makecell[t]{0.35 \\ (0.29)} & 21.49 & 99\% \\
\bottomrule
\end{tabular*}

\par\smallskip\begin{minipage}{\linewidth}\footnotesize
\emph{Notes:} Peer estimates are means over 500 replications, with standard deviations across replications in parentheses. We report the median and mean $\hat\beta$, since extreme tail draws distort the mean. Likelihood Ratio (LR) test columns report the mean test statistic and the rejection rate for $H_0:\beta=1$ at 10\% using $\chi^2_1$. Adoption is the mean share of adopting households in the simulated data.
\end{minipage}
\end{singlespacing}
\end{table}

\begin{table}[H]
\begin{singlespacing}\centering\small
\caption{Simulation Results with Intercept $=-1.70$ and Extreme $\beta$: Bias and mean squared error of the peer-effect estimates under the CES and linear norms.}
\label{tab:sim_biasmse_ext_CL_a}
\renewcommand{\arraystretch}{0.95}\setlength{\tabcolsep}{3pt}
\begin{tabular*}{\linewidth}{@{\extracolsep{\fill}}ccccccccc@{}}
\toprule
& \multicolumn{4}{c}{$\theta_1$} & \multicolumn{4}{c}{$\theta_2$} \\
\cmidrule(lr){2-5}\cmidrule(lr){6-9}
& \multicolumn{2}{c}{CES} & \multicolumn{2}{c}{Linear} & \multicolumn{2}{c}{CES} & \multicolumn{2}{c}{Linear} \\
\cmidrule(lr){2-3}\cmidrule(lr){4-5}\cmidrule(lr){6-7}\cmidrule(lr){8-9}
$N$ & Bias & MSE & Bias & MSE & Bias & MSE & Bias & MSE \\
\midrule
\multicolumn{9}{c}{\textbf{$\beta = -386.0$\quad(adoption $=11.8\%$)}} \\
\midrule
500 & -0.16 & 0.18 & -0.17 & 0.10 & -0.88 & 4.34 & -0.84 & 4.02 \\
1000 & -0.07 & 0.08 & -0.12 & 0.04 & -0.53 & 3.32 & -0.63 & 3.20 \\
2000 & -0.08 & 0.04 & -0.12 & 0.02 & -0.40 & 3.01 & -0.49 & 2.70 \\
4000 & -0.06 & 0.02 & -0.11 & 0.02 & -0.33 & 2.33 & -0.44 & 1.99 \\
\midrule
\multicolumn{9}{c}{\textbf{$\beta = 371.8$\quad(adoption $=31.6\%$)}} \\
\midrule
500 & 0.08 & 0.02 & 0.18 & 0.03 & -0.09 & 0.74 & -0.63 & 1.16 \\
1000 & 0.04 & 0.00 & 0.16 & 0.03 & 0.00 & 0.33 & -0.63 & 0.74 \\
2000 & 0.02 & 0.00 & 0.15 & 0.02 & 0.01 & 0.18 & -0.63 & 0.56 \\
4000 & 0.02 & 0.00 & 0.15 & 0.02 & -0.03 & 0.08 & -0.68 & 0.56 \\
\bottomrule
\end{tabular*}

\par\smallskip\begin{minipage}{\linewidth}\footnotesize
\emph{Notes:} Bias $=\mathrm{mean}(\hat\theta-\theta)$ and MSE $=\mathrm{mean}((\hat\theta-\theta)^2)$ over 500 replications. True $\theta_1=0.15$ and $\theta_2=1$. Adoption is the mean share of adopting households in the simulated data.
\end{minipage}
\end{singlespacing}
\end{table}

\begin{table}[H]
\begin{singlespacing}\centering\small
\caption{Simulation Results with Intercept $=-1.30$ and Extreme $\beta$: Bias and mean squared error of the peer-effect estimates under the CES and linear norms.}
\label{tab:sim_biasmse_ext_CL_b}
\renewcommand{\arraystretch}{0.95}\setlength{\tabcolsep}{3pt}
\begin{tabular*}{\linewidth}{@{\extracolsep{\fill}}ccccccccc@{}}
\toprule
& \multicolumn{4}{c}{$\theta_1$} & \multicolumn{4}{c}{$\theta_2$} \\
\cmidrule(lr){2-5}\cmidrule(lr){6-9}
& \multicolumn{2}{c}{CES} & \multicolumn{2}{c}{Linear} & \multicolumn{2}{c}{CES} & \multicolumn{2}{c}{Linear} \\
\cmidrule(lr){2-3}\cmidrule(lr){4-5}\cmidrule(lr){6-7}\cmidrule(lr){8-9}
$N$ & Bias & MSE & Bias & MSE & Bias & MSE & Bias & MSE \\
\midrule
\multicolumn{9}{c}{\textbf{$\beta = -386.0$\quad(adoption $=17.6\%$)}} \\
\midrule
500 & -0.09 & 0.06 & -0.13 & 0.04 & -0.64 & 3.64 & -0.75 & 3.51 \\
1000 & -0.06 & 0.03 & -0.11 & 0.02 & -0.31 & 2.46 & -0.44 & 2.31 \\
2000 & -0.07 & 0.02 & -0.11 & 0.02 & -0.17 & 1.95 & -0.26 & 1.65 \\
4000 & -0.05 & 0.01 & -0.10 & 0.01 & -0.03 & 1.14 & -0.19 & 1.07 \\
\midrule
\multicolumn{9}{c}{\textbf{$\beta = 371.8$\quad(adoption $=41.9\%$)}} \\
\midrule
500 & 0.06 & 0.01 & 0.12 & 0.02 & -0.07 & 0.82 & -0.59 & 1.04 \\
1000 & 0.03 & 0.00 & 0.11 & 0.01 & -0.03 & 0.34 & -0.65 & 0.71 \\
2000 & 0.02 & 0.00 & 0.11 & 0.01 & 0.01 & 0.22 & -0.64 & 0.56 \\
4000 & 0.01 & 0.00 & 0.11 & 0.01 & -0.01 & 0.09 & -0.65 & 0.50 \\
\bottomrule
\end{tabular*}

\par\smallskip\begin{minipage}{\linewidth}\footnotesize
\emph{Notes:} Bias $=\mathrm{mean}(\hat\theta-\theta)$ and MSE $=\mathrm{mean}((\hat\theta-\theta)^2)$ over 500 replications. True $\theta_1=0.15$ and $\theta_2=1$. Adoption is the mean share of adopting households in the simulated data.
\end{minipage}
\end{singlespacing}
\end{table}

\begin{table}[H]
\begin{singlespacing}\centering\small
\caption{Simulation Results with Intercept $=-1.70$ and Extreme $\beta$: 90\% Wald Coverage of Peer Effect Estimates under CES Norm.}
\label{tab:sim_coverage_ext_CL_a}
\renewcommand{\arraystretch}{0.95}\setlength{\tabcolsep}{3pt}
\begin{tabular*}{\linewidth}{@{\extracolsep{\fill}}cccccccc@{}}
\toprule
& & \multicolumn{3}{c}{$\theta_1$} & \multicolumn{3}{c}{$\theta_2$} \\
\cmidrule(lr){3-5}\cmidrule(lr){6-8}
$N$ & \% tail & All & Interior & Tail & All & Interior & Tail \\
\midrule
\multicolumn{8}{c}{\textbf{$\beta = -386.0$\quad(adoption $=11.8\%$)}} \\
\midrule
500 & 68\% & 92\% & 94\% & 91\% & 99\% & 100\% & 99\% \\
1000 & 70\% & 87\% & 87\% & 86\% & 99\% & 100\% & 98\% \\
2000 & 63\% & 80\% & 86\% & 77\% & 97\% & 99\% & 95\% \\
4000 & 62\% & 69\% & 70\% & 68\% & 90\% & 94\% & 88\% \\
\midrule
\multicolumn{8}{c}{\textbf{$\beta = 371.8$\quad(adoption $=31.6\%$)}} \\
\midrule
500 & 44\% & 69\% & 54\% & 89\% & 92\% & 92\% & 91\% \\
1000 & 44\% & 75\% & 61\% & 93\% & 94\% & 93\% & 95\% \\
2000 & 46\% & 75\% & 59\% & 95\% & 93\% & 94\% & 91\% \\
4000 & 51\% & 77\% & 62\% & 91\% & 90\% & 88\% & 92\% \\
\bottomrule
\end{tabular*}

\par\smallskip\begin{minipage}{\linewidth}\footnotesize
\emph{Notes:}  Coverage over 500 replications, split by the estimator's tail flag. \% tail is the share flagged as tail solutions where the profile likelihood is flat in $\beta$. Adoption is the mean share of adopting households in the simulated data.
\end{minipage}
\end{singlespacing}
\end{table}

\begin{table}[H]
\begin{singlespacing}\centering\small
\caption{Simulation Results with Intercept $=-1.30$ and Extreme $\beta$: 90\% Wald Coverage of Peer Effect Estimates under CES Norm.}
\label{tab:sim_coverage_ext_CL_b}
\renewcommand{\arraystretch}{0.95}\setlength{\tabcolsep}{3pt}
\begin{tabular*}{\linewidth}{@{\extracolsep{\fill}}cccccccc@{}}
\toprule
& & \multicolumn{3}{c}{$\theta_1$} & \multicolumn{3}{c}{$\theta_2$} \\
\cmidrule(lr){3-5}\cmidrule(lr){6-8}
$N$ & \% tail & All & Interior & Tail & All & Interior & Tail \\
\midrule
\multicolumn{8}{c}{\textbf{$\beta = -386.0$\quad(adoption $=17.6\%$)}} \\
\midrule
500 & 68\% & 83\% & 90\% & 79\% & 98\% & 99\% & 97\% \\
1000 & 59\% & 75\% & 78\% & 73\% & 92\% & 96\% & 90\% \\
2000 & 57\% & 64\% & 64\% & 63\% & 89\% & 93\% & 85\% \\
4000 & 57\% & 66\% & 59\% & 71\% & 89\% & 92\% & 86\% \\
\midrule
\multicolumn{8}{c}{\textbf{$\beta = 371.8$\quad(adoption $=41.9\%$)}} \\
\midrule
500 & 47\% & 72\% & 58\% & 88\% & 92\% & 91\% & 94\% \\
1000 & 52\% & 76\% & 58\% & 92\% & 94\% & 91\% & 96\% \\
2000 & 50\% & 77\% & 60\% & 93\% & 92\% & 91\% & 92\% \\
4000 & 53\% & 76\% & 62\% & 88\% & 89\% & 86\% & 92\% \\
\bottomrule
\end{tabular*}

\par\smallskip\begin{minipage}{\linewidth}\footnotesize
\emph{Notes:}  Coverage over 500 replications, split by the estimator's tail flag. \% tail is the share flagged as tail solutions where the profile likelihood is flat in $\beta$. Adoption is the mean share of adopting households in the simulated data.
\end{minipage}
\end{singlespacing}
\end{table}

\newpage

\subsection{Microfinance Estimation Tables}

In this section we present the tables related to the empirical exercise using the microfinance dataset of \citet{banerjee2013}. We report the summary statistics, the full pooled estimates (with and without the isolated-household dummy), the segment composition and the choice of wealth proxy, the full segment estimates for both partitions, and the subsidy-recipient characteristics for the illustrative heterogenous and rich segments.

\begin{table}[H]
\begin{singlespacing}\centering\small
\caption{Summary statistics for the Indian Microfinance Dataset of \citet{banerjee2013}.}
\label{tab:summary_stats}
\begin{tabular}{lrrrr}
\toprule
Variable & Mean & Std.\ dev. & Min & Max \\
\midrule
\multicolumn{5}{l}{\emph{Outcome}}\\
Adoption (took up a BSS loan) & 0.17 & 0.38 & 0 & 1 \\
\addlinespace
\multicolumn{5}{l}{\emph{Social network}}\\
Network degree & 5.75 & 4.73 & 0 & 48 \\
Isolated household ($\deg=0$) & 0.09 & 0.29 & 0 & 1 \\
\addlinespace
\multicolumn{5}{l}{\emph{Household characteristics}}\\
Number of rooms & 2.44 & 1.27 & 0 & 18 \\
Number of beds & 0.76 & 1.11 & 0 & 20 \\
Non-Hindu head of household & 0.05 & 0.22 & 0 & 1 \\
Scheduled caste or tribe & 0.33 & 0.47 & 0 & 1 \\
Owns a latrine & 0.26 & 0.44 & 0 & 1 \\
Village leader household & 0.12 & 0.33 & 0 & 1 \\
Rents (rather than owns) home & 0.06 & 0.23 & 0 & 1 \\
\bottomrule
\end{tabular}
\\[0.3em]\footnotesize\parbox{0.80\textwidth}{\emph{Notes:} 6,733 households in 29 villages from the 2006 baseline survey of \citet{banerjee2013}, measured before the BSS microfinance programme entered any village. Adoption is subsequent participation in the programme. Network degree is the number of households linked by a reported visiting tie; binary indicators take values in $\{0,1\}$, so their standard deviation is $\sqrt{\bar p(1-\bar p)}$. Covariate coding follows the household survey. All controls enter the private payoff alongside village fixed effects.}
\end{singlespacing}
\end{table}

\begin{table}[H]
\begin{singlespacing}\centering\small
\caption{Microfinance pooled sample estimates: With and Without the isolated-household dummy, under the CES and linear ($\beta=1$) norms.}
\label{tab:emp_full_app}
\setlength{\tabcolsep}{6pt}
\begin{tabular}{lcccc}
\toprule
 & \multicolumn{2}{c}{With isolated dummy} & \multicolumn{2}{c}{Without isolated dummy} \\
\cmidrule(lr){2-3}\cmidrule(lr){4-5}
 & Non-linear & Linear & Non-linear & Linear \\
\midrule
\multicolumn{5}{l}{\emph{Endogenous peer effects}} \\
Strategic complementarity ($\theta_1$) & 0.053$^{**}$ & 0.180$^{***}$ & 0.417$^{***}$ & 0.245$^{***}$ \\
 & (0.023) & (0.032) & (0.068) & (0.028) \\
Conformity ($\theta_2$) & 2.064$^{***}$ & 1.538$^{***}$ & -0.725$^{**}$ & -0.182 \\
 & (0.445) & (0.501) & (0.351) & (0.334) \\
CES elasticity ($\beta$) & 491.5$^{\dagger}$ & $1$ & $\to-\infty^{\dagger}$ & $1$ \\
 & (722.2) & &  & \\
\addlinespace
\multicolumn{5}{l}{\emph{Own effects}} \\
Number of rooms & -0.064$^{*}$ & -0.063$^{**}$ & -0.065$^{**}$ & -0.066$^{**}$ \\
 & (0.033) & (0.032) & (0.033) & (0.033) \\
Number of beds & -0.087$^{**}$ & -0.082$^{*}$ & -0.085$^{*}$ & -0.088$^{**}$ \\
 & (0.044) & (0.042) & (0.043) & (0.043) \\
Non-Hindu head of household & 0.869$^{***}$ & 0.877$^{***}$ & 1.221$^{***}$ & 1.147$^{***}$ \\
 & (0.149) & (0.154) & (0.163) & (0.154) \\
Scheduled caste or tribe & 0.375$^{***}$ & 0.348$^{***}$ & 0.465$^{***}$ & 0.444$^{***}$ \\
 & (0.069) & (0.072) & (0.077) & (0.077) \\
Owns a latrine & -0.365$^{***}$ & -0.348$^{***}$ & -0.381$^{***}$ & -0.375$^{***}$ \\
 & (0.091) & (0.090) & (0.093) & (0.093) \\
Village leader household & 0.562$^{***}$ & 0.567$^{***}$ & 0.588$^{***}$ & 0.580$^{***}$ \\
 & (0.095) & (0.094) & (0.097) & (0.097) \\
Rents (rather than owns) home & 0.026 & 0.028 & 0.019 & 0.010 \\
 & (0.143) & (0.138) & (0.145) & (0.143) \\
Isolated-household dummy & -1.033$^{***}$ & -0.894$^{***}$ & -- & -- \\
 & (0.189) & (0.214) &  &  \\
\midrule
Village fixed effects & Yes & Yes & Yes & Yes \\
Num.\ obs. & 6,733 & 6,733 & 6,733 & 6,733 \\
LR test $\beta=1$ (stat, $p$) & \multicolumn{2}{c}{4.63 (0.031)} & \multicolumn{2}{c}{8.74 (0.003)} \\
\bottomrule
\end{tabular}
\\[0.3em]\footnotesize\parbox{0.9\textwidth}{\emph{Notes:} All 29 villages pooled (6,733 households) with village fixed effects. CES columns leave the CES curvature $\beta$ free; linear columns impose $\beta=1$. $^{\dagger}$ marks a flat-tail fit, where $\beta$ is numerically indistinguishable with a flat profile likelihood and the presented value is the refined optimum in the flat region. $\theta_1$ is complementarity and $\theta_2$ conformity. Standard errors in parentheses, stars are Wald tests against zero, $^{*}p<0.10$, $^{**}p<0.05$, $^{***}p<0.01$. The LR statistic tests $\beta=1$ against a $\chi^2_1$ reference. CES SEs without the isolated-household dummy condition on the fixed minimum norm; uncertainty about $\beta$ and norm selection is excluded.} 
\end{singlespacing}
\end{table}

\begin{table}[H]
\begin{singlespacing}\centering\small
\caption{Segment composition for the Caste $\times$ Beds and Caste $\times$ Latrine
partitions.}
\label{tab:cluster_composition}
\setlength{\tabcolsep}{4pt}
\begin{tabular}{lcc|cccc|cc}
\toprule
Segment & Villages & $N$ & Caste min.\ share & Beds & Latrine & Rooms & Avg.\ degree & Adoption \\
\midrule
\multicolumn{9}{c}{\textbf{Caste $\times$ Beds partition}} \\
\midrule
Het/P & 8 & 1,616 & 0.514 & 0.529 & 0.190 & 2.271 & 5.55 & 19.0\% \\
Het/R & 6 & 1,387 & 0.397 & 0.865 & 0.252 & 2.539 & 6.11 & 20.3\% \\
Hom/P & 7 & 1,728 & 0.230 & 0.490 & 0.282 & 2.396 & 5.51 & 15.0\% \\
Hom/R & 8 & 2,002 & 0.223 & 1.110 & 0.308 & 2.557 & 5.86 & 16.0\% \\
\midrule
\multicolumn{9}{c}{\textbf{Caste $\times$ Latrine partition}} \\
\midrule
Het/P & 8 & 1,691 & 0.480 & 0.655 & 0.157 & 2.428 & 6.26 & 16.8\% \\
Het/R & 6 & 1,312 & 0.433 & 0.723 & 0.298 & 2.352 & 5.24 & 23.2\% \\
Hom/P & 7 & 1,718 & 0.228 & 0.767 & 0.175 & 2.480 & 6.00 & 12.4\% \\
Hom/R & 8 & 2,012 & 0.224 & 0.870 & 0.399 & 2.485 & 5.44 & 18.3\% \\
\bottomrule
\end{tabular}
\\[0.3em]\footnotesize\parbox{0.9\textwidth}{\emph{Notes:} Het/Hom = above/below the median village caste-minority share (low minority share = homogeneous); P/R = below/above the median on the wealth variable naming the partition. Columns 4--7 are village-population means of the household covariates; ``Avg.\ degree'' is mean network degree and ``Adoption'' the microfinance take-up rate.}
\end{singlespacing}
\end{table}

\begin{table}[H]
\begin{singlespacing}\centering\small
\caption{Separation of the two candidate wealth proxies.}
\label{tab:cluster_variant_choice}
\begin{tabular}{lcccc}
\toprule
Partition & Min.\ centroid dist. & Mean centroid dist. & Min.\ dist.\ (all $X$) & Smallest Seg. \\
\midrule
Caste $\times$ Beds & 1.26 & 1.72 & 1.37 & 1,387 \\
Caste $\times$ Latrine & 1.19 & 1.64 & 1.51 & 1,312 \\
\bottomrule
\end{tabular}
\\[0.3em]\footnotesize\parbox{0.9\textwidth}{\emph{Notes:} Centroid distances are between segment means, standardised. Larger is better separated. The two proxies are close with beds giving the larger minimum and mean separation on the segmenting variables and the larger smallest segment, latrine the larger minimum separation over the full covariate vector.}
\end{singlespacing}
\end{table}

\begin{table}[H]
\begin{singlespacing}\centering\small
\caption{Segment estimates Caste $\times$ Beds partition: CES (NL, $\beta$ free) and linear (L, $\beta=1$) with the isolated-household dummy.}
\label{tab:emp_caste_beds_full_app}
\setlength{\tabcolsep}{4pt}
\resizebox{\linewidth}{!}{%
\begin{tabular}{lcccccccc}
\toprule
 & \multicolumn{2}{c}{Het/P} & \multicolumn{2}{c}{Het/R} & \multicolumn{2}{c}{Hom/P} & \multicolumn{2}{c}{Hom/R} \\
\cmidrule(lr){2-3}\cmidrule(lr){4-5}\cmidrule(lr){6-7}\cmidrule(lr){8-9}
 & NL & L & NL & L & NL & L & NL & L \\
\midrule
\multicolumn{9}{l}{\emph{Endogenous peer effects}} \\
Strategic complementarity ($\theta_1$) & 0.287$^{**}$ & 0.150$^{**}$ & 0.013 & 0.136$^{**}$ & 0.506$^{***}$ & 0.209$^{***}$ & 0.076 & 0.230$^{***}$ \\
 & (0.125) & (0.067) & (0.038) & (0.061) & (0.176) & (0.065) & (0.064) & (0.062) \\
Conformity ($\theta_2$) & -0.317 & 0.638 & 2.680$^{***}$ & 1.488 & -1.192 & 0.224 & 1.742 & 1.390 \\
 & (1.565) & (1.337) & (0.833) & (0.945) & (1.045) & (1.068) & (1.589) & (1.128) \\
CES elasticity ($\beta$) & -83.1$^{\dagger}$ & $1$ & 491.5$^{\dagger}$ & $1$ & -296.7$^{\dagger}$ & $1$ & 787.7$^{\dagger}$ & $1$ \\
 & (126.2) &  & (1182.7) &  & (436.7) &  & (13855.2) &  \\
\addlinespace
\multicolumn{9}{l}{\emph{Own effects}} \\
Number of rooms & -0.058 & -0.058 & -0.063 & -0.061 & -0.061 & -0.058 & -0.035 & -0.038 \\
 & (0.069) & (0.069) & (0.060) & (0.060) & (0.086) & (0.086) & (0.057) & (0.054) \\
Number of beds & -0.182 & -0.176 & -0.069 & -0.072 & -0.242$^{*}$ & -0.250$^{*}$ & -0.057 & -0.055 \\
 & (0.116) & (0.114) & (0.087) & (0.084) & (0.130) & (0.130) & (0.058) & (0.054) \\
Non-Hindu head of household & 1.085$^{**}$ & 0.950$^{**}$ & 0.645$^{***}$ & 0.700$^{***}$ & 2.110$^{***}$ & 1.824$^{***}$ & 0.640$^{**}$ & 0.633$^{**}$ \\
 & (0.441) & (0.396) & (0.201) & (0.252) & (0.439) & (0.519) & (0.321) & (0.271) \\
Scheduled caste or tribe & 0.117 & 0.110 & 0.581$^{***}$ & 0.567$^{***}$ & 0.943$^{***}$ & 0.842$^{***}$ & -0.013 & -0.023 \\
 & (0.152) & (0.141) & (0.155) & (0.176) & (0.164) & (0.183) & (0.145) & (0.135) \\
Owns a latrine & -0.153 & -0.148 & -0.475$^{**}$ & -0.451$^{**}$ & -0.485$^{**}$ & -0.481$^{**}$ & -0.343$^{**}$ & -0.331$^{**}$ \\
 & (0.198) & (0.194) & (0.200) & (0.197) & (0.227) & (0.227) & (0.148) & (0.146) \\
Village leader household & 0.485$^{**}$ & 0.489$^{**}$ & 0.445$^{**}$ & 0.464$^{**}$ & 0.657$^{***}$ & 0.651$^{***}$ & 0.701$^{***}$ & 0.684$^{***}$ \\
 & (0.193) & (0.191) & (0.208) & (0.206) & (0.211) & (0.209) & (0.166) & (0.171) \\
Rents (rather than owns) home & 0.185 & 0.169 & 0.002 & 0.009 & -0.374 & -0.328 & 0.174 & 0.188 \\
 & (0.303) & (0.299) & (0.277) & (0.281) & (0.348) & (0.333) & (0.243) & (0.243) \\
Isolated-household dummy & -0.510 & -0.840$^{*}$ & -1.135$^{***}$ & -0.870$^{**}$ & 0.323 & -0.229 & -1.061$^{**}$ & -0.946$^{**}$ \\
 & (0.517) & (0.443) & (0.340) & (0.403) & (0.494) & (0.450) & (0.497) & (0.447) \\
\midrule
Village fixed effects & \multicolumn{2}{c}{Yes} & \multicolumn{2}{c}{Yes} & \multicolumn{2}{c}{Yes} & \multicolumn{2}{c}{Yes} \\
Num.\ obs. & \multicolumn{2}{c}{1,616} & \multicolumn{2}{c}{1,387} & \multicolumn{2}{c}{1,728} & \multicolumn{2}{c}{2,002} \\
LR test $\beta=1$ (stat, $p$) & \multicolumn{2}{c}{1.81 (0.179)} & \multicolumn{2}{c}{3.39 (0.066)} & \multicolumn{2}{c}{1.56 (0.212)} & \multicolumn{2}{c}{0.75 (0.387)} \\
\bottomrule
\end{tabular}%
}
\\[0.3em]\footnotesize\parbox{0.95\textwidth}{\emph{Notes:} Segments are Het/Hom (heterogeneous vs homogeneous on caste-minority share) $\times$ P/R (poor vs rich), split at village-level medians of the Caste $\times$ Beds partition. NL leaves the CES curvature $\beta$ free; L imposes $\beta=1$. $^{\dagger}$ marks a flat-tail fit ($\beta$ is numerically indistinguishable; the value is the refined estimate in the plateau). $\theta_1$ is complementarity, $\theta_2$ conformity. Standard errors in parentheses where computed; stars are Wald tests vs zero, $^{*}p<0.10$, $^{**}p<0.05$, $^{***}p<0.01$. The LR statistic tests $\beta=1$ against a $\chi^2_1$ reference. A dash marks a control dropped for lack of within-segment variation.}
\end{singlespacing}
\end{table}

\begin{table}[H]
\begin{singlespacing}\centering\small
\caption{Segment estimates Caste $\times$ Latrine partition: CES (NL, $\beta$ free) and linear (L, $\beta=1$) with the isolated-household dummy.}
\label{tab:emp_caste_latrine_full_app}
\setlength{\tabcolsep}{4pt}
\resizebox{\linewidth}{!}{%
\begin{tabular}{lcccccccc}
\toprule
 & \multicolumn{2}{c}{Het/P} & \multicolumn{2}{c}{Het/R} & \multicolumn{2}{c}{Hom/P} & \multicolumn{2}{c}{Hom/R} \\
\cmidrule(lr){2-3}\cmidrule(lr){4-5}\cmidrule(lr){6-7}\cmidrule(lr){8-9}
 & NL & L & NL & L & NL & L & NL & L \\
\midrule
\multicolumn{9}{l}{\emph{Endogenous peer effects}} \\
Strategic complementarity ($\theta_1$) & 0.155 & 0.105$^{*}$ & 0.321$^{**}$ & 0.142$^{**}$ & 0.351$^{*}$ & 0.168$^{**}$ & 0.232$^{***}$ & 0.248$^{***}$ \\
 & (0.135) & (0.060) & (0.155) & (0.066) & (0.209) & (0.075) & (0.070) & (0.057) \\
Conformity ($\theta_2$) & 2.747$^{***}$ & 3.083$^{***}$ & -0.057 & 1.019 & 4.035$^{*}$ & 3.376$^{***}$ & 0.779 & 0.671 \\
 & (1.062) & (0.837) & (1.997) & (1.172) & (2.155) & (1.190) & (1.126) & (0.895) \\
CES elasticity ($\beta$) & -0.97 & $1$ & -282.8$^{\dagger}$ & $1$ & -4.05 & $1$ & 2.00 & $1$ \\
 & (4.3) &  & (756.4) &  & (5.9) &  & (2.1) &  \\
\addlinespace
\multicolumn{9}{l}{\emph{Own effects}} \\
Number of rooms & -0.010 & -0.010 & -0.123$^{*}$ & -0.116$^{*}$ & 0.016 & 0.017 & -0.114$^{*}$ & -0.114$^{*}$ \\
 & (0.060) & (0.060) & (0.067) & (0.067) & (0.067) & (0.069) & (0.058) & (0.058) \\
Number of beds & -0.143$^{*}$ & -0.140$^{*}$ & -0.075 & -0.074 & -0.021 & -0.021 & -0.083 & -0.082 \\
 & (0.084) & (0.084) & (0.115) & (0.111) & (0.062) & (0.063) & (0.069) & (0.068) \\
Non-Hindu head of household & 1.354$^{**}$ & 1.065$^{***}$ & 0.734$^{**}$ & 0.638$^{**}$ & -- & -- & 1.063$^{***}$ & 1.082$^{***}$ \\
 & (0.644) & (0.323) & (0.351) & (0.261) &  &  & (0.411) & (0.327) \\
Scheduled caste or tribe & 0.303$^{**}$ & 0.290$^{**}$ & 0.281 & 0.254$^{*}$ & 0.384$^{**}$ & 0.377$^{**}$ & 0.291$^{**}$ & 0.293$^{**}$ \\
 & (0.153) & (0.127) & (0.189) & (0.153) & (0.159) & (0.148) & (0.146) & (0.139) \\
Owns a latrine & -0.268 & -0.276 & -0.340$^{*}$ & -0.325$^{*}$ & -0.459$^{*}$ & -0.537$^{**}$ & -0.347$^{**}$ & -0.347$^{**}$ \\
 & (0.222) & (0.225) & (0.186) & (0.172) & (0.241) & (0.233) & (0.144) & (0.144) \\
Village leader household & 0.422$^{**}$ & 0.420$^{**}$ & 0.546$^{***}$ & 0.542$^{***}$ & 0.653$^{***}$ & 0.629$^{***}$ & 0.649$^{***}$ & 0.650$^{***}$ \\
 & (0.195) & (0.189) & (0.205) & (0.203) & (0.216) & (0.204) & (0.162) & (0.166) \\
Rents (rather than owns) home & 0.463 & 0.436 & -0.220 & -0.220 & 0.438 & 0.394 & -0.162 & -0.161 \\
 & (0.311) & (0.290) & (0.279) & (0.273) & (0.396) & (0.391) & (0.220) & (0.218) \\
Isolated-household dummy & -1.837$^{***}$ & -1.957$^{***}$ & -0.245 & -0.592 & -1.713$^{*}$ & -1.508$^{***}$ & -0.572 & -0.547 \\
 & (0.503) & (0.453) & (0.502) & (0.377) & (0.900) & (0.568) & (0.410) & (0.370) \\
\midrule
Village fixed effects & \multicolumn{2}{c}{Yes} & \multicolumn{2}{c}{Yes} & \multicolumn{2}{c}{Yes} & \multicolumn{2}{c}{Yes} \\
Num.\ obs. & \multicolumn{2}{c}{1,691} & \multicolumn{2}{c}{1,312} & \multicolumn{2}{c}{1,718} & \multicolumn{2}{c}{2,012} \\
LR test $\beta=1$ (stat, $p$) & \multicolumn{2}{c}{0.67 (0.412)} & \multicolumn{2}{c}{3.16 (0.075)} & \multicolumn{2}{c}{5.23 (0.022)} & \multicolumn{2}{c}{0.35 (0.556)} \\
\bottomrule
\end{tabular}%
}
\\[0.3em]\footnotesize\parbox{0.95\textwidth}{\emph{Notes:} Segments are Het/Hom (heterogeneous vs homogeneous on caste-minority share) $\times$ P/R (poor vs rich), split at village-level medians of the Caste $\times$ Latrine partition. NL leaves the CES curvature $\beta$ free; L imposes $\beta=1$. $^{\dagger}$ marks a flat-tail fit ($\beta$ is numerically indistinguishable; the value is the refined estimate in the plateau). $\theta_1$ is complementarity, $\theta_2$ conformity. Standard errors in parentheses where computed; stars are Wald tests vs zero, $^{*}p<0.10$, $^{**}p<0.05$, $^{***}p<0.01$. The LR statistic tests $\beta=1$ against a $\chi^2_1$ reference. A dash marks a control dropped for lack of within-segment variation.}
\end{singlespacing}
\end{table}

\begin{table}[H]
\begin{singlespacing}\centering\small
\caption{Segment estimates Caste $\times$ Beds partition: CES (NL, $\beta$ free) and linear (L, $\beta=1$) without the isolated-household dummy.}
\label{tab:emp_caste_beds_dummy_off_full_app}
\setlength{\tabcolsep}{4pt}
\resizebox{\linewidth}{!}{%
\begin{tabular}{lcccccccc}
\toprule
 & \multicolumn{2}{c}{Het/P} & \multicolumn{2}{c}{Het/R} & \multicolumn{2}{c}{Hom/P} & \multicolumn{2}{c}{Hom/R} \\
\cmidrule(lr){2-3}\cmidrule(lr){4-5}\cmidrule(lr){6-7}\cmidrule(lr){8-9}
 & NL & L & NL & L & NL & L & NL & L \\
\midrule
\multicolumn{9}{l}{\emph{Endogenous peer effects}} \\
Strategic complementarity ($\theta_1$) & 0.316$^{**}$ & 0.207$^{***}$ & 0.353$^{**}$ & 0.208$^{***}$ & 0.354 & 0.226$^{***}$ & 0.286$^{***}$ & 0.275$^{***}$ \\
 & (0.144) & (0.061) & (0.162) & (0.050) & (0.286) & (0.056) & (0.077) & (0.060) \\
Conformity ($\theta_2$) & -1.354 & -1.179 & -0.854 & 0.255 & -0.557 & -0.189 & -0.781 & -0.718 \\
 & (0.973) & (0.861) & (0.808) & (0.626) & (0.963) & (0.634) & (0.840) & (0.722) \\
CES elasticity ($\beta$) & -491.5$^{\dagger}$ & $1$ & -80.6$^{\dagger}$ & $1$ & -5.06 & $1$ & 0.00 & $1$ \\
 & (531.4) &  & (109.9) &  & (25.1) &  & (2.4) &  \\
\addlinespace
\multicolumn{9}{l}{\emph{Own effects}} \\
Number of rooms & -0.061 & -0.064 & -0.064 & -0.069 & -0.059 & -0.056 & -0.040 & -0.040 \\
 & (0.071) & (0.070) & (0.061) & (0.061) & (0.086) & (0.086) & (0.057) & (0.056) \\
Number of beds & -0.183 & -0.178 & -0.078 & -0.072 & -0.244$^{*}$ & -0.252$^{*}$ & -0.057 & -0.057 \\
 & (0.116) & (0.114) & (0.086) & (0.085) & (0.138) & (0.130) & (0.056) & (0.056) \\
Non-Hindu head of household & 1.216$^{***}$ & 1.221$^{***}$ & 0.880$^{***}$ & 0.771$^{***}$ & 2.030$^{***}$ & 1.951$^{***}$ & 0.894$^{***}$ & 0.893$^{***}$ \\
 & (0.398) & (0.380) & (0.312) & (0.278) & (0.490) & (0.429) & (0.307) & (0.281) \\
Scheduled caste or tribe & 0.142 & 0.150 & 0.776$^{***}$ & 0.673$^{***}$ & 0.919$^{***}$ & 0.876$^{***}$ & -0.013 & -0.011 \\
 & (0.162) & (0.161) & (0.185) & (0.167) & (0.180) & (0.171) & (0.162) & (0.161) \\
Owns a latrine & -0.170 & -0.170 & -0.484$^{**}$ & -0.466$^{**}$ & -0.483$^{**}$ & -0.490$^{**}$ & -0.344$^{**}$ & -0.344$^{**}$ \\
 & (0.199) & (0.197) & (0.208) & (0.200) & (0.222) & (0.227) & (0.154) & (0.154) \\
Village leader household & 0.485$^{**}$ & 0.496$^{***}$ & 0.481$^{**}$ & 0.460$^{**}$ & 0.658$^{***}$ & 0.657$^{***}$ & 0.717$^{***}$ & 0.716$^{***}$ \\
 & (0.192) & (0.192) & (0.214) & (0.212) & (0.211) & (0.209) & (0.169) & (0.172) \\
Rents (rather than owns) home & 0.177 & 0.147 & 0.020 & 0.034 & -0.360 & -0.342 & 0.173 & 0.170 \\
 & (0.305) & (0.302) & (0.299) & (0.292) & (0.343) & (0.333) & (0.254) & (0.253) \\
\midrule
Village fixed effects & \multicolumn{2}{c}{Yes} & \multicolumn{2}{c}{Yes} & \multicolumn{2}{c}{Yes} & \multicolumn{2}{c}{Yes} \\
Num.\ obs. & \multicolumn{2}{c}{1,616} & \multicolumn{2}{c}{1,387} & \multicolumn{2}{c}{1,728} & \multicolumn{2}{c}{2,002} \\
LR test $\beta=1$ (stat, $p$) & \multicolumn{2}{c}{4.12 (0.042)} & \multicolumn{2}{c}{2.13 (0.144)} & \multicolumn{2}{c}{1.06 (0.303)} & \multicolumn{2}{c}{0.15 (0.695)} \\
\bottomrule
\end{tabular}%
}
\\[0.3em]\footnotesize\parbox{0.95\textwidth}{\emph{Notes:} Segments are Het/Hom (heterogeneous vs homogeneous on caste-minority share) $\times$ P/R (poor vs rich), split at village-level medians of the Caste $\times$ Beds partition. NL leaves the CES curvature $\beta$ free; L imposes $\beta=1$. The isolated-household dummy is excluded. $^{\dagger}$ marks a flat-tail fit ($\beta$ is numerically indistinguishable; the value is the refined estimate in the plateau). $\theta_1$ is complementarity, $\theta_2$ conformity. Standard errors in parentheses where computed; stars are Wald tests vs zero, $^{*}p<0.10$, $^{**}p<0.05$, $^{***}p<0.01$. The LR statistic tests $\beta=1$ against a $\chi^2_1$ reference. A dash marks a control dropped for lack of within-segment variation.}
\end{singlespacing}
\end{table}

\begin{table}[H]
\begin{singlespacing}\centering\small
\caption{Segment estimates Caste $\times$ Latrine partition: CES (NL, $\beta$ free) and linear (L, $\beta=1$) without the isolated-household dummy.}
\label{tab:emp_caste_latrine_dummy_off_full_app}
\setlength{\tabcolsep}{4pt}
\resizebox{\linewidth}{!}{%
\begin{tabular}{lcccccccc}
\toprule
 & \multicolumn{2}{c}{Het/P} & \multicolumn{2}{c}{Het/R} & \multicolumn{2}{c}{Hom/P} & \multicolumn{2}{c}{Hom/R} \\
\cmidrule(lr){2-3}\cmidrule(lr){4-5}\cmidrule(lr){6-7}\cmidrule(lr){8-9}
 & NL & L & NL & L & NL & L & NL & L \\
\midrule
\multicolumn{9}{l}{\emph{Endogenous peer effects}} \\
Strategic complementarity ($\theta_1$) & 0.338$^{***}$ & 0.248$^{***}$ & 0.319$^{*}$ & 0.188$^{***}$ & 0.560$^{**}$ & 0.228$^{***}$ & 0.294$^{***}$ & 0.291$^{***}$ \\
 & (0.075) & (0.045) & (0.169) & (0.059) & (0.247) & (0.076) & (0.065) & (0.051) \\
Conformity ($\theta_2$) & 0.001 & 0.104 & -0.732 & -0.377 & -0.017 & -0.061 & -0.294 & -0.287 \\
 & (0.753) & (0.582) & (1.229) & (0.828) & (0.735) & (0.711) & (0.807) & (0.612) \\
CES elasticity ($\beta$) & -2.28 & $1$ & -491.5$^{\dagger}$ & $1$ & -298.2$^{\dagger}$ & $1$ & 0.76 & $1$ \\
 & (2.0) &  & (545.4) &  & (191.6) &  & (2.8) &  \\
\addlinespace
\multicolumn{9}{l}{\emph{Own effects}} \\
Number of rooms & -0.027 & -0.027 & -0.125$^{*}$ & -0.119$^{*}$ & 0.020 & 0.026 & -0.114$^{*}$ & -0.114$^{*}$ \\
 & (0.064) & (0.063) & (0.069) & (0.069) & (0.073) & (0.073) & (0.059) & (0.059) \\
Number of beds & -0.157$^{*}$ & -0.155$^{*}$ & -0.078 & -0.078 & -0.023 & -0.025 & -0.084 & -0.085 \\
 & (0.086) & (0.086) & (0.114) & (0.113) & (0.068) & (0.067) & (0.070) & (0.070) \\
Non-Hindu head of household & 1.640$^{***}$ & 1.505$^{***}$ & 0.795$^{***}$ & 0.775$^{***}$ & -- & -- & 1.281$^{***}$ & 1.280$^{***}$ \\
 & (0.455) & (0.417) & (0.302) & (0.266) &  &  & (0.357) & (0.290) \\
Scheduled caste or tribe & 0.460$^{***}$ & 0.482$^{***}$ & 0.314$^{*}$ & 0.323$^{*}$ & 0.575$^{***}$ & 0.589$^{***}$ & 0.337$^{**}$ & 0.335$^{**}$ \\
 & (0.150) & (0.145) & (0.181) & (0.166) & (0.155) & (0.170) & (0.149) & (0.143) \\
Owns a latrine & -0.307 & -0.325 & -0.353$^{**}$ & -0.348$^{**}$ & -0.548$^{**}$ & -0.576$^{**}$ & -0.357$^{**}$ & -0.357$^{**}$ \\
 & (0.232) & (0.238) & (0.179) & (0.175) & (0.244) & (0.261) & (0.144) & (0.145) \\
Village leader household & 0.384$^{*}$ & 0.385$^{*}$ & 0.550$^{***}$ & 0.555$^{***}$ & 0.666$^{***}$ & 0.665$^{***}$ & 0.665$^{***}$ & 0.664$^{***}$ \\
 & (0.202) & (0.200) & (0.206) & (0.206) & (0.219) & (0.217) & (0.165) & (0.166) \\
Rents (rather than owns) home & 0.482 & 0.462 & -0.234 & -0.245 & 0.433 & 0.400 & -0.186 & -0.187 \\
 & (0.305) & (0.298) & (0.285) & (0.281) & (0.429) & (0.421) & (0.222) & (0.221) \\
\midrule
Village fixed effects & \multicolumn{2}{c}{Yes} & \multicolumn{2}{c}{Yes} & \multicolumn{2}{c}{Yes} & \multicolumn{2}{c}{Yes} \\
Num.\ obs. & \multicolumn{2}{c}{1,691} & \multicolumn{2}{c}{1,312} & \multicolumn{2}{c}{1,718} & \multicolumn{2}{c}{2,012} \\
LR test $\beta=1$ (stat, $p$) & \multicolumn{2}{c}{4.29 (0.038)} & \multicolumn{2}{c}{4.50 (0.034)} & \multicolumn{2}{c}{3.90 (0.048)} & \multicolumn{2}{c}{0.07 (0.796)} \\
\bottomrule
\end{tabular}%
}
\\[0.3em]\footnotesize\parbox{0.95\textwidth}{\emph{Notes:} Segments are Het/Hom (heterogeneous vs homogeneous on caste-minority share) $\times$ P/R (poor vs rich), split at village-level medians of the Caste $\times$ Latrine partition. NL leaves the CES curvature $\beta$ free; L imposes $\beta=1$. The isolated-household dummy is excluded. $^{\dagger}$ marks a flat-tail fit ($\beta$ is numerically indistinguishable; the value is the refined estimate in the plateau). $\theta_1$ is complementarity, $\theta_2$ conformity. Standard errors in parentheses where computed; stars are Wald tests vs zero, $^{*}p<0.10$, $^{**}p<0.05$, $^{***}p<0.01$. The LR statistic tests $\beta=1$ against a $\chi^2_1$ reference. A dash marks a control dropped for lack of within-segment variation.}
\end{singlespacing}
\end{table}

\begin{table}[H]
\begin{singlespacing}\centering\small
\caption{Household characteristics by subsidy-recipient group in the Het/R segments.}
\label{tab:het_r_targeting_characteristics}
\setlength{\tabcolsep}{5pt}
\begin{tabular}{lrrrrrrrr}
\toprule
Recipient group & $N$ & Degree & Rooms & Beds & Non-Hindu & SC/ST & Latrine & Leader \\
\midrule
\multicolumn{9}{c}{\textbf{Caste $\times$ beds: Het/R}} \\
\midrule
CES only    & 10    & 16.10 & 2.00 & 1.10 & 20.0\% & 60.0\% & 10.0\% & 40.0\% \\
Linear only & 18    & 15.72 & 3.22 & 1.33 & 0.0\%  & 61.1\% & 33.3\% & 5.6\% \\
Both        & 5     & 23.40 & 3.00 & 1.00 & 0.0\%  & 80.0\% & 0.0\%  & 40.0\% \\
Neither     & 1,354 & 5.85  & 2.53 & 0.86 & 5.8\%  & 39.1\% & 25.3\% & 11.2\% \\
\midrule
\multicolumn{9}{c}{\textbf{Caste $\times$ latrine: Het/R}} \\
\midrule
CES only    & 92    & 6.04  & 3.40 & 1.16 & 13.0\% & 51.1\% & 57.6\% & 3.3\% \\
Linear only & 23    & 13.61 & 2.30 & 0.57 & 21.7\% & 69.6\% & 34.8\% & 26.1\% \\
Both        & 1     & 12.00 & 4.00 & 2.00 & 0.0\%  & 100.0\% & 100.0\% & 0.0\% \\
Neither     & 1,196 & 5.01  & 2.27 & 0.69 & 10.1\% & 42.1\% & 27.5\% & 12.5\% \\
\bottomrule
\end{tabular}

\par\smallskip\begin{minipage}{0.98\linewidth}\footnotesize
\emph{Notes:} Entries are group means; binary characteristics are percentages. A household is a recipient if its subsidy exceeds $10^{-6}$. Each segment receives a budget of 20 utility-index units, and each policy uses its own estimates. The CES estimates are flat-tail fits ($\hat\beta=491.5$ for beds and $\hat\beta=-282.8$ for latrine).
\end{minipage}
\end{singlespacing}
\end{table}

\newpage

\subsection{Microfinance Estimation Figures}

This section presents the figures for the empirical exercise, plotting the profiled log-likelihood in $\beta$ by segment for both partitions and the CES-versus-linear optimal-subsidy comparison for the illustrative Het/R segments.

\begin{figure}[H]\centering
\includegraphics[width=\textwidth]{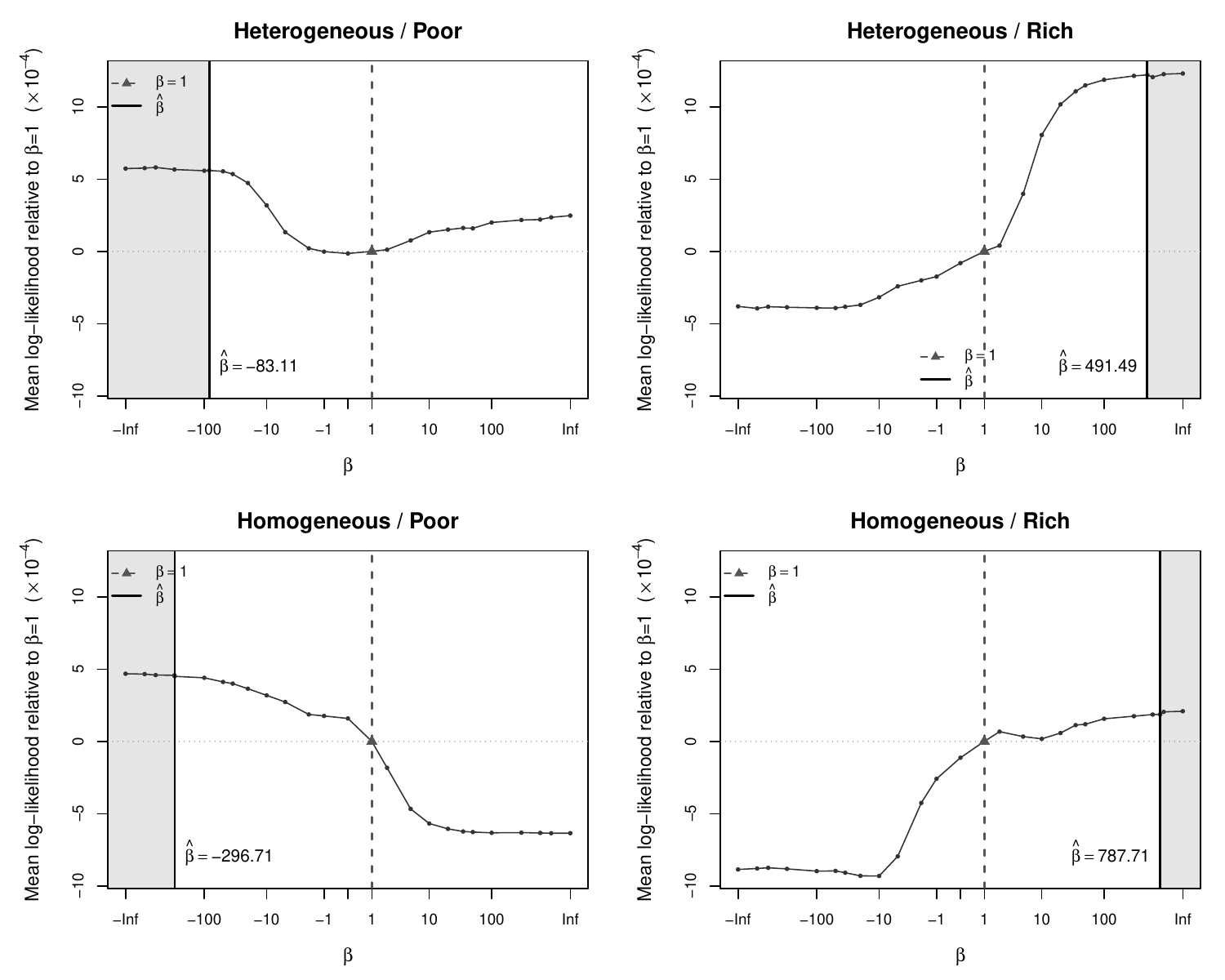}
\caption{Profiled log-likelihood in $\beta$ by segment, Caste $\times$ Beds.}
\par\smallskip\begin{minipage}{0.98\linewidth}\footnotesize
\emph{Notes:} The y-axis is the Log-Likelihood of the unrestricted CES model relative to the linear case. Shaded region denotes the flat tail-region where $\beta$ is numerically indistinguishable.
\label{fig:lik_caste_beds}
\end{minipage}
\end{figure}

\begin{figure}[H]\centering
\includegraphics[width=\textwidth]{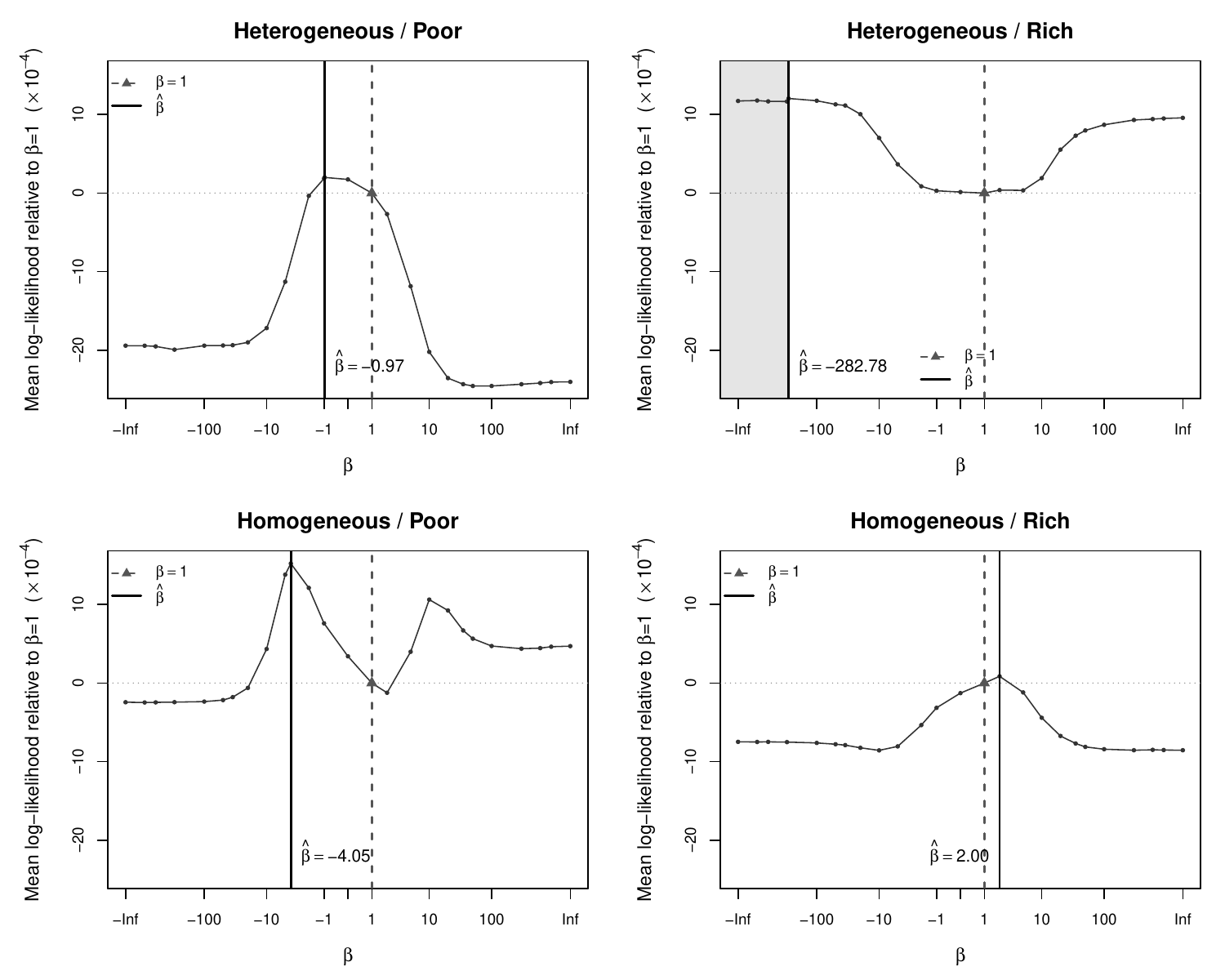}
\caption{Profiled log-likelihood in $\beta$ by segment, Caste $\times$ Latrine.}
\par\smallskip\begin{minipage}{0.98\linewidth}\footnotesize
\emph{Notes:} The y-axis is the Log-Likelihood of the unrestricted CES model relative to the linear case. Shaded region denotes the flat tail-region where $\beta$ is numerically indistinguishable.
\label{fig:lik_caste_latrine}
\end{minipage}
\end{figure}

\begin{figure}[p]
  \centering
  \captionsetup[subfigure]{position=top,font=small,labelfont=bf,textfont=bf,justification=centering,width=\linewidth,skip=3pt}
  \begin{subfigure}[t]{0.96\linewidth}
    \centering
    \includegraphics[width=\linewidth]{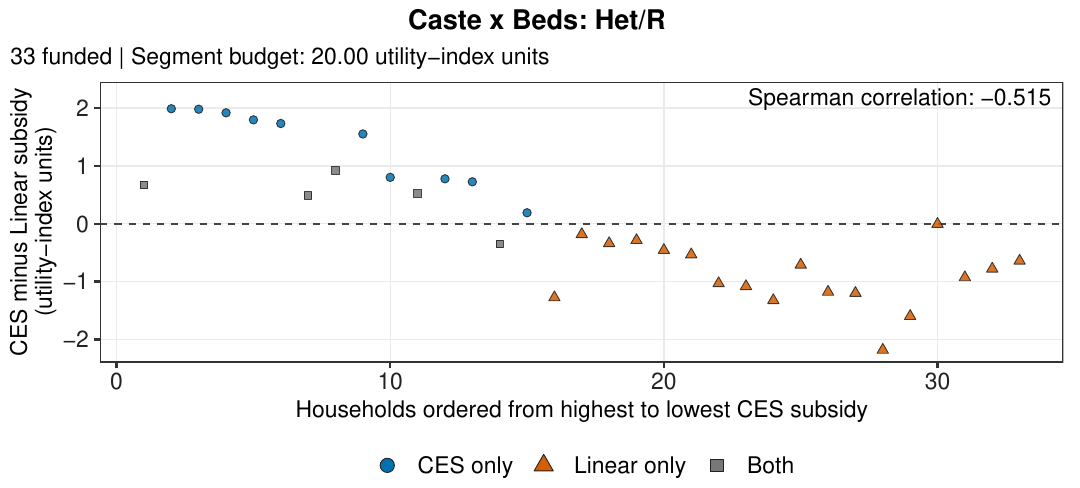}\par
    \begin{minipage}{\linewidth}\footnotesize
      \textit{Notes:} Linear adoption rises by 1.35 percentage points (0.33 direct, 1.01 indirect); CES adoption rises by 13.29 points (0.33 direct, 12.96 indirect).
    \end{minipage}
  \end{subfigure}\par\medskip
  \begin{subfigure}[t]{0.96\linewidth}
    \centering
    \includegraphics[width=\linewidth]{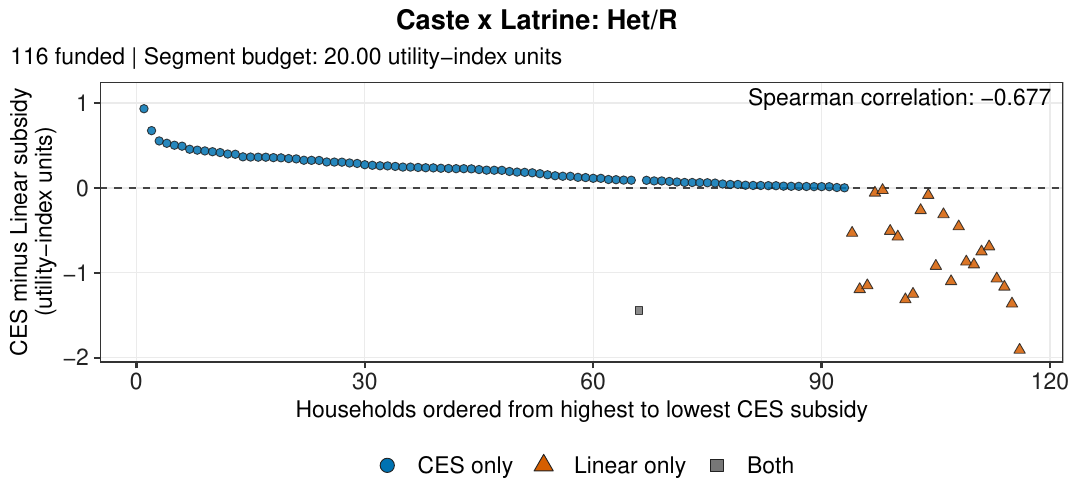}\par
    \begin{minipage}{\linewidth}\footnotesize
      \textit{Notes:} Linear adoption rises by 1.05 percentage points (0.37 direct, 0.68 indirect); CES adoption rises by 0.91 points (0.28 direct, 0.63 indirect).
    \end{minipage}
  \end{subfigure}
  \caption{Household subsidy differences under CES and Linear norms for the two Het/R segmentations. Each model uses its own estimates and stationary allocation. Within each segment, households funded by either model are ordered by CES subsidy; Spearman correlations are computed over this funded union, excluding households unfunded under both. Each segment receives $B=20$ utility-index units. Direct gains hold baseline peer probabilities fixed; indirect gains are the remaining equilibrium response.}
  \label{fig:het-r-subsidy-ranks}
  
\end{figure}

\end{document}